\UseRawInputEncoding
\documentclass[trackchanges,twocolumn]{aastex7}
\usepackage{verbatim}

\begin{document}

\title{Investigating the physical properties of dusty star-forming galaxies at $z\gtrsim1.5$ in the GOODS-South field using JWST}

\author{Dipanjan Mitra}
\affiliation{School of Physics and Astronomy, Cardiff University, The Parade, CF24 3AA, UK}
\email[show]{MitraD@cardiff.ac.uk, dipanmitra1@gmail.com}  

\author{Mattia Negrello} 
\affiliation{School of Physics and Astronomy, Cardiff University, The Parade, CF24 3AA, UK}
\email{negrellomattia@gmail.com}

\author{Gianfranco De Zotti}
\affiliation{INAF, Osservatorio Astronomico di Padova, Vicolo Osservatorio 5, I-35122, Padova, Italy}
\email{gianfranco.dezotti@inaf.it}

\begin{abstract}
We investigated how well the physical properties of progenitors of present-day massive spheroidal galaxies (proto-spheroids) can be constrained by the \textit{JWST} Advanced Deep Extragalactic Survey (JADES) in the GOODS-South field, which benefits from extensive photometric and spectroscopic data, including those from the \textit{Hubble}, \textit{Spitzer} and \textit{Herschel}. We adopted a physical model for the evolution of proto-spheroidal galaxies, which form the bulk of dusty star-forming galaxies (DSFGs) at $z\gtrsim1.5$ 
and confirmed its consistency with recent 
mid-infrared high$-z$ galaxy luminosity functions. 
Using the model and the JADES survey strategy, we simulated a sample of proto-spheroids over $87.5\,\hbox{arcmin}^2$, matching the JADES/GOODS-S survey area. 
Photometric redshifts estimated from simulated \textit{JWST} photometry showed $\gtrsim 95$ $\%$ accuracy
and were used in SED fitting with CIGALE. 
We demonstrated that \textit{JWST} will provide reliable stellar mass estimates up to $0.1$ dex for the majority of proto-spheroids at $z\gtrsim1.5$ and can detect low-mass systems during cosmic noon that were inaccessible in the pre-{\it JWST} era. 
Focusing on the active star-forming phase of the proto-spheroid evolution, we defined a sub-sample flux limited at $250\,\mu$m (DSFG sample) and derived SFR, dust luminosity and dust mass complementing the \textit{JWST} photometry with that from \textit{Spitzer}/MIPS and \textit{Herschel}. 
We also constructed a {\it JWST}-selected DSFG catalog from ASTRODEEP data using NIRCam colour criteria and demonstrated strong consistency between the observed and simulated DSFG populations.
%
\end{abstract}



\section{Introduction}

The discovery of dusty star-forming galaxies (DSFGs), also known as sub-millimetre galaxies \citep[SMGs;][]{smail_deep_1997, Barger1998, Hughes1998} revolutionised our understanding of galaxy evolution and the study of these high-$z$ dusty sources became a crucial area of extragalactic astronomy.
It is now widely agreed, based on several observational evidences, that high-$z$ SMGs can be interpreted as  being the progenitors of present-day massive spheroidal galaxies \citep{Lilly1999, Swinbank2006, Hainline2011, Toft2014, Simpson2014, Simpson2017, Liao2024, Tan2024, Birkin2024, Amvrosiadis2025}; in the following, these will be referred to as proto-spheroidal galaxies or proto-spheroids. This interpretation is consistent with the relatively old ages of stellar populations of nearby early-type galaxies \citep{Thomas2010, Bernardi2010, Lu2023}, implying that these galaxies formed the bulk of their stars at early times ($\log(\hbox{age}/\hbox{yr})\gtrsim 9.5$ corresponding to $z\gtrsim 1.5$), on a relatively short timescale, with high SFRs, and were quenched afterwards. 

The fact that SMGs are frequently found to be gas-rich rotation-dominated disks \citep[e.g.,][]{Gullberg2019, Gillman2023} is not in conflict with this interpretation. The extensive analysis of ALMA sub-millimeter surface brightness profiles of SMGs by \citet{Tan2024} has highlighted that most of these galaxies are fully triaxial rather than flat disks. Other studies have pointed out that disky SMGs are dynamically hotter than local disc galaxies (i.e., have lower ratios of rotation to random velocity) and with a larger fraction of irregular morphologies \citep{Lelli2016, Swinbank2017, Turner2017, Wisnioski2019, ForsterSchreiberWuyts2020, Kartaltepe2023, Birkin2024}. 

The pathway from SMGs to evolved spheroidal galaxies, implying kinematic and morphological evolution, is, however, still not completely clear. In our reference scenario \citep{granato_physical_2004, cai_hybrid_2013, Lapi2014}, high-$z$ SMGs evolve to the galaxy main sequence,  with lower dust-to-stellar-mass ratios and substantially lower specific SFR compared to SMGs \citep{Mancuso2016}, and end in passive evolution. In this process their stellar mass grows while their dust obscuration decreases as a result of supernova and AGN feedback, which sweep off the interstellar medium. 

So far, the observational assessment of the evolutionary history of spheroidal galaxies has been hampered to limited depth and poor angular resolution of the instruments. The next generation instruments already launched like the \textit{James Webb Space Telescope} \citep[\textit{JWST};][]{Gardner_2006, Rigby_2023, McElwain_2023}, the \textit{Euclid} space observatory \citep{laureijs_euclid_2011} and those in preparation like the \textit{Vera C. Rubin Legacy Survey of Space and Time} \citep[\textit{LSST};][]{ivezic_lsst_2019} will be able to overcome the above limitations with unprecedented sensitivity and resolution.

In this paper we investigate the potential of the \textit{JWST} Advanced Deep Extragalactic Survey \citep[JADES;][]{eisenstein_overview_2023}  for shedding light on the proto-spheroid evolution. We adopt the standard approach, exploiting models to simulate the survey outcome. A phenomenological model specifically designed to simulate \textit{JWST} extragalactic surveys was presented by \citet{Williams2018} who applied it to make predictions for the JADES in GOODS-South and GOODS-North fields \citep{Giavalisco_2004}, aiding their optimization and the interpretation of the data. These fields are among the best-studied extragalactic deep fields with the availability of a large amount of photometric and spectroscopic data from the ultraviolet to the far-infrared, including data from the Hubble Space Telescope (\textit{HST}), \textit{Spitzer}, and \textit{Herschel} \citep{Liu_2018,Barro_2019}.

In this paper we exploit the model by \cite{cai_hybrid_2013} which co-evolves massive proto-spheroids and the central AGN at  $z\gtrsim1.5$. It is a physical model, therefore well suited to link the observational data to the physics driving the early galaxy evolution. Importantly, the model captures the full life-cycle of massive galaxies: starting with an initial UV-bright star-forming phase, followed by a heavily dust-enshrouded starburst phase (corresponding to the DSFG stage), then transitioning through a short-lived quasar phase, before settling into passive evolution. 

We adopt the version upgraded by \cite{mitra_euclid_2024}, which links the model outputs to the formalism by \cite{da_cunha_simple_2008} and by \cite{fritz_revisiting_2006} to compute the spectral energy distributions (SEDs) of DSFG and of AGN components, respectively. The model was used by \cite{mitra_euclid_2024} to study the ability of \textit{Euclid} in constraining the physical properties of DSFGs at $z\gtrsim1.5$ detected by the \textit{Herschel} Astrophysical Terahertz Large Area Survey \citep[H-ATLAS;][]{eales_herschel_2010}.

In the pre-JWST era, several studies have been conducted to estimate the physical properties of DSFGs in the GOODS-S field. \cite{Franco_2020} studied a sample of 35 galaxies detected with \textit{ALMA} at 1.1 mm in the GOODS-\textit{ALMA} field \citep{franco_goods-alma_2018}. To derive their physical properties, they performed SED fitting using the available multi-wavelength data. The sample comprised massive galaxies having a median stellar mass estimate of $8.5\times10^{10}M_{\odot}$ in the redshift range $z\sim2-4$. \cite{Yamaguchi_2020} estimated the physical properties of a sample of 24 K-band selected galaxies from the Four Star Galaxy Evolution Survey \citep[ZFOURGE;][]{Straatman_2016} having deep 1.2 mm ALMA observations, as a part of the \textit{ALMA} twenty-six Arc minute$^2$ Survey of GOODS-S at One-millimeter (ASAGO)\footnote{\url{https://sites.google.com/view/asagao26/home}}. With a median redshift of $2.38\pm0.14$, these sources have median stellar mass estimates of $\log(M_{\star}/M_{\odot})\sim 10.75$ in both the redshift bins $1<z\leq2$ and $2<z\leq3$. In the same redshift bins, the obtained median $\log (\dot{M_{\star}}/M_{\odot}\hbox{yr}^{-1})$ were $2.14$ and $2.15$, respectively. \cite{Wiklind_2014} studied a sample of 10 sub-mm sources observed with \textit{ALMA} at 870 $\mu\rm m$ originally detected using \textit{LABOCA} \citep{siringo_large_2009} in the CANDELS survey of the GOODS-S. A median stellar mass estimate of $9.1\times10^{10}M_{\odot}$ was obtained. The average SFR obtained was $(0.8\pm0.7)\times10^3\,M_{\odot}\hbox{yr}^{-1}$. A panchromatic study of 11 DSFGs with spectroscopic redshift $1.5<z_{\rm spec}<3$ in the GOODS-S field was conducted by \cite{pantoni_unveiling_2021}. They performed SED fitting using CIGALE and estimated the median stellar mass and SFR of the sources to be $6.5\times 10^{10}M_{\odot}$ and $241\,M_{\odot}\hbox{yr}^{-1}$, respectively. As can be seen, the studies conducted so far on DSFGs in the GOODS-S could only probe median stellar masses down to about $\log(M_{\star}/M_{\odot})\approx10.8$. In this work, we also try to investigate the minimum stellar mass at which {\it JWST} can reliably detect DSFGs.


The paper is organised as follows. Section \ref{sec2} gives a brief description of the model. In Section \ref{secfor}, we provide a concise presentation of the relevant surveys. Section \ref{sec4} presents a  comparison of physical properties of ASTRODEEP-\textit{JWST} sources with model predictions. Finally, Section \ref{sec5} contains the summary and the conclusions of this paper. Here, we adopt a flat $\Lambda$-CDM cosmology with
present-day matter density (in units of the critical density), $\Omega_{m,0} = 0.3153$ and baryon density $\Omega_{b,0} = 0.0493$. We set the value of the Hubble-Lema\^itre constant to $h = H_0/100 = 0.6736$, the slope of the spectrum of primordial density perturbations to $n=0.9649$ and the normalization of the density fluctuations on a scale of $8h^{-1}$\,Mpc to $\sigma_{8} = 0.8111$ \citep{planck_collaboration_planck_2020}.

\section{Methodology}
\label{sec2}

\subsection{Model Outline}

The \citet{cai_hybrid_2013} model was inspired by the fact that the stellar content of present-day massive spheroidal galaxies is dominated by old populations, formed at $z\gtrsim1.5$. In contrast, the disc-shaped galaxies contain relatively young stellar populations, with luminosity-weighted age $\lesssim7$ Gyr, i.e. mostly formed at $z\lesssim1$. Therefore, at $z\gtrsim1.5$, the dominant star-forming galaxies are the proto-spheroidal galaxies, the progenitors of the present-day ellipticals. This physical treatment takes into account the co-evolution of the SFR of the proto-spheroids and of the supermassive black holes (SMBHs) residing at their centre. 

The star formation history of these proto-spheroids is determined by a set of equations that describe the gas cooling and condensation into stars, the accretion of the gas into the SMBH as well as the feedback from the supernova explosions and from the AGN. We refer the readers to \cite{cai_hybrid_2013} and \cite{mitra_euclid_2024} for more detailed explanations about the equations governing the co-evolution of the stellar and of the AGN components.

The SED of the photo-spheroids is modelled using the formalism put forward by \cite{da_cunha_simple_2008} which is based on the principle of energy balance, with a few modifications explained in detail in \cite{mitra_euclid_2024}. We adopt the smooth torus model introduced by \cite{fritz_revisiting_2006} to model the SED of the AGN component. In modelling the dust emission from the AGN, we do not incorporate the contribution from polar dust \citep{honig_dust_2013,yang_dust_2020} as it emits mostly in the mid-IR and has a negligible effect at UV/optical/near-IR wavelengths \citep[see their Figure 8]{honig_dust_2013}.

\begin{figure*}
\centering
\includegraphics[width=.3\textwidth]{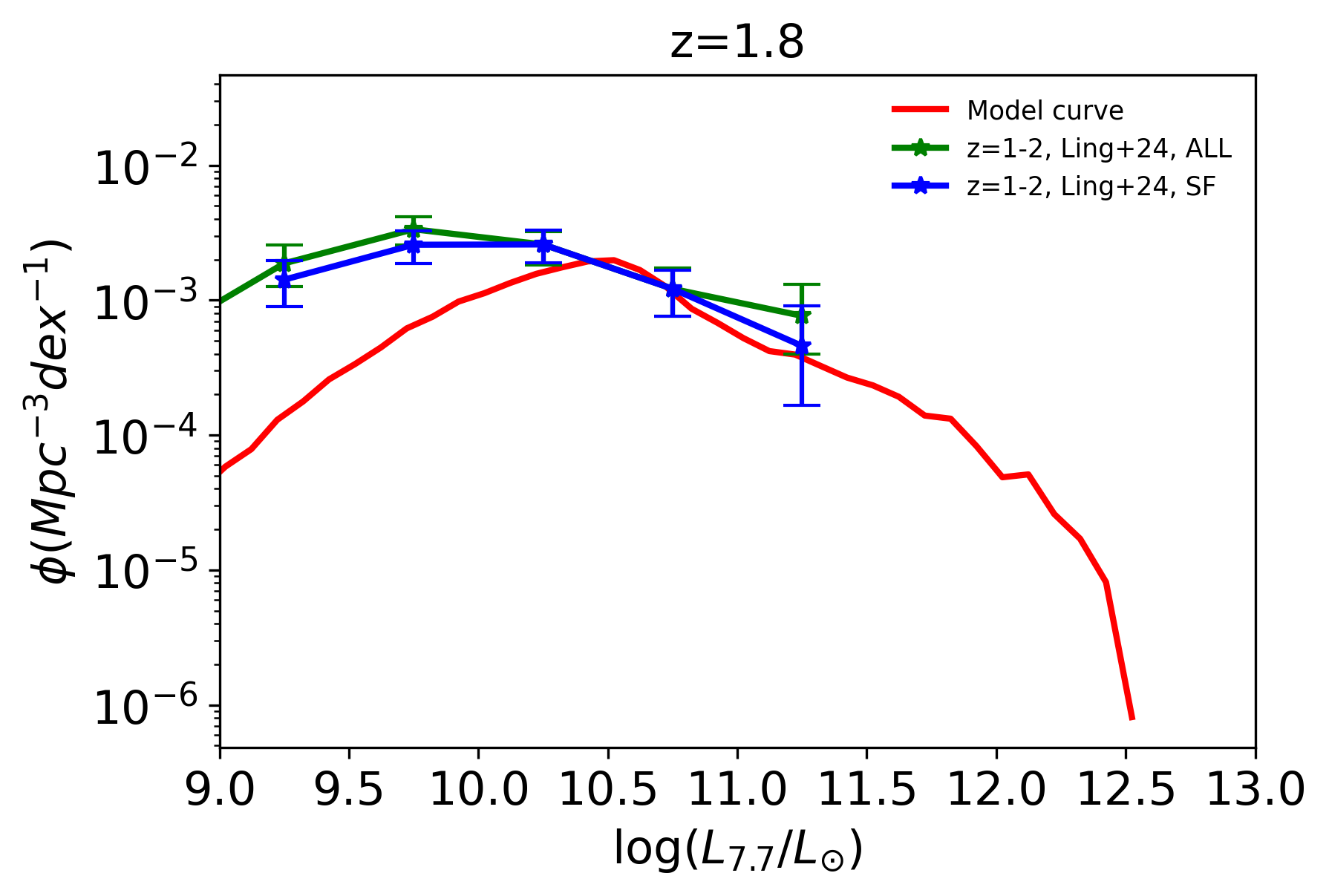}\hfill
\includegraphics[width=.3\textwidth]{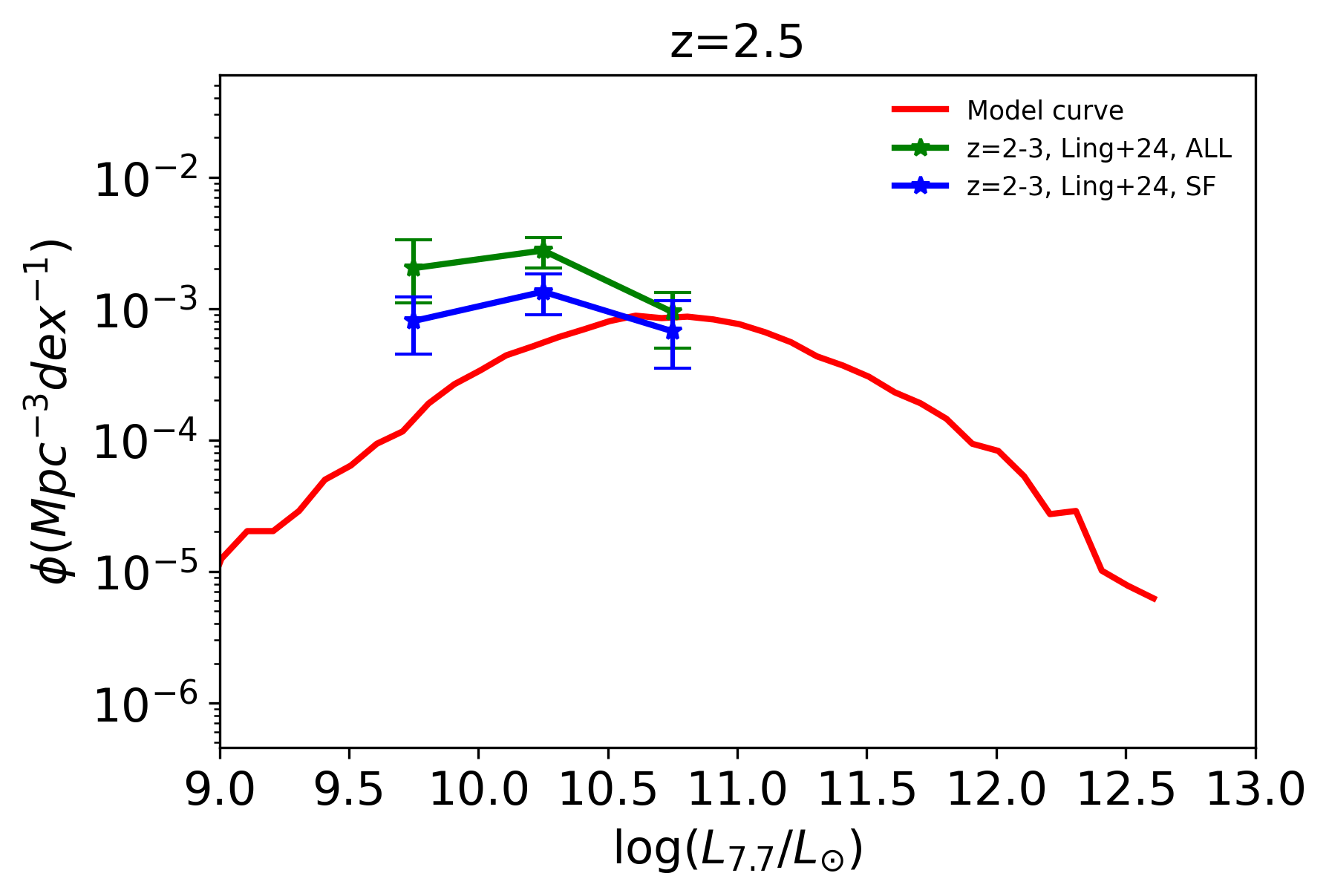}\hfill
\includegraphics[width=.3\textwidth]{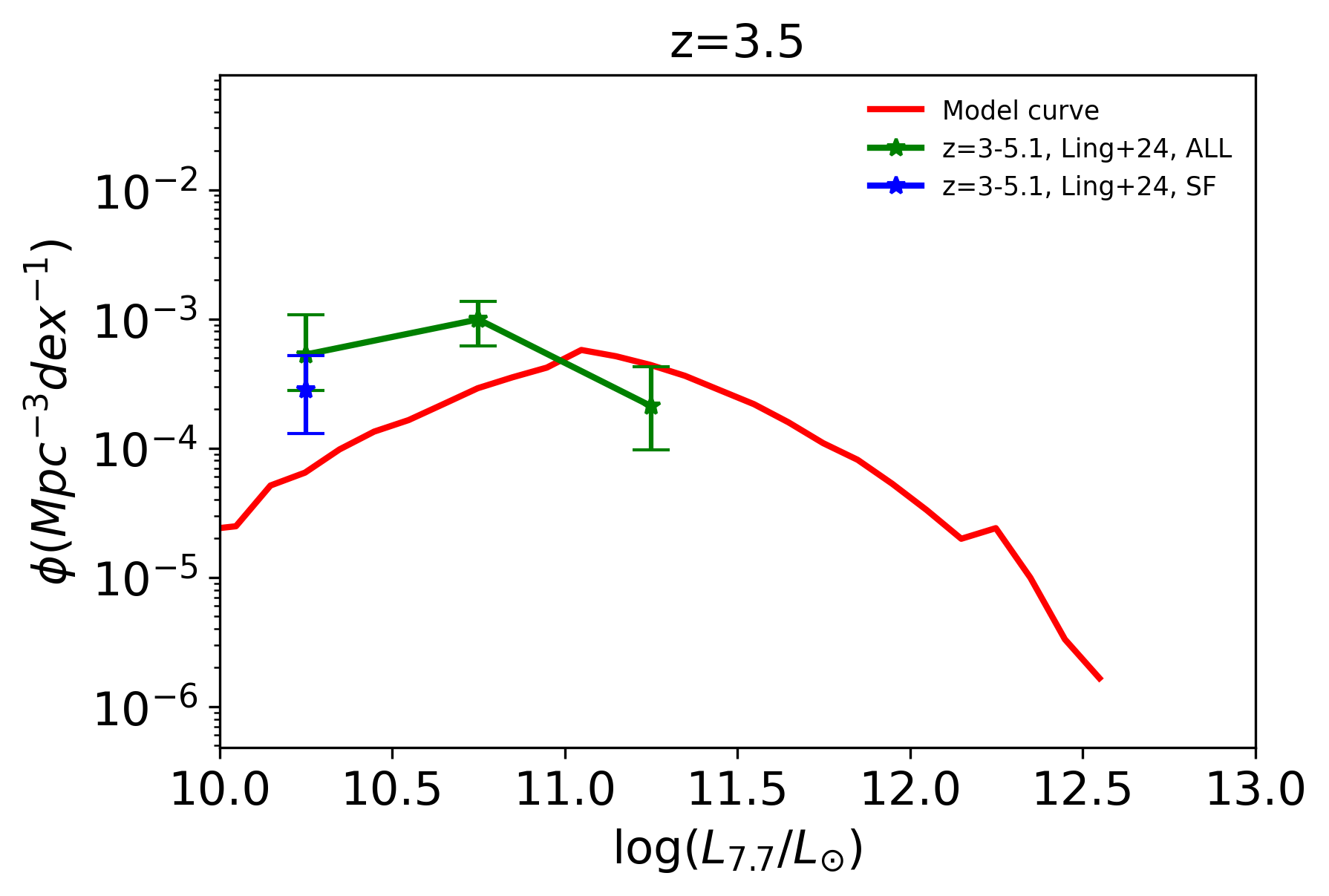}\hfill
\includegraphics[width=.3\textwidth]{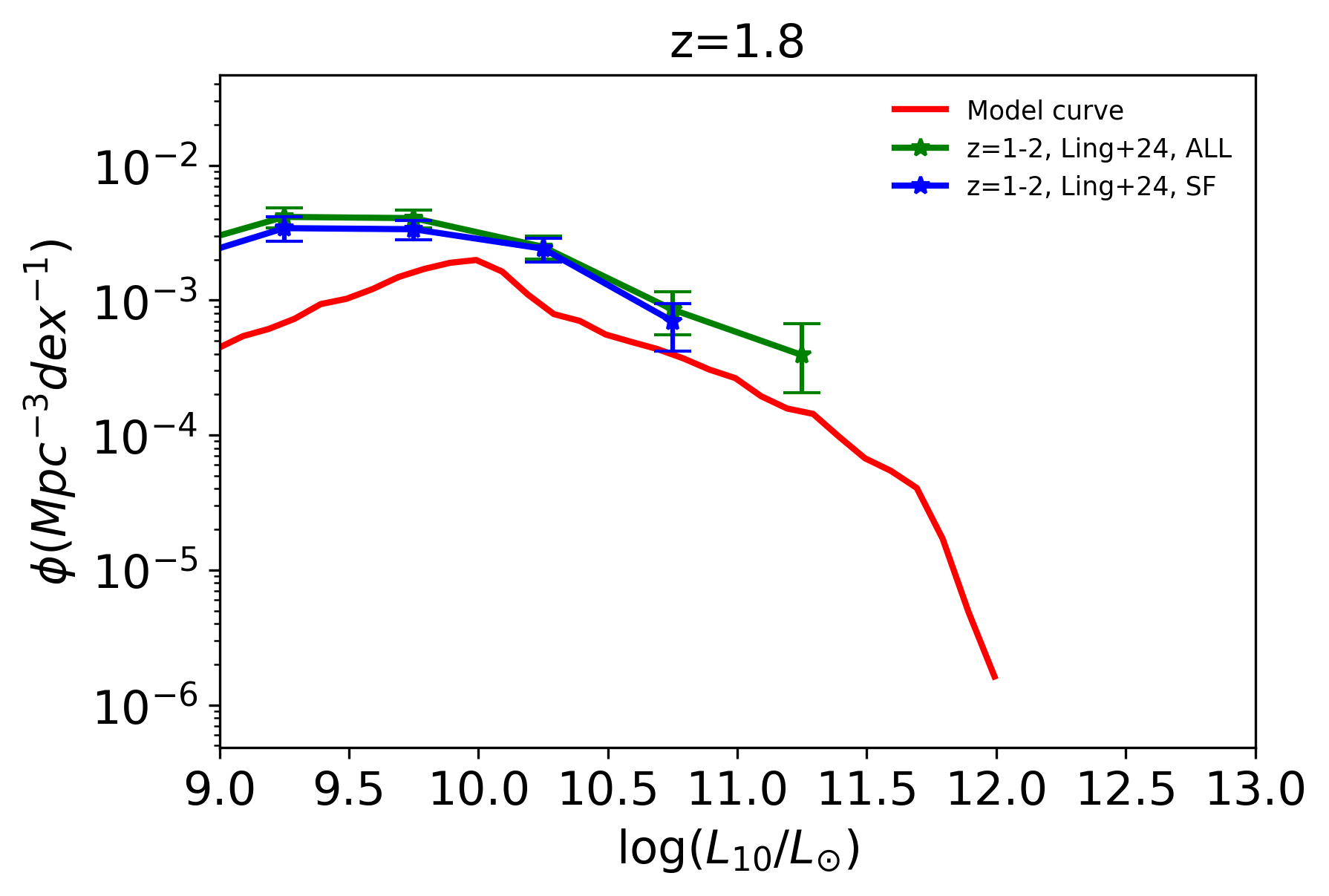}\hfill
\includegraphics[width=.3\textwidth]{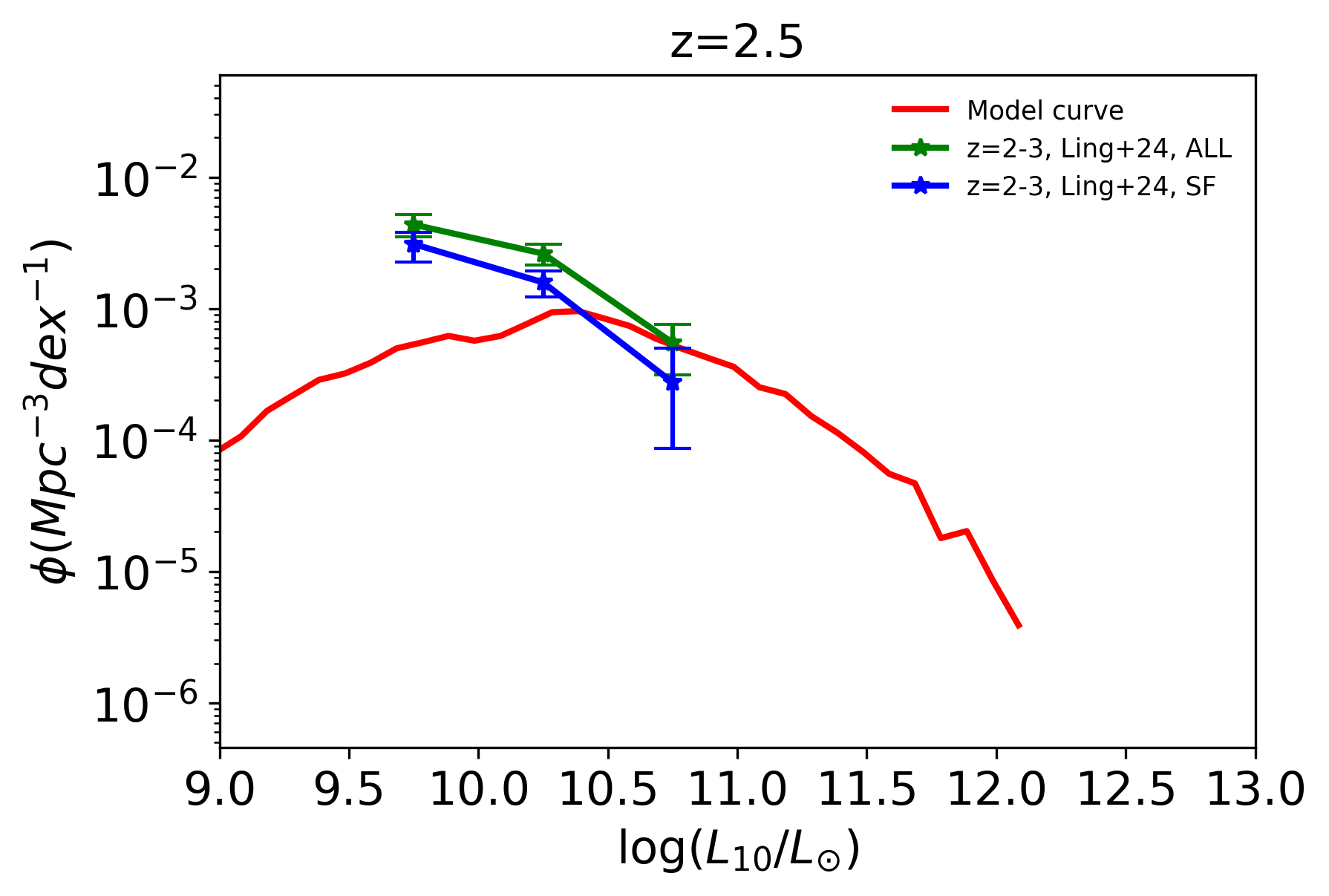}\hfill
\includegraphics[width=.3\textwidth]{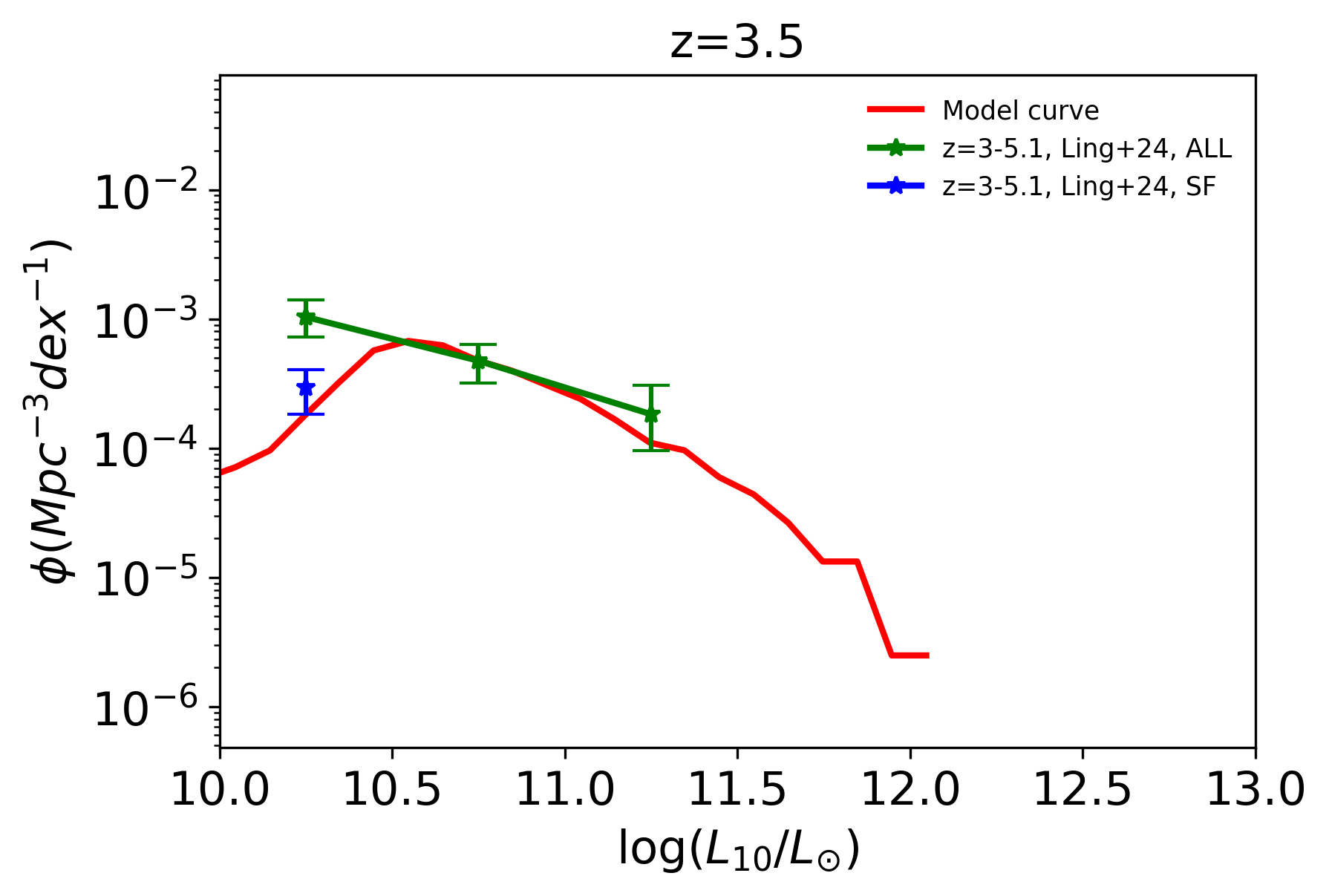}\hfill
\includegraphics[width=.3\textwidth]{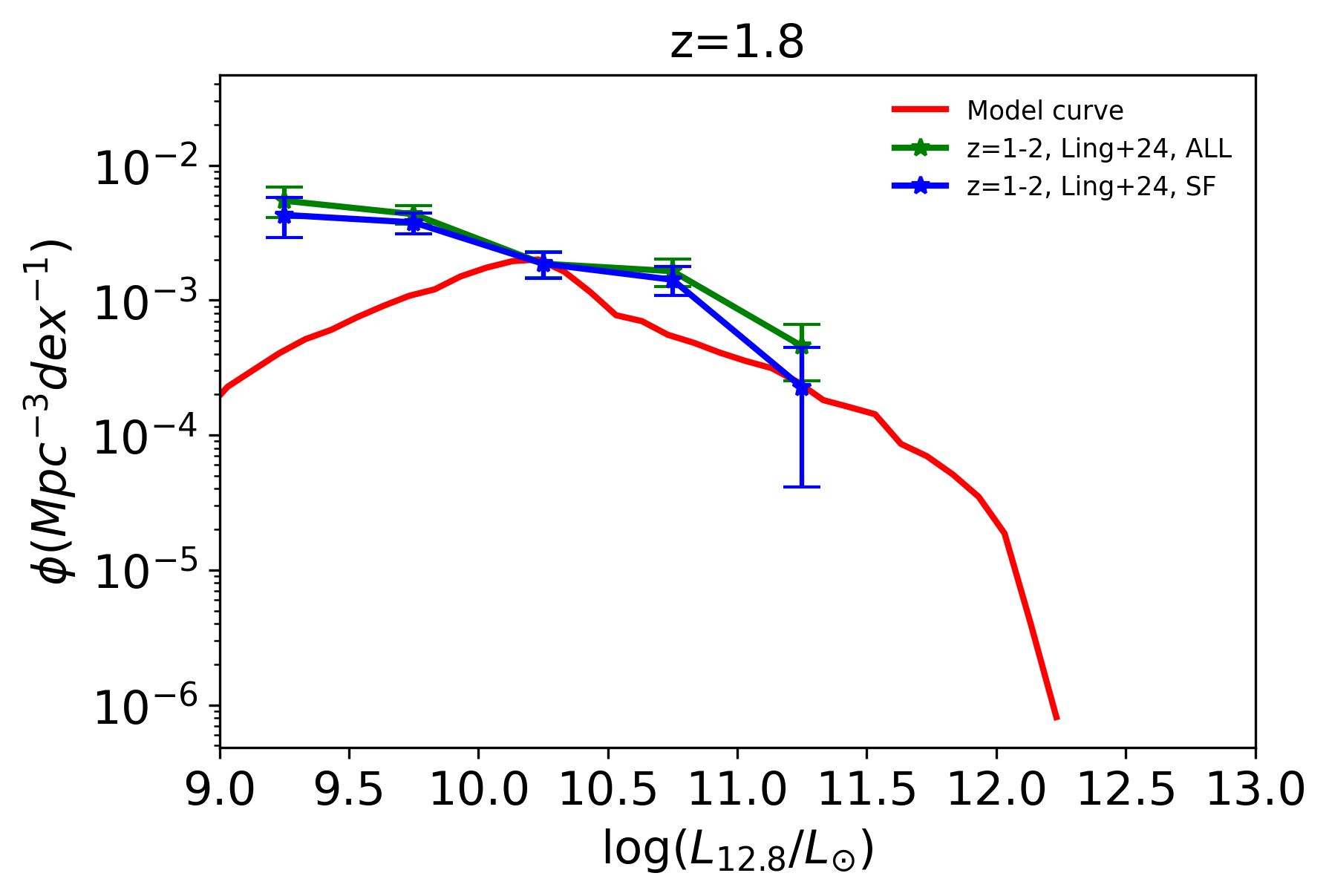}\hfill
\includegraphics[width=.3\textwidth]{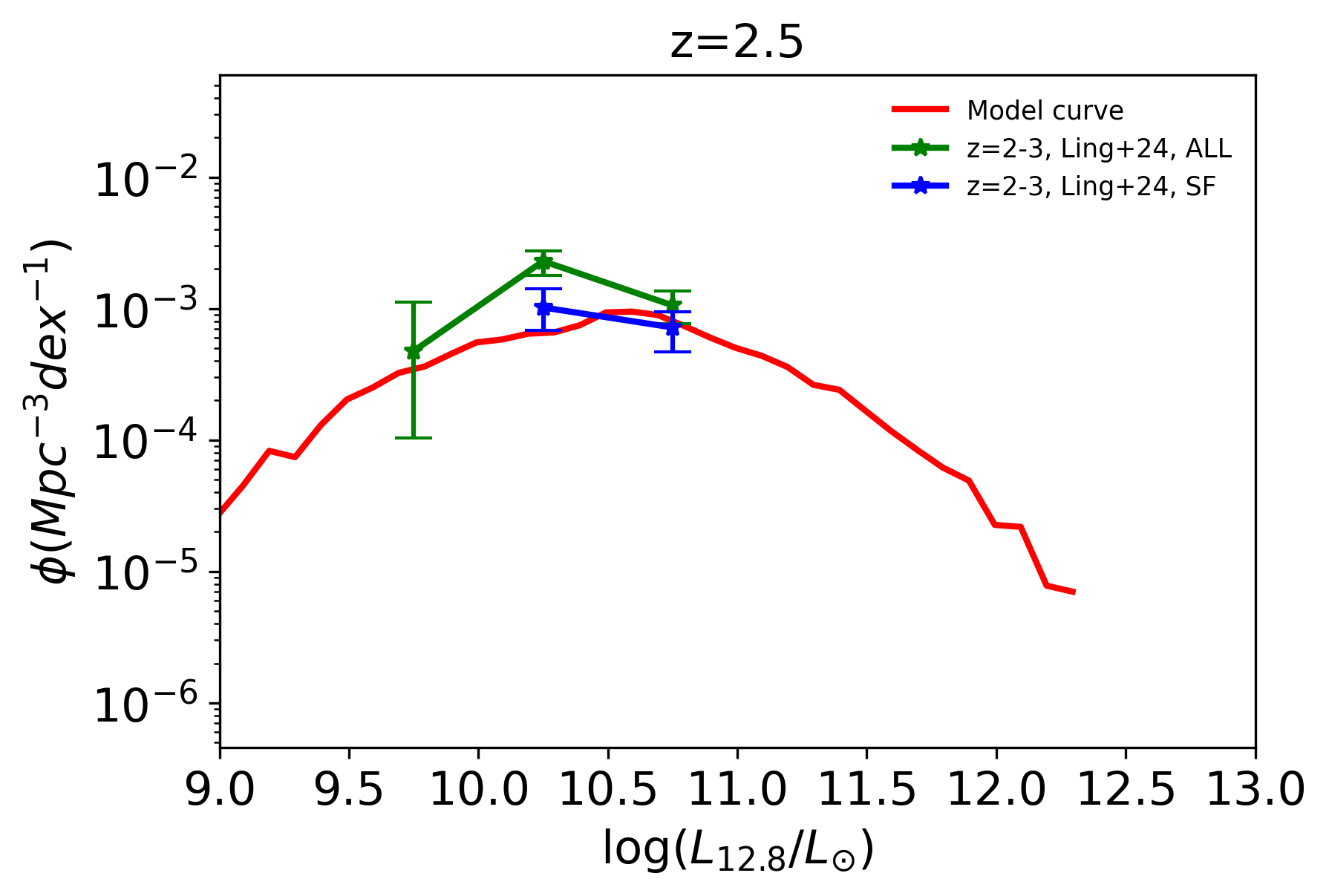}\hfill
\includegraphics[width=.3\textwidth]{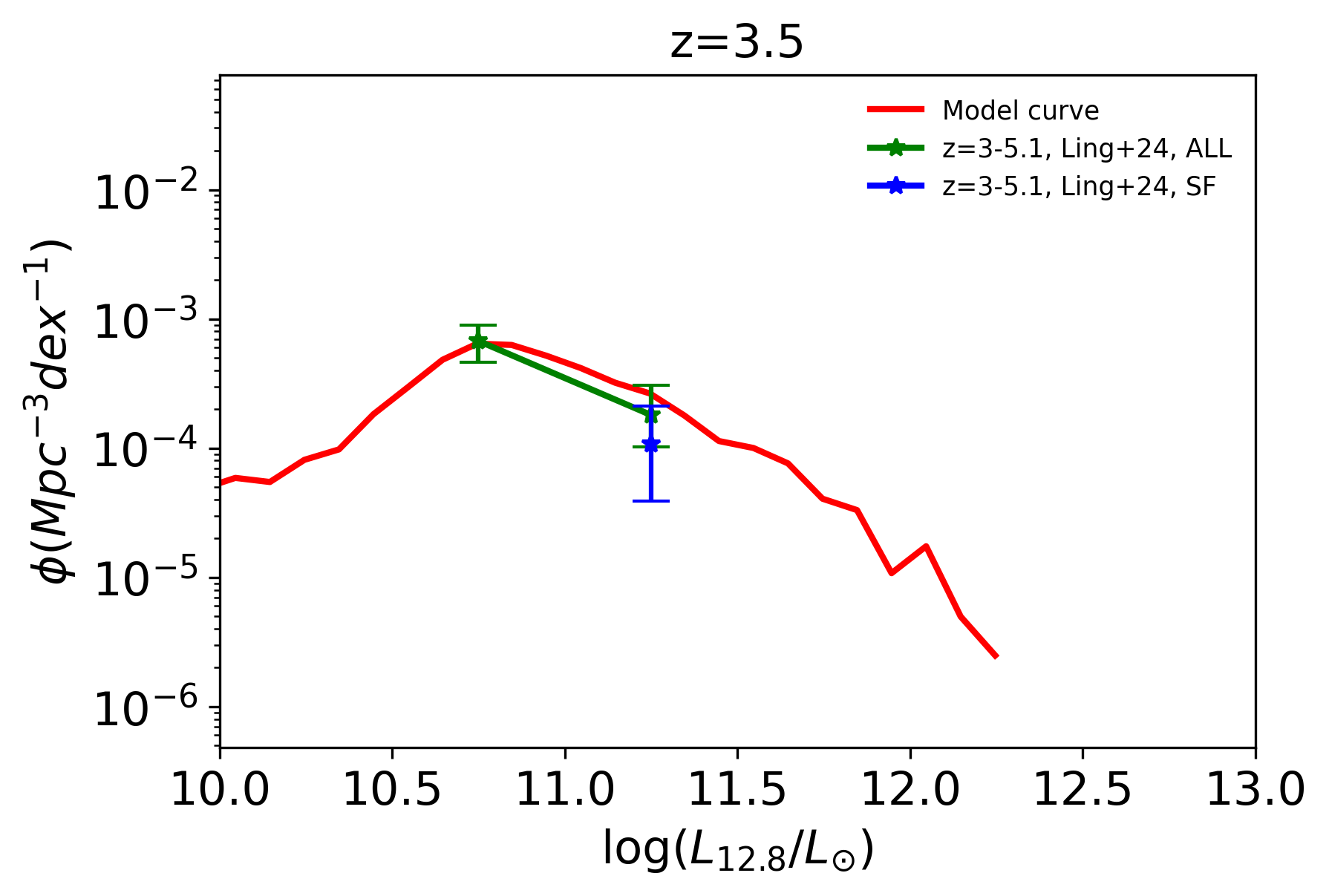}\hfill
\includegraphics[width=.3\textwidth]{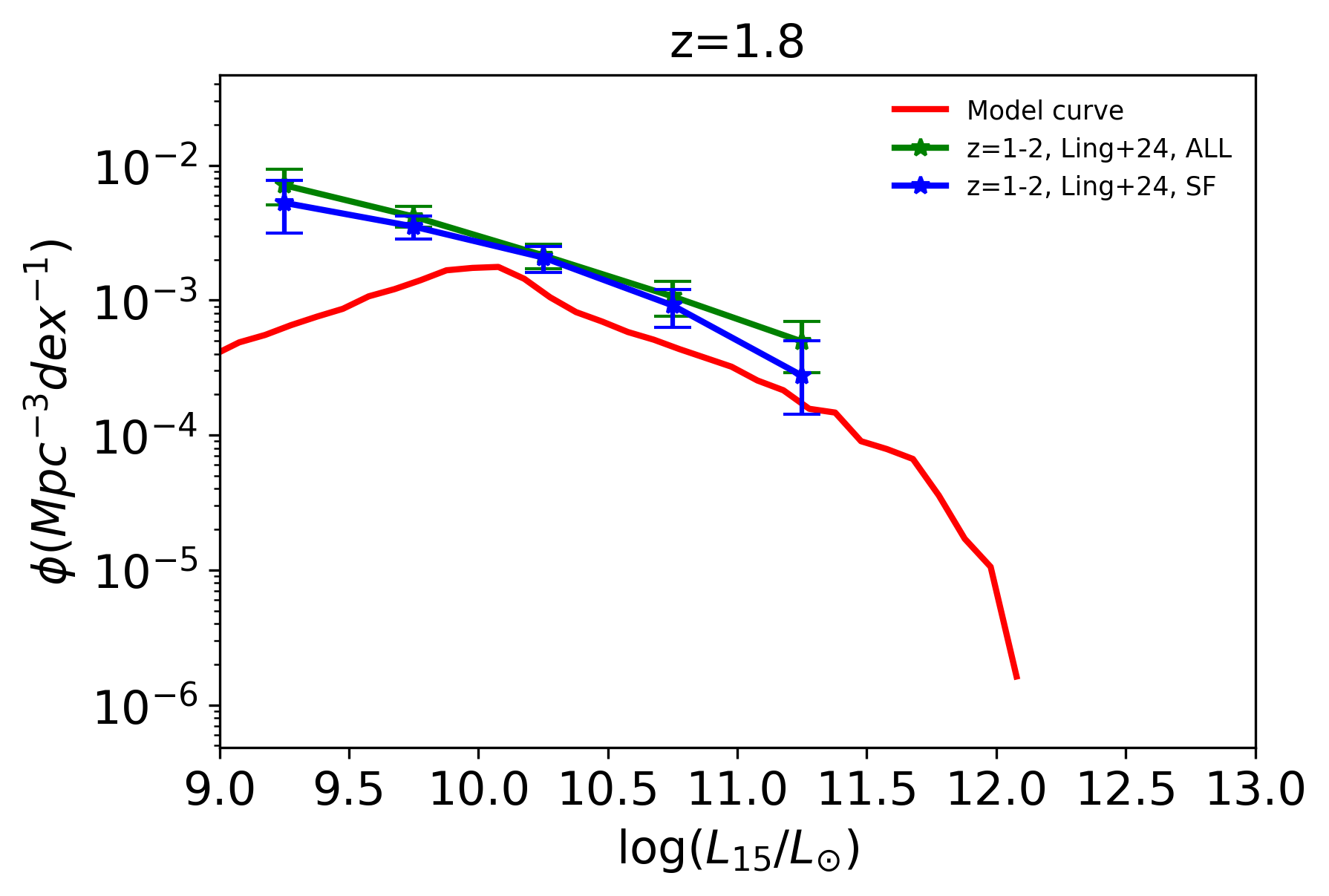}\hfill
\includegraphics[width=.3\textwidth]{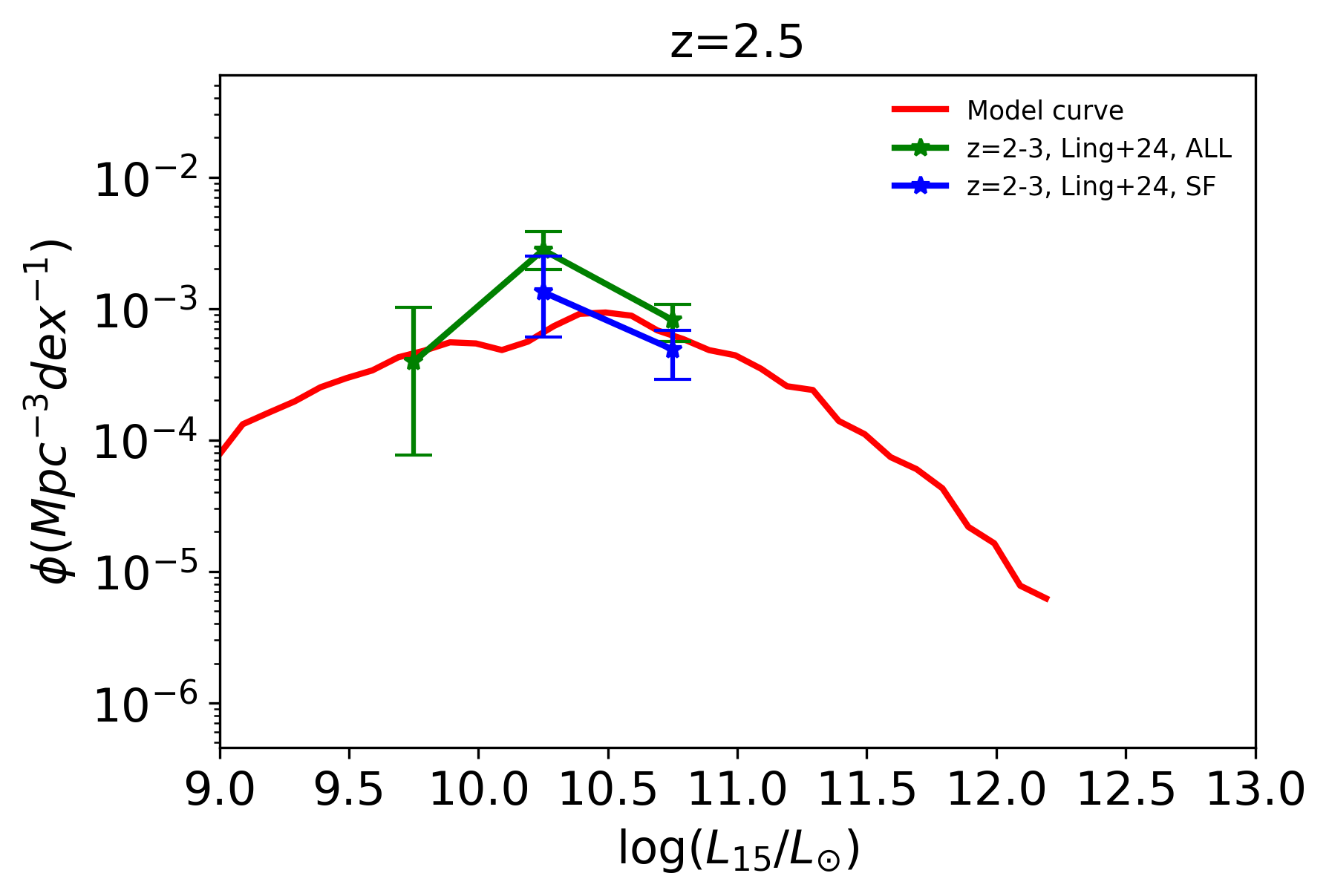}\hfill
\includegraphics[width=.3\textwidth]{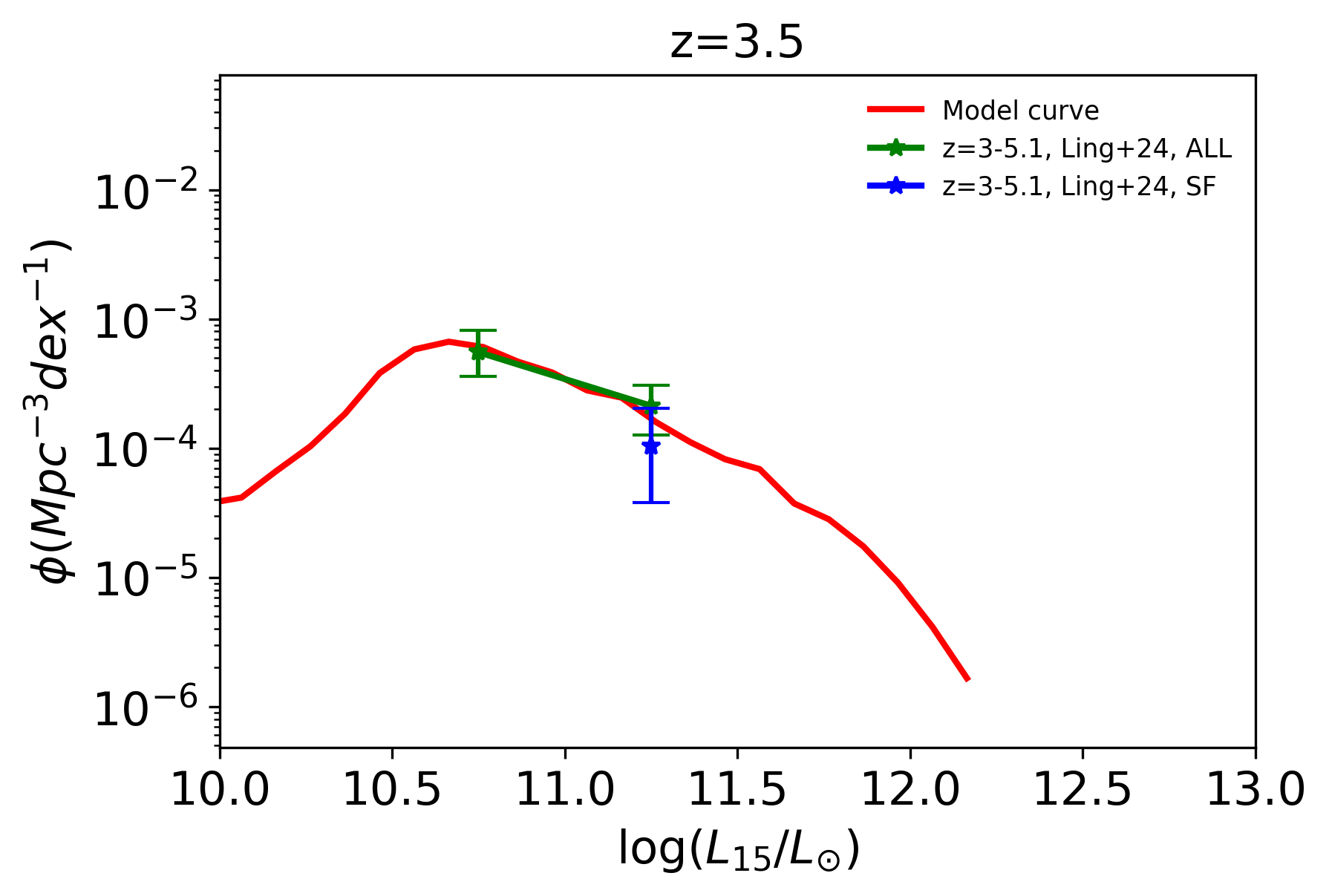}\hfill
\includegraphics[width=.3\textwidth]{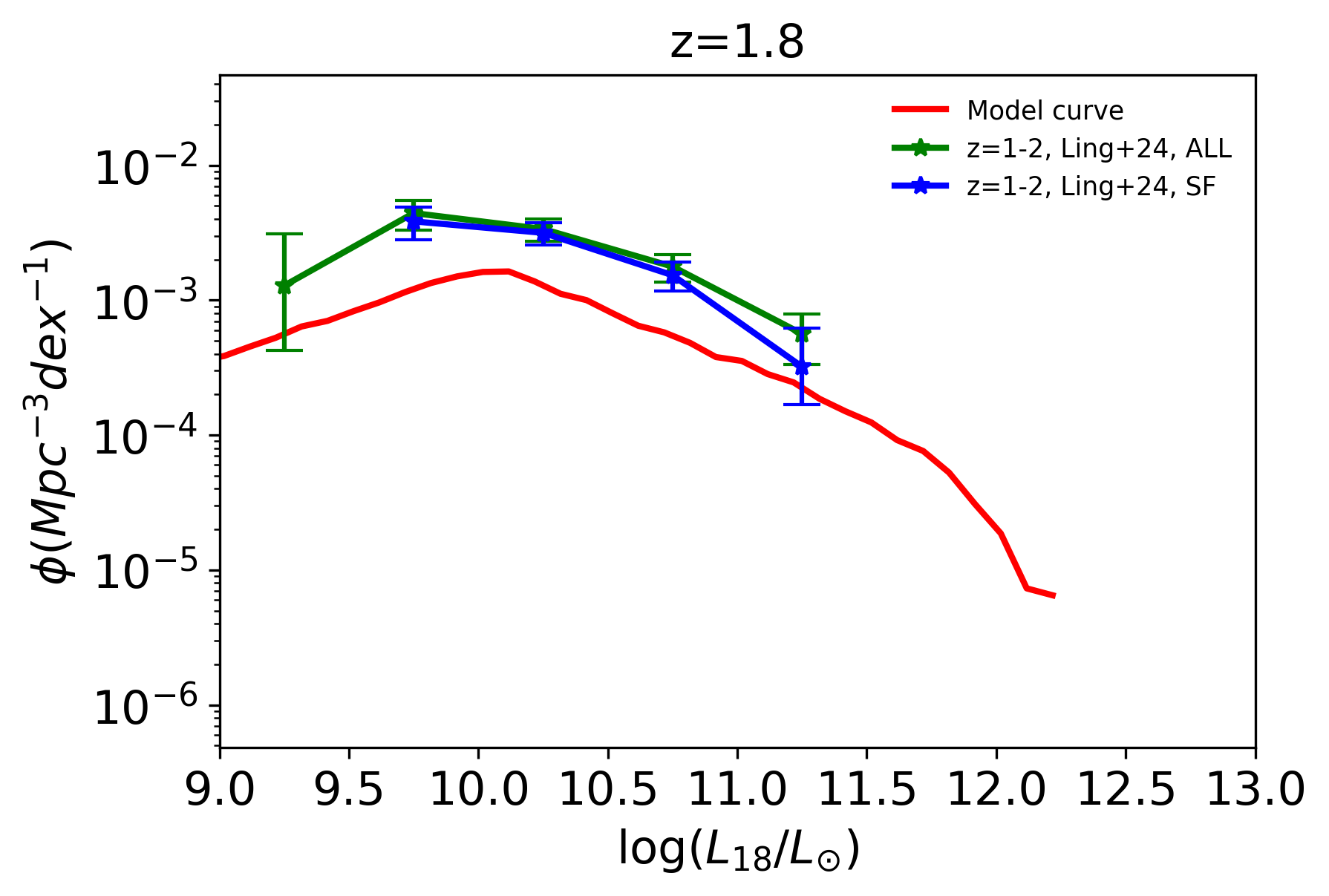}\hfill
\includegraphics[width=.3\textwidth]{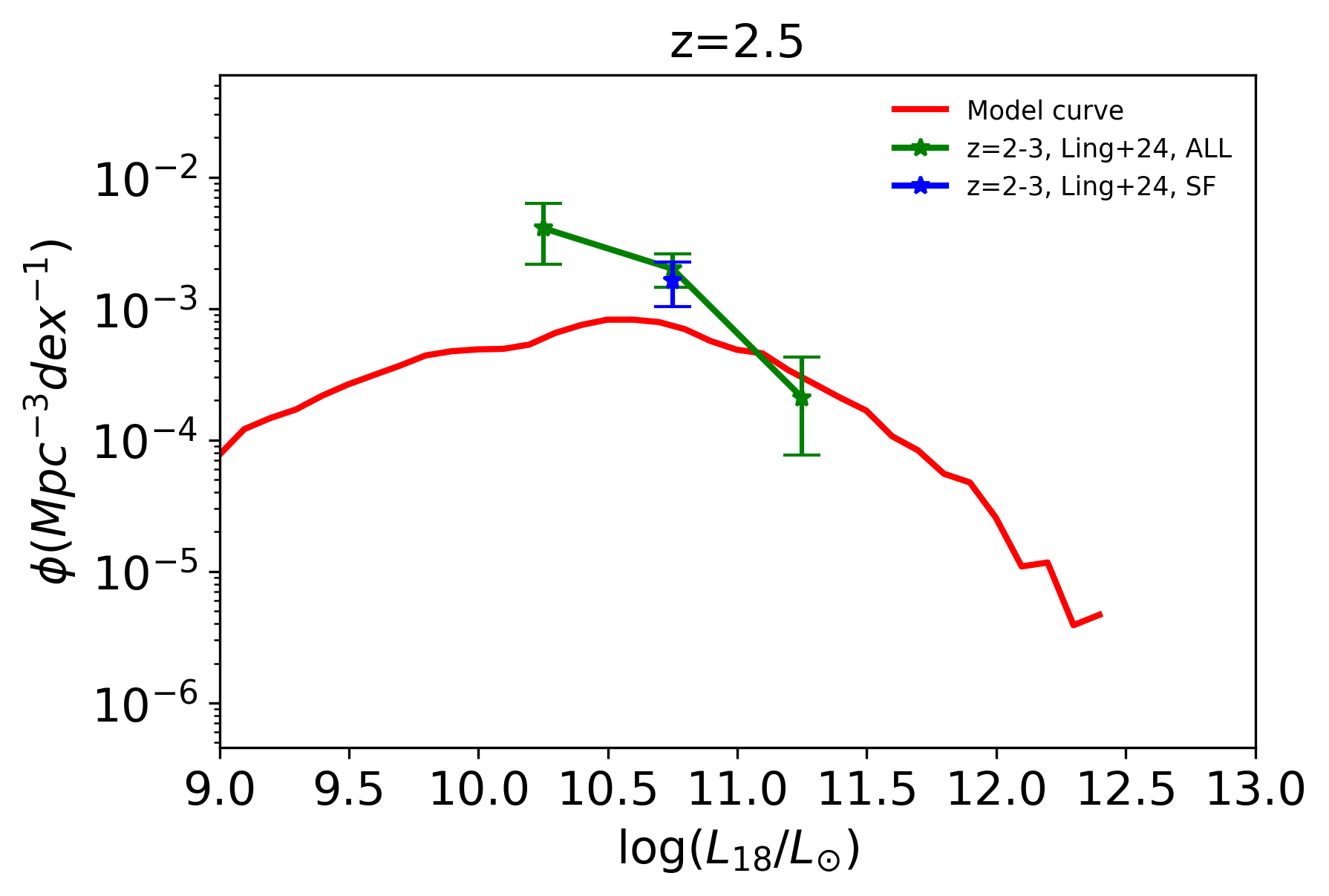}\hfill
\includegraphics[width=.3\textwidth]{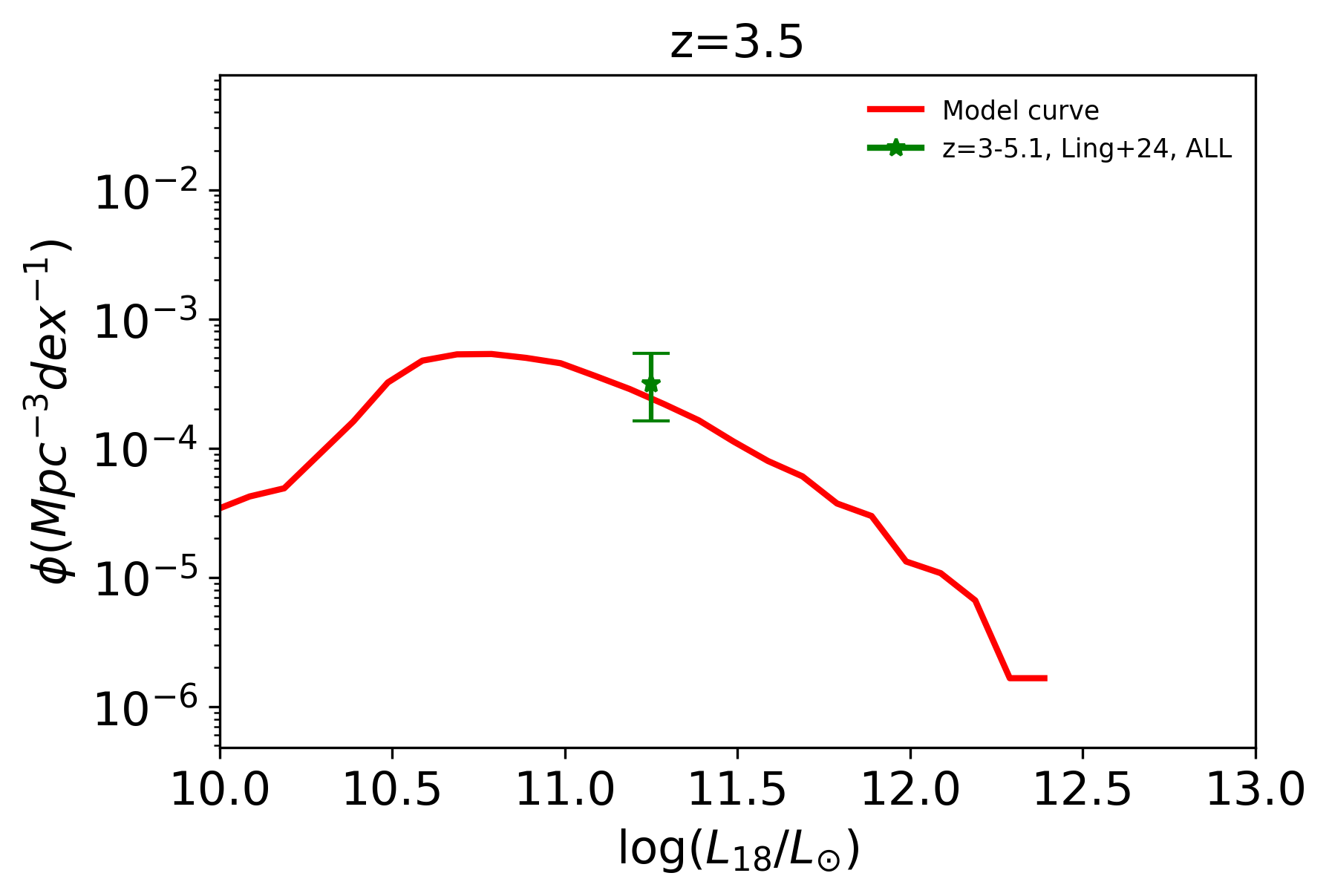}\hfill
\includegraphics[width=.3\textwidth]{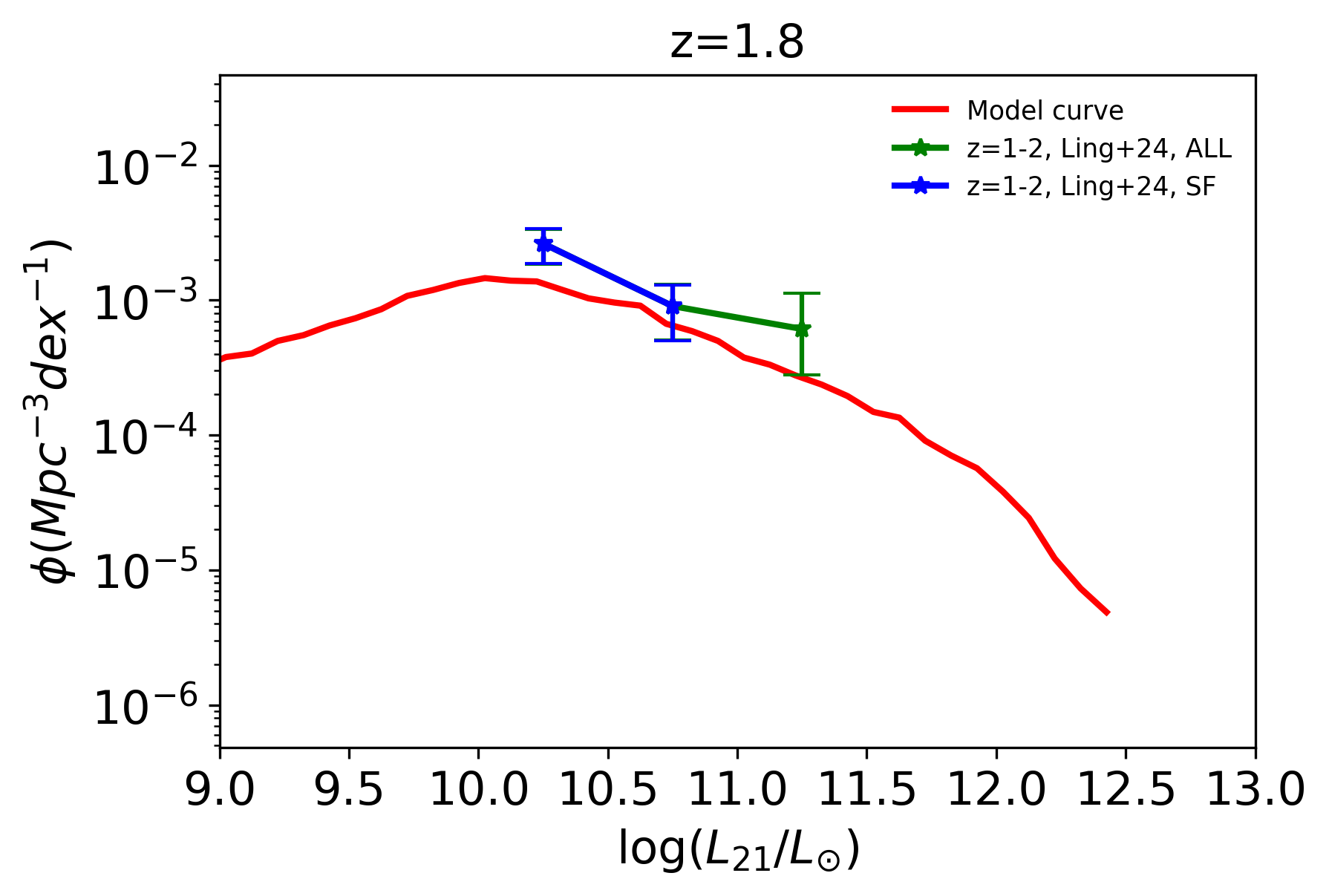}\hfill
\includegraphics[width=.3\textwidth]{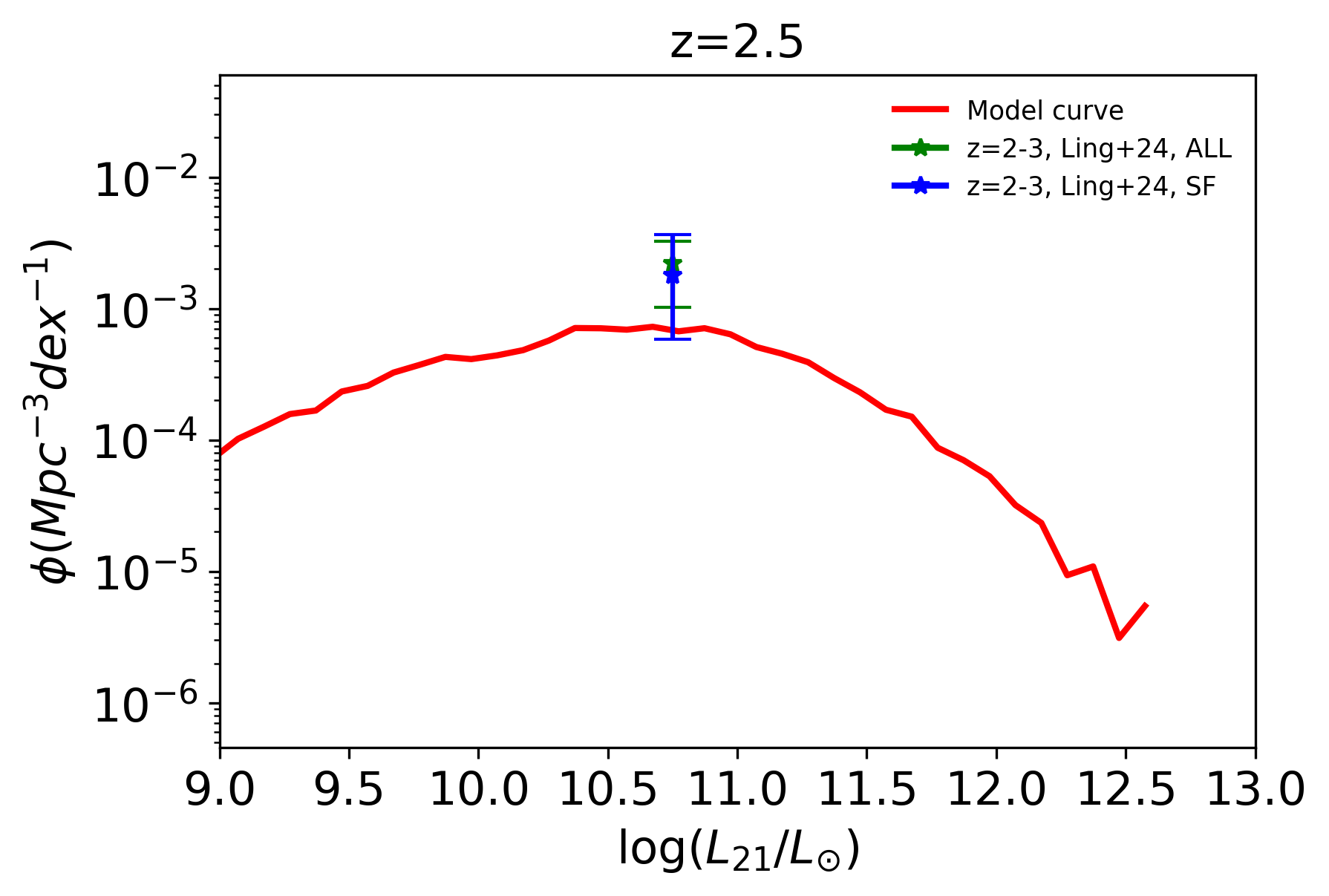}\hfill
\includegraphics[width=.3\textwidth]{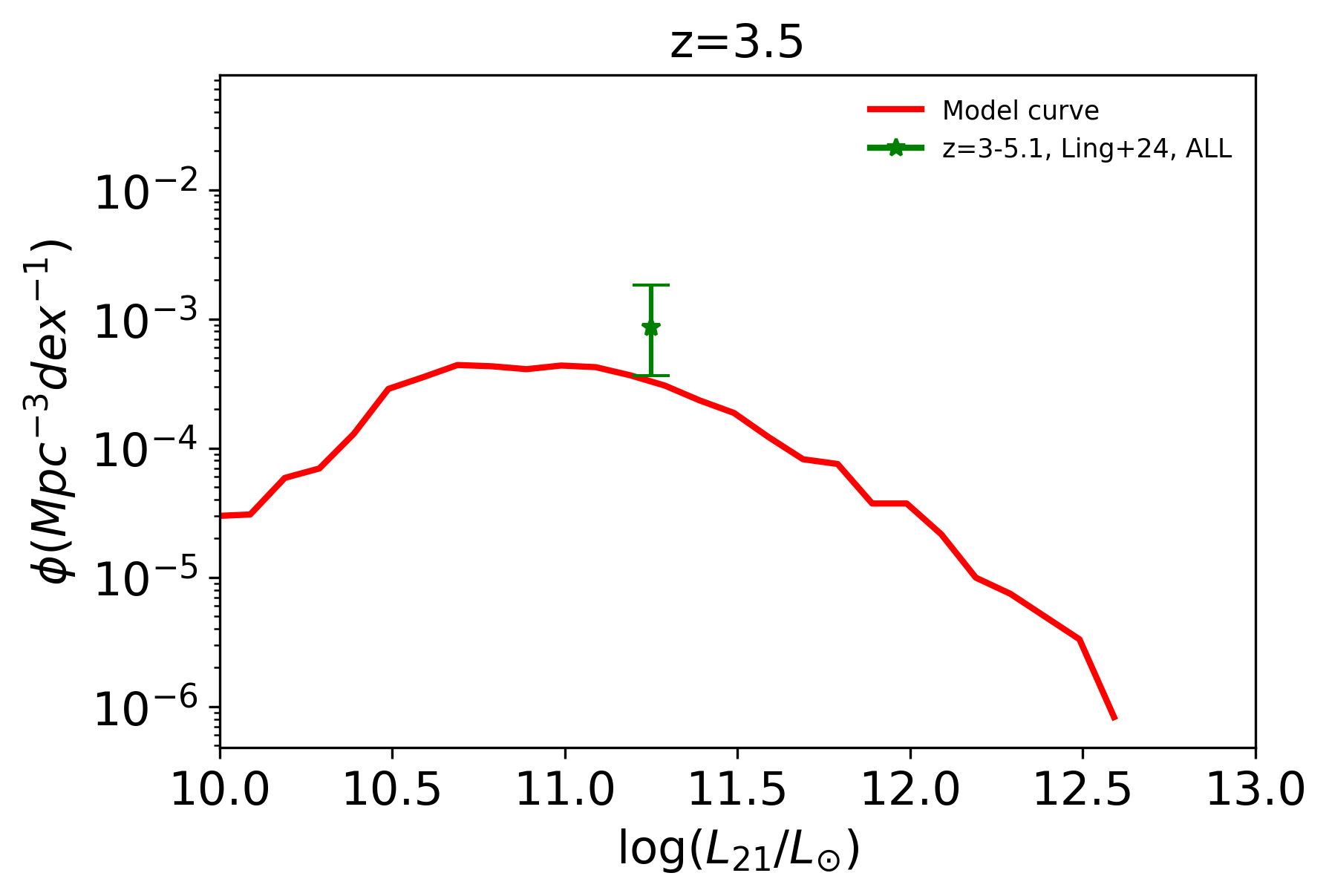}
\caption{Comparison of the measured monochromatic LFs with those predicted for the population of massive ($11.3\leq\log(M_{\rm vir}/M_{\odot})\leq13.3$) proto-spheroids at 7.7, 10, 12.8, 15, 18, and 21 $\mu$m for $z=1.8$, 2.5, and 3.5. The red line shows the model curve. Note that the model parameters were not optimized; no fit of the data was attempted. At low luminosity the dominant contribution to the LFs comes from galaxies with halo masses below the adopted lower limit. This leads to a fast decrease of the model curve at low luminosity, implying an under-prediction of the LFs. The green line shows the LF from \citet{ling2024} for all galaxy populations, while the blue line shows the same for the star-forming (SF) galaxies. The first column shows the data points for the redshift bin $z=1-2$, the middle column for $z=2-3$, and the last column for $z=3-5.1$. }
\label{figmidirlf}
\end{figure*}

To verify that this modified version of Cai's model correctly predicts the luminosity functions (LFs) of proto-spheroidal galaxies at mid-IR wavelengths, we computed the rest-frame LFs of these galaxies at 7.7, 10, 12.8, 15, 18 and $21\,\mu\rm m$ respectively, and compared them with the rest-frame LFs produced by \cite{ling2024} using a sample of 506 galaxies at $z=0-5.1$ from the \textit{JWST} Cosmic Evolution Early Release Science \citep[CEERS;][]{2017jwst.prop.1345F} survey. Figure \ref{figmidirlf} shows the plot of the mid-IR LFs for redshift $z=1.8, 2.5$ and $3.5$ respectively. 

It is to be noted that the values of model parameters are not optimized, i.e. no fit of the data was attempted. The model curve (red) shows the contribution of galaxies with virialized halo masses in the range $11.3\leq\log(M_{\rm vir}/M_{\odot})\leq13.3$ \citep[see][and sub-Sect.\,\ref{subsect:simul}]{mitra_euclid_2024}. The adopted lower limit to halo masses translates to a fast decrease of the model LFs at low luminosities, where less massive galaxies dominate. As a consequence, the model under-predicts the low-luminosity end of LFs. Decreasing the minimum halo mass would make the model predictions less solid. This is because a basic assumption of the model, namely that the halo formation rate is well approximated by the positive term of the cosmic time derivative of the halo mass function, is less accurate at low halo masses. 

At high luminosity, the consistency with observations is satisfactory, especially considering that the model is not optimized. At some combinations of rest-frame wavelength and redshift, the complexity of the SED at MIR wavelengths, where the PAH features are prominent, can also contribute,  to the small discrepancy between the predicted and the measured luminosity functions. In fact, such a complexity is not fully captured by our formalism, which adopts only a single template for the PAHs.

\subsection{NIRCam colour comparison between CEERS and the model}

\begin{figure}

\centering
\includegraphics[width=.47\textwidth]{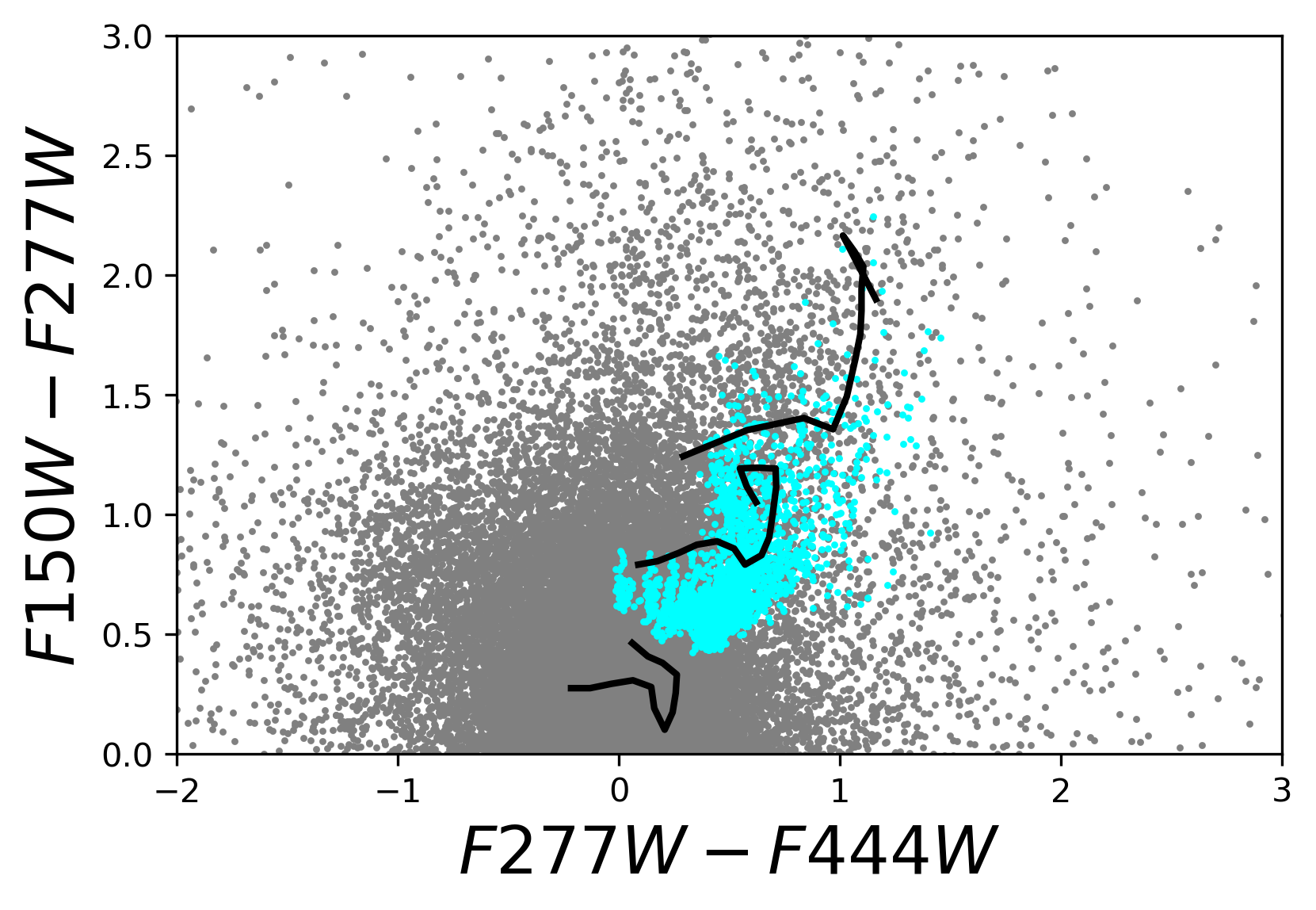}
\caption{{\it JWST}/NIRCam  $F150W - F277W$ versus $F277W - F444W$ colour-colour diagram. Grey points represent observed galaxies from the CEERS survey (taken from the ASTRODEEP-{\it JWST} catalog), and cyan points show simulated proto-spheroids from our model. The black curves represent evolutionary tracks from redshift $z \sim 1$ to 6 for galaxies with different dust attenuations: $A_V < 1$ (bottom), $A_V \sim 2$ (middle), and $A_V \gtrsim 3$ (top). The model galaxies align well with the moderately to heavily obscured tracks, indicating a good consistency between simulated and observed colour distributions.}
\label{figcolor}
\end{figure}

To further evaluate the realism of our simulated proto-spheroidal galaxies, we compare their near-infrared colours to galaxies observed in the CEERS survey using a $F150W - F277W$ vs. $F277W - F444W$ colour-colour diagram as shown in Figure \ref{figcolor}. The CEERS sources, shown in grey, are taken from the ASTRODEEP-{\it JWST} catalog \citep[][for more details see Section \ref{astrodeep}]{Merlin_2024} and span a broad range of redshifts, dust contents, and stellar populations, forming a wide distribution in the colour--colour plane. In contrast, the simulated proto-spheroids (cyan points), occupy a more confined region. Notably, these galaxies follow a trend consistent with high-redshift star-forming systems, with moderate to substantial dust obscuration. Overlaid on the plot are evolutionary tracks of dusty galaxies based on V-band dust attenuation ($A_V$) and redshifted from $z=1$ to $6$ by \cite{Long_2024}. These tracks illustrate how dust attenuation and redshift drive the observed colours. The lower, middle, and upper tracks correspond to $A_V < 1$ (little attenuation, $A_V \sim 2$ (average attenuation), and $A_V \gtrsim 3$ (high attenuation), respectively. They obtain these evolutionary tracks using the SEDs of ALESS submillimeter galaxies from \cite{da_cunha_alma_2015}. The majority of the simulated DSFGs cluster at moderate attenuation, consistent with the expected properties of proto-spheroids at $z \gtrsim 1.5$. 


\subsection{Building the Simulated Catalogue}\label{subsect:simul}

Here we give a brief explanation about the generation of a simulated sample of proto-spheroidal galaxies to be used for forecast and analysis. The building of the simulated catalogue starts by randomly sampling the formation rate function of dark matter haloes \citep[see Eq.\,(A25) of][]{mitra_euclid_2024} in both the virialisation mass ($M_{\rm vir}$) and the virialisation redshift ($z_{\rm vir}$) of the halos. We restrict ourselves to the ranges $11.3\leq\log(M_{\rm vir}/M_{\odot})\leq13.3$ and $1.5\leq z_{\rm vir}\leq 8$ respectively. 

The number of sampled halos depends on the size of the surveyed area of the sky. Here we consider a survey area of 87 square arcmin., like that covered by the JADES-Medium survey in the GOODS-S field. We then allow the baryons within the simulated halos to evolve into galaxies by solving the equations of the model by \cite{cai_hybrid_2013} to get the star formation history (SFH) and the black hole accretion rate (BHAR) as a function of the galaxy age. The galaxy is assumed to be born at $z_{\rm vir}$. Next, we generate the SEDs of both the stellar and the AGN components of simulated galaxies. We refer the readers to \cite{mitra_euclid_2024} for a detailed description of the steps involved and of the adopted values of the different parameters.

\section{Forecasts}
\label{secfor}

Here, we investigate to what extent \textit{JWST} can constrain the physical properties of high-$z$ proto-spheroids along with the minimum stellar mass at which JWST can reliably detect them in the GOOD-S field, the primary JADES field. Since the CANDELS survey studied the GOODS-S field using \textit{HST}, we also complement the \textit{JWST} data with the \textit{HST} data. {IR and sub-mm data from \textit{Spitzer} and \textit{Herschel} are also available for this field}. We examine the effect on the physical parameters by considering different combinations of data from UV to IR/sub-mm with \textit{JWST} photometry from JADES. To begin with, we provide a brief summary of the CANDELS and of the JADES survey programs and then proceed with the results and the analysis.

\subsection{A brief overview of the CANDELS and JADES survey}

The Cosmic Assembly Near-infrared Deep Extragalactic
Legacy Survey \citep[CANDELS;][]{Grogin_2011, Koekemoer_2011} was designed to study galaxy formation and evolution over the $1.5\leq z\leq 8$ range. The main idea behind it was to exploit the revolutionary NIR \textit{HST}/WFC3 camera for obtaining deep imaging of distinct faint objects. Parallel observations using \textit{HST}/ACS were also carried out, to provide multi-wavelength coverage of galaxies. The survey covered 800 sq. arcmin. of the sky in five fields: GOODS \citep[both North and South;][]{Giavalisco_2004}, UDS \citep{lawrence_2007}, EGS \citep{Davis_2007}, and COSMOS \citep{Scoville_2007}. The \textit{HST} filters used in CANDELS are F435W, F606W, F775W, F814W, and F850LP for \textit{HST}/ACS; and F098M, F105W, F125W, and F160W filters for \textit{HST}/WFC3. Table \ref{tabNIRCam} gives the $5\,\sigma$ point source sensitivity values of the \textit{HST} filters used, from \cite{Guo_2013}. The CANDELS/GOODS survey also has mid-/far-IR observations from \textit{Spitzer} (MIPS - 24 $\mu\rm m$, 70 $\mu\rm m$) and \textit{Herschel} (PACS - 100 $\mu\rm m$, 160 $\mu\rm m$, SPIRE - 250 $\mu\rm m$, 350 $\mu\rm m$ and 500 $\mu\rm m$). The table also provides the $5\,\sigma$ limits for the FIR filters, from \cite{Barro_2019}.  

The \textit{JWST} Advanced Deep Extragalactic Survey \citep[JADES;][]{eisenstein_overview_2023} is a joint project of the NIRCam \citep{Rieke_2023} and NIRSpec \citep{Jakobsen_2022} Instrument Development teams. JADES is the largest program being conducted in the Cycle 1 run of \textit{JWST} with an observing time of 770 hours; it observes two of the best-studied deep fields, namely GOODS-South and GOODS-North. The main focus is on GOODS-S, which includes the \textit{Chandra} Deep Field South \citep{Giacconi_2002} and the \textit{Hubble} Ultra Deep field \citep{Beckwith_2006}. Along with that, deep \textit{ALMA} \citep{franco_goods-alma_2018, hatsukade_alma_2018} and \textit{JVLA} data \citep{Alberts_2020} are available for this field. The GOODS-N, which contains the Hubble deep field and was covered by a deep Chandra survey, was chosen as the second field.

The design of JADES is like a two-layer wedding cake comprising deep portions of both imaging and spectroscopic data, along with medium-depth surveys over larger areas. The continuous portions of NIRCam imaging are named as ``prime'' while those with parallel NIRSpec exposures are designated as ``parallel''. JADES-Deep comprises a survey area of 36 sq. arcmin. while JADES-Medium covers 175 sq. arcmin.. The JADES-Deep survey will be carried out only on GOODS-S. The survey area of JADES-Medium is almost equally distributed between the GOODS-N and the GOODS-S fields. NIRCam imaging in JADES is done using nine filters namely F090W, F115W, F150W, F200W, F277W, F335M, F356W, F410M and F444W. The mean wavelength and the $5\,\sigma$ depth for each filter are given in Table \ref{tabNIRCam}.



\setlength{\tabcolsep}{1.5pt} 
\renewcommand{\arraystretch}{1} 
\begin{table}
\caption{\textit{HST}, \textit{JWST}, \textit{Spitzer} and \textit{Herschel} filters along with their central wavelengths and $5\,\sigma$ depths ($5\,\sigma$ depths are in AB magnitudes except for \textit{Spitzer} and \textit{Herschel}). The \textit{JWST} limiting magnitudes refer to JADES-Medium.}
\label{tabNIRCam}
\begin{tabular}{lcc}
\hline\hline
Filter & Central & $5\,\sigma$ depth\\
& Wavelength&\\
&$\hspace{1ex}(\mu \rm m)$&\\
\hline
F435W/ACS/\textit{HST} & 0.436 & 28.95\\
F606W/ACS/\textit{HST} & 0.603 & 29.35\\
F775W/ACS/\textit{HST} & 0.773 & 28.55\\
F814W/ACS/\textit{HST} & 0.813 & 28.84\\
F850LP/ACS/\textit{HST} & 0.908 &  28.55\\
F098M/WFC3/\textit{HST} & 0.990 &  28.77\\
F105W/WFC3/\textit{HST} & 1.065 &  28.45\\
F125W/WFC3/\textit{HST} & 1.257 &  28.34\\
F160W/WFC3/\textit{HST} & 1.543 &  28.16\\
F090W/NIRCam/\textit{JWST} & 0.98 &  28.85\\
F115W/NIRCam/\textit{JWST} & 1.162 &  28.98\\
F150W/NIRCam/\textit{JWST} & 1.510 &  28.93\\
F200W/NIRCam/\textit{JWST} & 2.002 &  29.02\\
F277W/NIRCam/\textit{JWST} & 2.784 &  29.32\\
F335M/NIRCam/\textit{JWST} & 3.367 &  28.89\\
F356W/NIRCam/\textit{JWST} & 3.593 &  29.32\\
F410M/NIRCam/\textit{JWST} & 4.088 &  28.87\\
F444W/NIRCam/\textit{JWST} & 4.439 &  29.03\\
\textit{Spitzer}/MIPS & 24 &  0.03 mJy\\
\textit{Spitzer}/MIPS & 70 &  2.5 mJy\\
\textit{Herschel}/PACS & 100 &  1.1 mJy\\
\textit{Herschel}/PACS & 160 &  3.4 mJy\\
\textit{Herschel}/SPIRE & 250 &  8.3 mJy\\
\textit{Herschel}/SPIRE & 350 &  11.5 mJy\\
\textit{Herschel}/SPIRE & 500 &  11.3 mJy\\
\hline
\end{tabular}
\end{table}

\subsection{Simulation set-up}
\label{simsetup}

A source in the simulated catalogue is said to be detected by \textit{JWST} if its flux density is $> 5\,\sigma$ in all the NIRCam bands. To incorporate \textit{HST} photometry, we apply the $5\,\sigma$ detection criteria to ACS/F435W and WFC3/F160W filters on \textit{JWST} detected sources. For {\it Herschel} we adopt the $5\,\sigma$ detection limit in the SPIRE $250\,\mu\rm m$ band. Among these sources, we say a source is detected by \textit{Spitzer}/MIPS if it has a flux density $>5\,\sigma$ at both 24 $\mu\rm m$ and 70 $\mu\rm m$. 


To investigate the detectability of proto-spheroids by \textit{JWST}, we have simulated galaxies having virialised halo mass in the range $11.3\leq\log(M_{\rm vir}/M_{\odot})\leq13.3$. We call this the ``parent sample''; it comprises 27748 galaxies detected at $5\,\sigma$ in all the 9 NIRCam bands.  Approximately $41\%$ of them are also detected by \textit{HST} above $5\,\sigma$ in the F435W and F160W bands. However, only $1.5\%$ and $1.7\%$ of these \textit{JWST} sources are detected above $5\,\sigma$ by \textit{Spitzer} and \textit{Herschel} respectively. While \textit{JWST} can help to constrain the stellar mass of galaxies, FIR observations are crucial in constraining the SFR of DSFGs. 

To get a larger fraction of detections by \textit{Herschel} and \textit{Spitzer}, we simulated all galaxies with $12\leq\log(M_{\rm vir}/M_{\odot})\leq13.3$; galaxies with $\log(M_{\rm vir}/M_{\odot})<12$ are undetected by {\it Herschel}. In this sub-sample there are 507 galaxies above the {\it Herschel} $5\sigma$ detection limit at $250\,\mu$m; these constitute our DSFG sample. 
Among them, 503 galaxies, approximately 99$\%$, are detected at $5\,\sigma$ in all the 9 NIRCam bands. {\it HST} detects $\approx98.8\%$ of these DSFGs, while 434 ($\approx 86\%$) are detected in both \textit{Spitzer}/MIPS bands at $>5\,\sigma$.

\subsection{Estimation of Photometric Redshifts and Galaxy Physical Properties}
\label{secsoftawres}

To estimate the physical properties of galaxies, the knowledge of redshift is essential. Therefore, the first thing that we estimated for our simulated sample was the redshift from the \textit{JWST} photometry. Since the \textit{HST}photometry is available for most of the sources, we also estimated the photometric redshift by complementing the \textit{JWST} photometry with that from \textit{HST}.

We used the template-based photo-$z$ estimation code EAZY \citep[Easy and Accurate Redshifts from Yale;][]{brammer_eazy_2008}. This code is applicable for the estimation of photo-$z$ using UV/optical/near-IR photometric data. We use "v1.3" of EAZY which uses 9 templates. The original 5 templates of "v1.0" of EAZY are from \cite{grazian_goods-music_2006}. The dusty galaxies are taken into account by adding a starburst template with $t=50$ Myr and $A_V=2.75$. The evolved simple stellar population (SSP) model by \cite{maraston_evolutionary_2005} is added to include massive old galaxies at $z<1$. The dust template is taken from \cite{erb_physical_2010}. The \cite{madau_radiative_1995} model for dust absorption by the intergalactic medium (IGM) is also taken into account in EAZY. A redshift range of $z=1$ to 8 in steps of $\Delta z=0.01$ was chosen.

After estimating the photo-$z$, we proceeded to estimate the physical properties of the simulated galaxies. For this, we used the SED fitting code CIGALE \citep[\textbf{C}ode \textbf{I}nvestigating \textbf{GAL}axy \textbf{E}mission;][]{burgarella_star_2005, noll_analysis_2009, boquien_cigale_2019} which uses the principle of ``energy balance'' i.e., the amount of stellar energy absorbed by dust in the UV/optical regime is entirely processed and re-emitted at far-IR/sub-mm wavelengths, to build the panchromatic model of the SEDs of galaxies. The SED templates generated by CIGALE depend on the modules and values of different parameters chosen by the user. In this work, we adopted a delayed SFH with an additional late burst of star formation. The \cite{bruzual_stellar_2003} SSP models along with a Chabrier initial mass function \citep[IMF;][]{chabrier_galactic_2003} and solar metallicity are chosen. The \cite{charlot_simple_2000} model is used for modelling the dust attenuation. The dust emission is modelled following \cite{draine_andromedas_2014}. To incorporate the contribution from the AGN, templates from \cite{fritz_revisiting_2006} are used. With the above-chosen modules and parameter values (see Table \ref{tabcigale}), CIGALE generated around 6 billion SED templates and then used a $\chi^2$ minimisation technique to get the best fit SED and estimate the physical properties \citep{noll_analysis_2009}. 

\begin{table*}
\caption{Parameter values given as input to CIGALE for SED fitting}
\label{tabcigale}
\begin{tabular}{ll}
\hline\hline
\multicolumn{2}{l}{\bf{SFH : sfhdelayed - delayed SFH with optional exponential burst}}                                                                                                \\ \hline\hline
\multicolumn{1}{l}{e-folding time of the main stellar population (Myr)}    & 500, 1000, 2000, 3000, 4000, 5000,\\
 & 6000, 7000                                    \\ 
\multicolumn{1}{l}{e-folding time of the late starburst population (Myr)}  & 1000, 5000, 10000                                          \\ 
\multicolumn{1}{l}{mass fraction of the late starburst population}   & 0.0, 0.001, 0.01, 0.1, 0.15, 0.3, 0.5                  \\ 
\multicolumn{1}{l}{age of the main stellar population (Myr)}               & 500, 1000, 2000, 3000, 5000, 6000,\\
 & 7000, 8000, 9000, 10000 \\ 
\multicolumn{1}{l}{age of the late starburst (Myr)}                        & 10, 30, 50, 70, 100, 150, 300                                   \\ \hline\hline
\multicolumn{2}{l}{\bf{SSP : bc03} \citep{bruzual_stellar_2003}}                                                                                                   \\ \hline\hline
\multicolumn{1}{l}{Initial mass function (IMF)}                            & Chabrier \citep{chabrier_galactic_2003}                                                 \\ 
\multicolumn{1}{l}{Metallicity}                                      & 0.02 ($Z_{\odot}$) , 0.008                                                    \\ \hline\hline
\multicolumn{2}{l}{\bf{Dust attenuation :  dustatt\_modified\_CF00} \citep{charlot_simple_2000}}                                                                  \\ \hline\hline
\multicolumn{1}{l}{V-band attenuation in ISM ($A_{V}^{\rm ISM}$)}                        & 0.3, 0.5, 0.9, 1.1, 1.7, 2.0                                \\ 
\multicolumn{1}{l}{$\mu$}                                            & 0.44                                                     \\ 
\multicolumn{1}{l}{power law slope of attenuation in the ISM}        & $-0.7$, $-1.3$                                                     \\ 
\multicolumn{1}{l}{power law slope of attenuation in the stellar birth clouds (BCs)}         & $-0.7$, $-1.3$                                                     \\ \hline\hline
\multicolumn{2}{l}{\bf{Dust emission : dl2014} \citep{draine_andromedas_2014}}                                                                                       \\ \hline\hline
\multicolumn{1}{l}{PAH mass fraction ($q_{\rm \rm PAH}$, in $\%$)}                             & 2.5, 3.9, 4.58, 7.32                                        \\ 
\multicolumn{1}{l}{minimum radiation field ($U_{\rm min}$) (Habing)}                          & 20, 25, 50                                           \\ 
\multicolumn{1}{l}{dust emission power law slope ($\alpha$)}                          & 2, 3                                                      \\ 
\multicolumn{1}{l}{fraction illuminated from $U_{\rm min}$ to $U_{\rm max}$ ($\gamma$)} & 0.02                                                       \\ \hline\hline
\multicolumn{2}{l}{\bf{AGN : fritz2006} \citep{fritz_revisiting_2006}}                                                                                       \\ \hline\hline
\multicolumn{1}{l}{ratio of the maximum to minimum radii of the dusty torus ($r_{\rm ratio}$)}                             &  60, 100, 150                                          \\ 
\multicolumn{1}{l}{equatorial optical depth at $9.7$ $\mu$m ($\tau$)}                          & 0.6, 10.0                                             \\ 
\multicolumn{1}{l}{radial dust distribution within the torus ($\beta$)}                          &  0.0                                                       \\ 
\multicolumn{1}{l}{angular dust distribution within the torus ($\gamma$)} & 6                                                       \\
\multicolumn{1}{l}{full opening angle of the dusty torus (Opening angle)} & 100                                                       \\
\multicolumn{1}{l}{Angle between the equatorial axis and line of sight ($\Psi$)} & 0.001, 89.990                                                       \\
\multicolumn{1}{l}{AGN fraction ($f_{\rm AGN}$)} & 0.0, 0.1, 0.15, 0.25, 0.5                                                      \\\hline\hline\end{tabular}
\end{table*}


\begin{figure*}
    \centering
    \includegraphics[width=16cm,height=7.5cm]{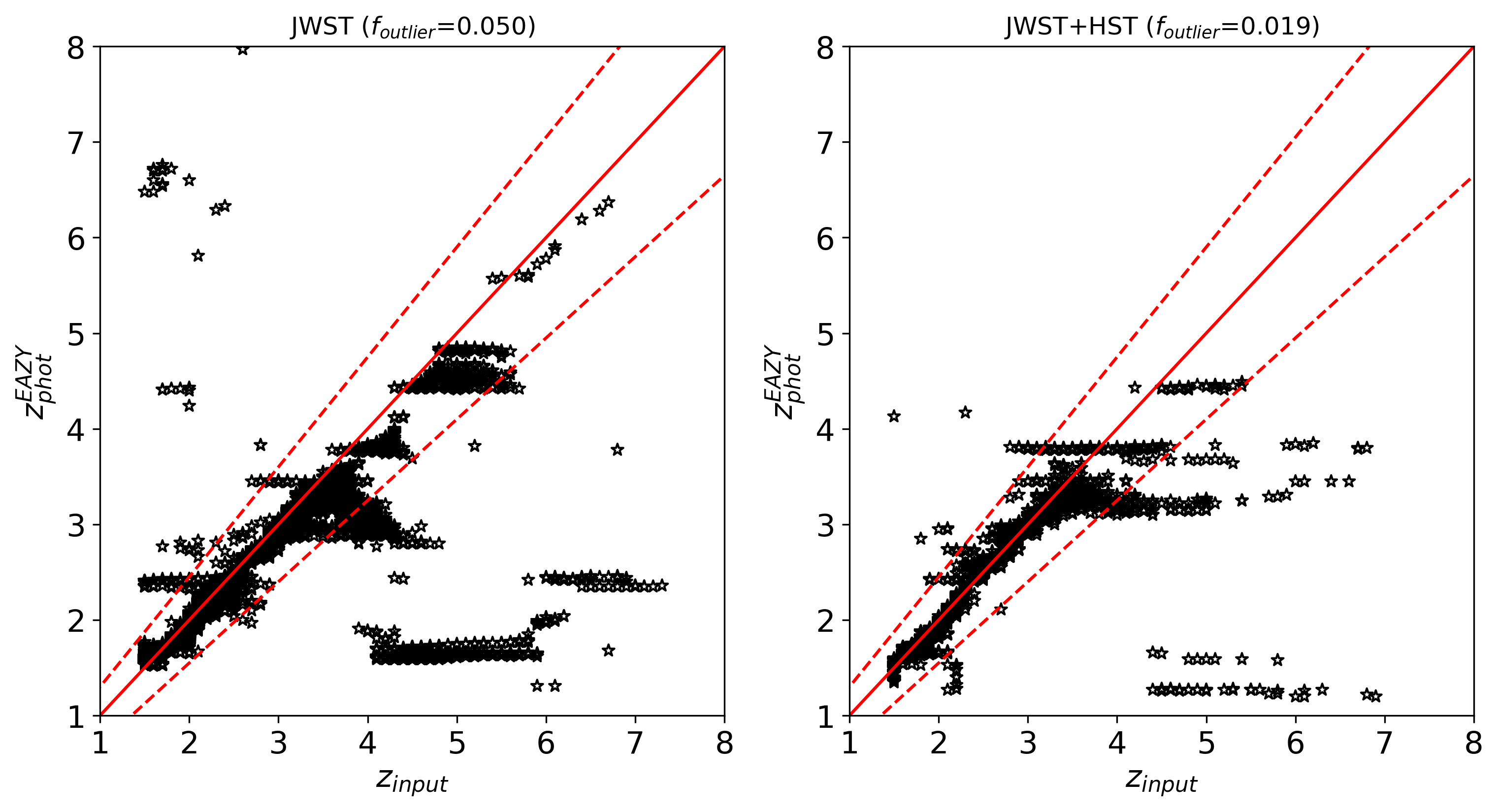}
      \includegraphics[width=16cm,height=7.5cm]{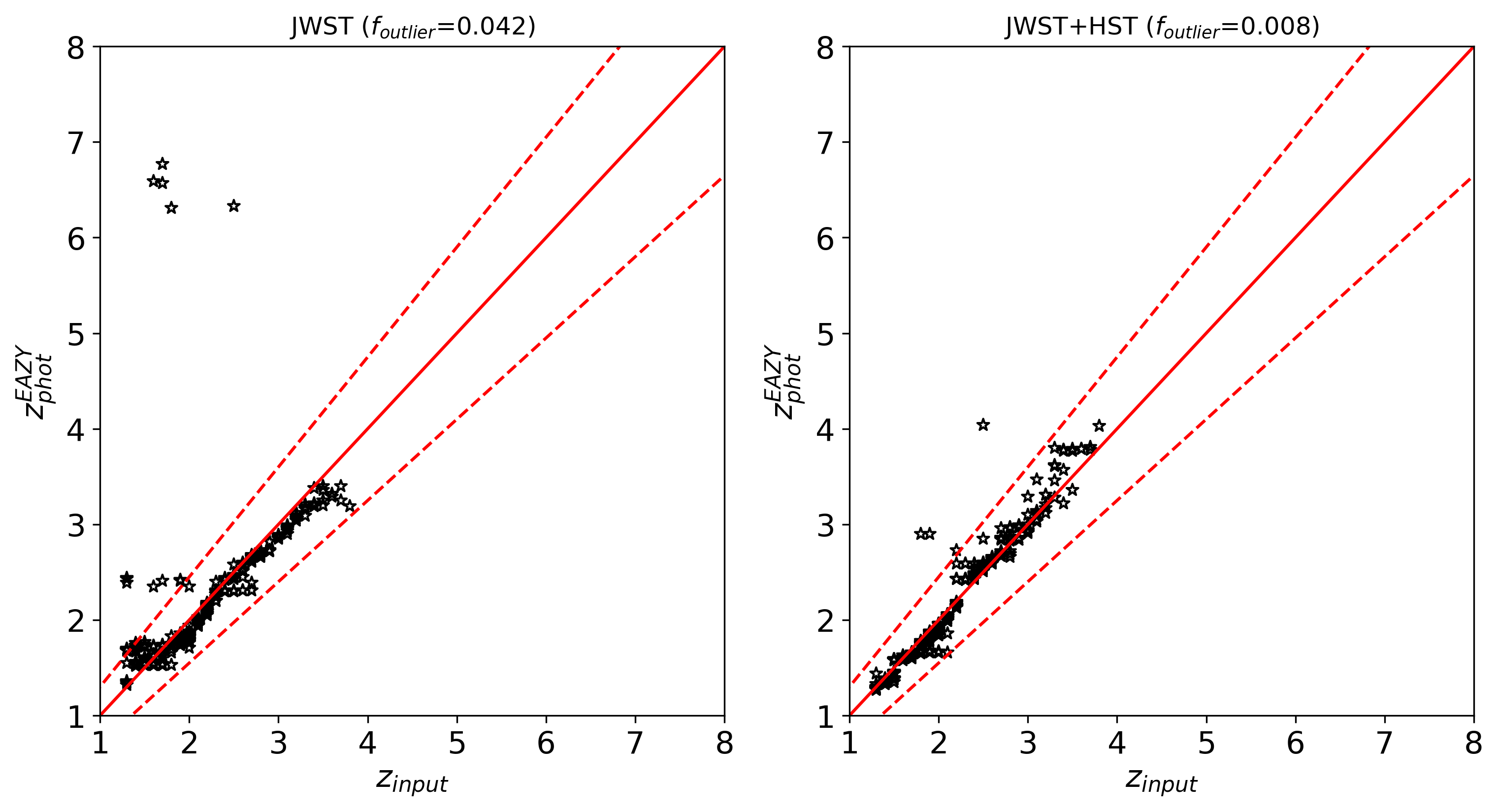}
      
      \caption{Derived photometric redshift using EAZY vs input redshift of the galaxies detected by \textit{JWST} at $>5\,\sigma$ in all 9 NIRCam bands, also detected by the \textit{HST} at $\ge 5\,\sigma$ in the F435W and F160W bands. The solid red line denotes $z_{\rm input}=z_{\rm phot}^{\rm EAZY}$ while the dashed red lines define the region where $|\Delta z| \leq 0.15(1+z_{\rm input})$.The top row refers to the parent sample while the bottom row is for DSFG sub-sample (detected at $\geq 5\,\sigma$ at $250\,\mu$m, i.e. having $S_{250\mu\rm m}\ge 8.3\,$mJy). Most catastrophic outliers (black patches towards the bottom right of the top panel), i.e. sources having $z_{\rm input}\gtrsim4$ and low $z_{\rm phot}^{\rm EAZY}$, are AGN dominated. In other cases, catastrophic errors are due to the degeneracy between the Lyman-$\alpha$ break at $912$ Å and the 4000\,Å break.}
      \label{figeazyphot}
\end{figure*}


\section{Results}
\label{sec4}

In this section, we discuss the results obtained for the photometric redshifts and for the physical properties of the \textit{JWST}-detected simulated galaxies.

\subsection{Derived photometric redshift}

The scatter plot of the photometric redshifts estimated from the \textit{JWST} photometry alone versus the input redshifts is shown in the left-hand panel of Figure \ref{figeazyphot} for both the parent catalogue (top panel) and the DSFG sub-sample (bottom panel) of galaxies. We observe that \textit{JWST} provides a good redshift estimate up to $z\sim4$ for both the parent sample and the sub-sample. The outlier fraction, defined as $|\Delta z|/(1+z_{\rm input}) > 0.15$ where $|\Delta z| = |z_{\rm input}-z_{\rm phot}^{\rm EAZY}|$ \citep{laigle_cosmos2015_2016}, is $f_{\rm outlier}=0.05$ for the parent sample and $f_{\rm outlier}=0.042$ for the DSFG sub-sample respectively. The most catastrophic errors on $z$ occur in the range $4\leq z\leq5-6$, where there is a leakage of high-$z$ sources to the low redshift regime. 
Also, a few galaxies at $z\sim2$ are wrongly assigned a redshift $z>4$. These wrong estimations of redshift are mostly due to the degeneracy between the Lyman-$\alpha$ break at $912$ Å and the 4000\,Å break. The right-hand panels of Figure \ref{figeazyphot} show the estimated versus true photo-$z$ values for the \textit{JWST} sources which are also detected by \textit{HST}. 
We observe that, on complementing the \textit{JWST} photometry with that from HST, most of the low-$z$ contaminants are removed and we get a very good level of agreement between the input and the estimated photo-$z$ values at all redshifts up to $z\leq5$. Thanks to the availability of the HST photometry, EAZY can properly sample the Lyman break and the degeneracy with the 4000\,Å break is rectified. This is because, at $z\sim4-5$, the Lyman break is missed by the \textit{JWST} NIRCam filters, but it falls within the \textit{HST} filters. However, for the parent sample, there are still some catastrophic redshift determinations. These are mostly AGN-dominated sources, the SEDs of which are featureless and power-law shaped, which makes it difficult for EAZY to give a correct estimation of redshift. The outlier fraction, for the \textit{JWST}+\textit{HST} photometry, is reduced to $f_{\rm outlier}=0.019$ and $f_{\rm outlier}=0.008$ for the parent sample and the DSFG sub-sample, respectively. Overall, the recovery of redshifts for DSFGs using {\it JWST} photometry demonstrates excellent performance, with high accuracy and remarkably low outlier fractions. The broad wavelength coverage and the high sensitivity of NIRCam enable reliable photometric redshift estimates even for heavily dust-obscured systems. This highlights {\it JWST}'s unique capability to precisely constrain the redshift distribution of DSFGs, which has been a major challenge for previous infrared and optical surveys.

For the DSFG sample, we attempted to estimate photometric redshifts using \textit{Herschel}/SPIRE photometry based on the \cite{pearson_h-atlas_2013} template set. However, this approach resulted in a high outlier fraction of approximately 33\%. Even after restricting the analysis to sources with additional $5\sigma$ detections at both 350 and 500 $\mu$m, the outlier fraction remained high, at 28\,\%. Due to this significant level of inaccuracy, we do not adopt the FIR-based redshift estimates in our analysis. Instead, we rely on photometric redshifts estimated using EAZY applied to \textit{JWST} photometry for the rest of our analysis.

\subsection{Derived physical properties}

The estimation of physical properties like stellar mass ($M_{\star}$), SFR ($\dot{M}_{\star}$), dust luminosity ($L_{\rm dust}$) and dust mass ($M_{\rm dust}$), of the \textit{JWST}-detected simulated galaxies is done using CIGALE by setting the redshift to the value derived by EAZY. Also, we discuss the improvement that \textit{JWST} NIRCam photometry brings to the constraints on these physical parameters when complemented with the existing multiwavelength data from \textit{HST}, \textit{Spitzer} and \textit{Herschel}. We make a comparative study of different combinations of photometry and analyse how \textit{JWST} improves on those values. For each of the physical quantity, say $P$, we define the ratio of estimated ($P^{\rm CIGALE}$) and true value ($P^{\rm input}$) as 
\begin{equation}
Q_{\log P}=\log\left(\frac{P^{\rm CIGALE}}{P^{\rm input}}\right)
\end{equation}
We plot this ratio as a function of the true value. Moreover, we plot the histogram of this ratio. For each of the recovered physical properties, the median value along with the $1\,\sigma$ dispersion are reported in Table \ref{tabmedianvalues}. Moreover, the median and RMS values of 
$Q_{\log P}$ for the different band combinations are explicitly reported in Table \ref{tabQvalues}.

As far as the best-fit SED is concerned, we get a median reduced $\chi^2$ value of $\sim 0.9-1.08$ for {\it JWST} and {\it JWST}+{\it HST} photometry, but when adding FIR photometry the value increases to $\approx 5$. The increase in reduced $\chi^2$ when adding {\it Spitzer} and {\it Herschel} photometry to {\it JWST} and {\it HST} data in CIGALE betrays the fact that the UV-NIR and the far-IR/sub-mm bands preferentially see the emission from low- and high-obscuration regions, respectively. Also, while the {\it JWST} and the {\it HST} photometry are high-resolution and well-matched in aperture, the far-IR/sub-mm data have larger beam sizes and potential blending issues. Our model does not allow us to deal with these issues because it yields a single, spatially-integrated SED for each source, i.e. the UV/NIR and FIR/sub-mm emission are co-located by construction.

CIGALE enforces energy balance between absorbed stellar light and dust re-emission. If the UV/optical fit suggests low dust attenuation, the model may struggle to match the observed IR fluxes, especially when the FIR SED shape or amplitude deviates from the assumed templates. Despite this higher reduced $\chi^2$, the uncertainties (dispersions) in key physical parameters like stellar mass and SFR, which are the main focus of our study, remain low. This indicates that while the exact SED shape may not be perfectly reproduced across all bands, the overall physical interpretation is robust.

\begin{table*}
\caption{Median values along with the median absolute deviation of the estimated physical quantities for the parent sample (where the simulated galaxies have virialised halo mass in the range $11.3\leq\log(M_{\rm vir}/M_{\odot})\leq13.3$) and the DSFG sub-sample (having flux density $\gtrsim8.3$ mJy at $250\,\mu$m), respectively.}
\label{tabmedianvalues}

\begin{tabular}{lc|c|c|c|c}
\hline
\hline
\multicolumn{1}{l}{}                          & \multicolumn{1}{l}{$\log(M_{\star}/M_{\odot})$}  & \multicolumn{1}{l}{$\log(\dot{M}_{\star}/M_{\odot}\hbox{yr}^{-1})$} & \multicolumn{1}{l}{$\log(L_{\rm dust}/L_{\odot})$} & \multicolumn{1}{l}{$\log(M_{\rm dust}/M_{\odot})$} \\ \hline
\multicolumn{5}{c}{Parent Sample}    \\ \hline
\multicolumn{1}{l}{\textit{JWST}}                      & \multicolumn{1}{c}{$9.81\pm0.16$}                & \multicolumn{1}{c}{-}                                                                                  & \multicolumn{1}{c}{-}                          & \multicolumn{1}{c}{-}                                               \\ 
\multicolumn{1}{l}{\textit{JWST}+\textit{HST}}                  & \multicolumn{1}{c}{$9.89\pm0.23$}                & \multicolumn{1}{c}{-}                                                                                   & \multicolumn{1}{c}{-}                          & \multicolumn{1}{c}{-}                                                \\ \hline
\multicolumn{5}{c}{Sub-sample}       \\ \hline
\multicolumn{1}{l}{\textit{JWST}}                      & \multicolumn{1}{c}{$10.8\pm0.23$} & \multicolumn{1}{c}{-}                                                                                   & \multicolumn{1}{c}{-}                          & \multicolumn{1}{c}{-}                                                \\ 
\multicolumn{1}{l}{\textit{JWST}+\textit{HST}}                  & \multicolumn{1}{c}{$10.86\pm0.23$}               & \multicolumn{1}{c}{-}                                                                                   & \multicolumn{1}{c}{-}                          & \multicolumn{1}{c}{-}                                                \\ 
\multicolumn{1}{l}{\textit{Spitzer}+\textit{Herschel}}          & \multicolumn{1}{c}{-}                            & \multicolumn{1}{c}{$2.21\pm0.3$}                                                             & \multicolumn{1}{c}{$12.34\pm0.26$}              & \multicolumn{1}{c}{$8.5\pm0.26$}                                    \\ 
\multicolumn{1}{l}{\textit{HST}+\textit{Spitzer}+\textit{Herschel}}      & \multicolumn{1}{c}{$10.8\pm0.3$}               & \multicolumn{1}{c}{$2.5\pm0.21$}                                                           & \multicolumn{1}{c}{$12.5\pm0.23$}              & \multicolumn{1}{c}{$8.5\pm0.23$}                                   \\ 
\multicolumn{1}{l}{\textit{JWST}+\textit{Spitzer}+\textit{Herschel}}     & \multicolumn{1}{c}{$10.83\pm0.27$}               & \multicolumn{1}{c}{$2.52\pm0.22$}                                                           & \multicolumn{1}{c}{$12.48\pm0.22$}              & \multicolumn{1}{c}{$8.6\pm0.22$}                                    \\ 
\multicolumn{1}{l}{\textit{JWST}+\textit{HST}+\textit{Spitzer}+\textit{Herschel}} & \multicolumn{1}{c}{$10.83\pm0.27$}               & \multicolumn{1}{c}{$2.52\pm0.23$}                                                           & \multicolumn{1}{c}{$12.5\pm0.23$}              & \multicolumn{1}{c}{$8.6\pm0.23$}                                    \\ \hline\hline
\end{tabular}
\end{table*}

\begin{table*}
\caption{Mean \textbf{(median)} values of $Q_{\log P}$ for the parent sample (where the simulated galaxies have virialised halo mass in the range $11.3\leq\log(M_{\rm vir}/M_{\odot})\leq13.3$) and the DSFG sub-sample (having $5\,\sigma$ flux density $\gtrsim8.3$ mJy at $250\,\mu$m), respectively. The corresponding plots are shown in Figure \ref{figboxsmall}, \ref{figboxsfrsub} and \ref{figboxdustsub} respectively.}
\label{tabQvalues}

\begin{tabular}{lc|c|c|c|c}
\hline
\hline
\multicolumn{1}{l}{}                          & \multicolumn{1}{l}{$Q_{\log(M_{\star})}$}  & \multicolumn{1}{l}{$Q_{\log(\dot{M}_{\star})}$} & \multicolumn{1}{l}{$Q_{\log(L_{\rm dust})}$} & \multicolumn{1}{l}{$Q_{\log(M_{\rm dust})}$} \\ \hline
\multicolumn{5}{c}{Parent Sample}    \\ \hline
\multicolumn{1}{l}{\textit{JWST}}                      & \multicolumn{1}{c}{$-0.14 \mathbf{(-0.132)}$}                & \multicolumn{1}{c}{-}                                                                                  & \multicolumn{1}{c}{-}                          & \multicolumn{1}{c}{-}                                               \\ 
\multicolumn{1}{l}{\textit{JWST}+\textit{HST}}                  & \multicolumn{1}{c}{$-0.068 \mathbf{(-0.083)}$}                & \multicolumn{1}{c}{-}                                                                                   & \multicolumn{1}{c}{-}                          & \multicolumn{1}{c}{-}                                                \\ \hline
\multicolumn{5}{c}{Sub-sample}       \\ \hline
\multicolumn{1}{l}{\textit{JWST}}                      & \multicolumn{1}{c}{$-0.004 \mathbf{(-0.017)}$} & \multicolumn{1}{c}{-}                                                                                   & \multicolumn{1}{c}{-}                          & \multicolumn{1}{c}{-}                                                \\ 
\multicolumn{1}{l}{\textit{JWST}+\textit{HST}}                  & \multicolumn{1}{c}{$0.035 \mathbf{(0.055)}$}               & \multicolumn{1}{c}{-}                                                                                   & \multicolumn{1}{c}{-}                          & \multicolumn{1}{c}{-}                                                \\ 
\multicolumn{1}{l}{\textit{Spitzer}+\textit{Herschel}}          & \multicolumn{1}{c}{-}                            & \multicolumn{1}{c}{$-0.196 \mathbf{(-0.24)}$}                                                             & \multicolumn{1}{c}{$0.075 \mathbf{(0.085)}$}              & \multicolumn{1}{c}{$-0.012 \mathbf{(-0.033)}$}                                    \\ 
\multicolumn{1}{l}{\textit{HST}+\textit{Spitzer}+\textit{Herschel}}      & \multicolumn{1}{c}{$0.01 \mathbf{(-0.02)}$}               & \multicolumn{1}{c}{$0.037 \mathbf{(0.056)}$}                                                           & \multicolumn{1}{c}{$-0.006 \mathbf{(-0.077)}$}              & \multicolumn{1}{c}{$0.124 \mathbf{(0.12)}$}                                   \\ 
\multicolumn{1}{l}{\textit{JWST}+\textit{Spitzer}+\textit{Herschel}}     & \multicolumn{1}{c}{$-0.0169 \mathbf{(-0.039)}$}               & \multicolumn{1}{c}{$0.044 \mathbf{(0.032)}$}                                                           & \multicolumn{1}{c}{$-0.049 \mathbf{(-0.045)}$}              & \multicolumn{1}{c}{$0.15 \mathbf{(0.13)}$}                                    \\ 
\multicolumn{1}{l}{\textit{JWST}+\textit{HST}+\textit{Spitzer}+\textit{Herschel}} & \multicolumn{1}{c}{$-0.04 \mathbf{(-0.052)}$}               & \multicolumn{1}{c}{$0.053 \mathbf{(0.066)}$}                                                           & \multicolumn{1}{c}{$-0.066 \mathbf{(-0.077)}$}              & \multicolumn{1}{c}{$0.13 \mathbf{(0.13)}$}                                    \\ \hline\hline
\end{tabular}
\end{table*}

\subsubsection{Stellar Mass}

\begin{figure*}
    \centering
\includegraphics[width=.42\textwidth]{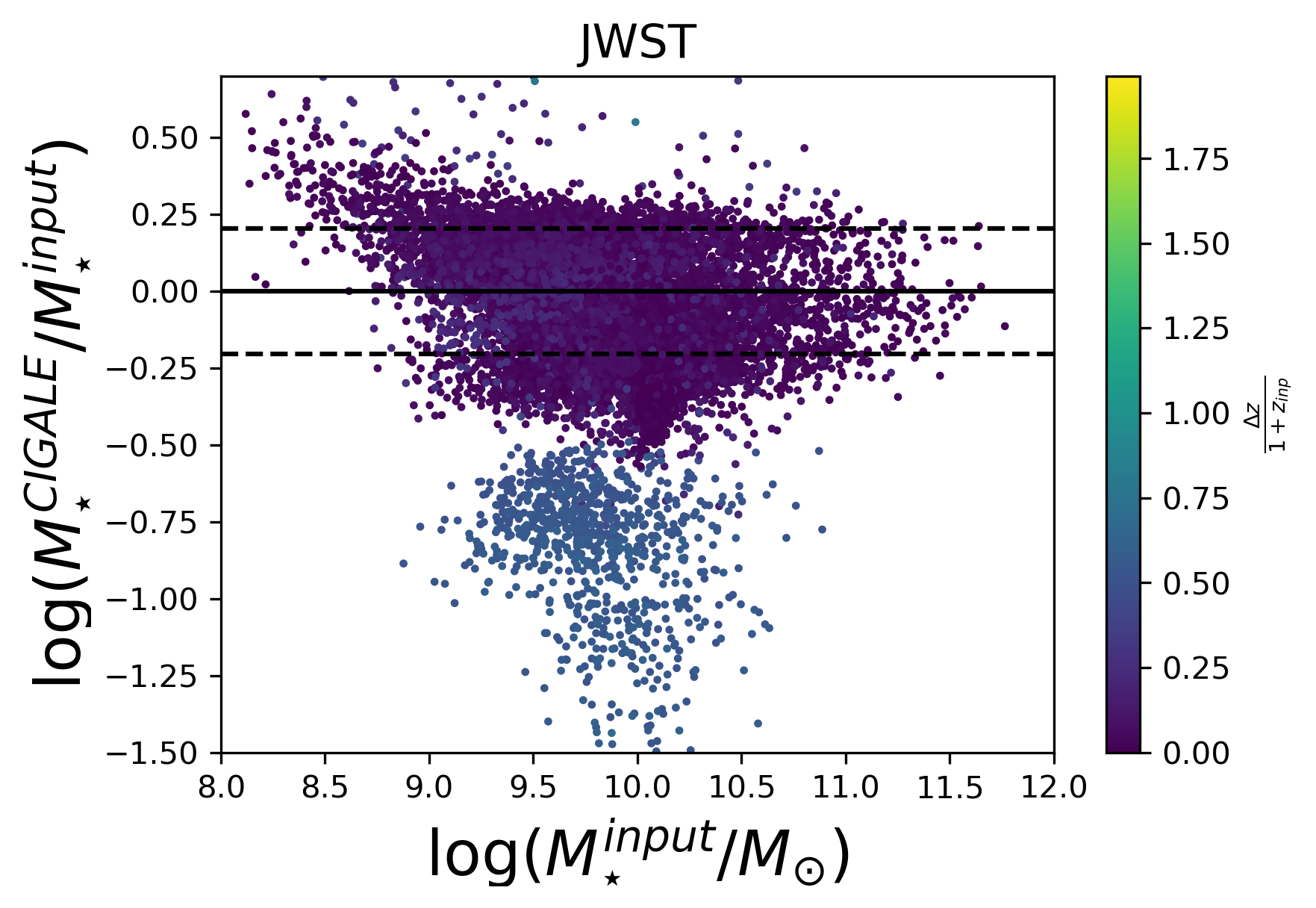}
\includegraphics[width=.4\textwidth]{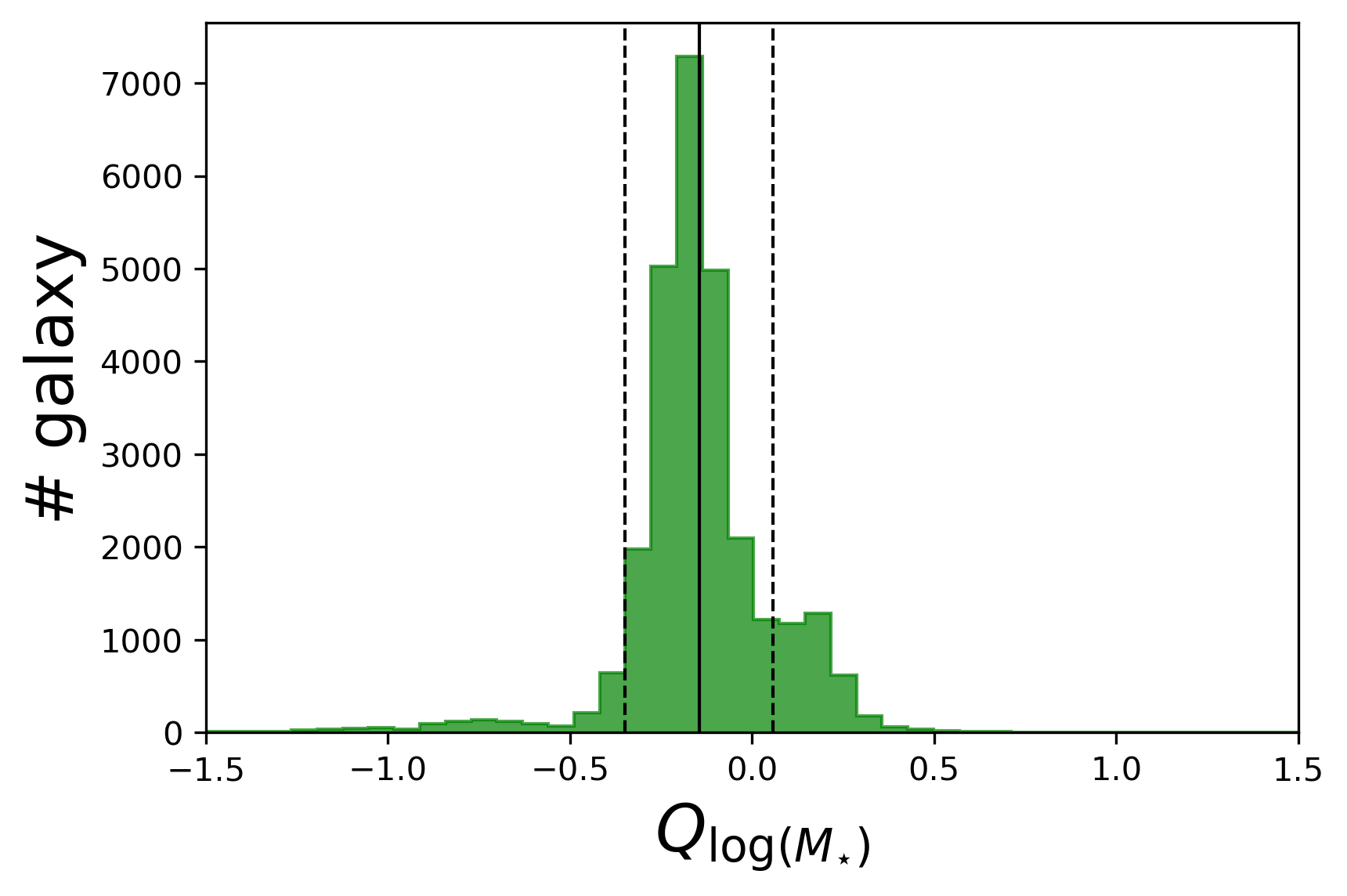}
\includegraphics[width=.42\textwidth]{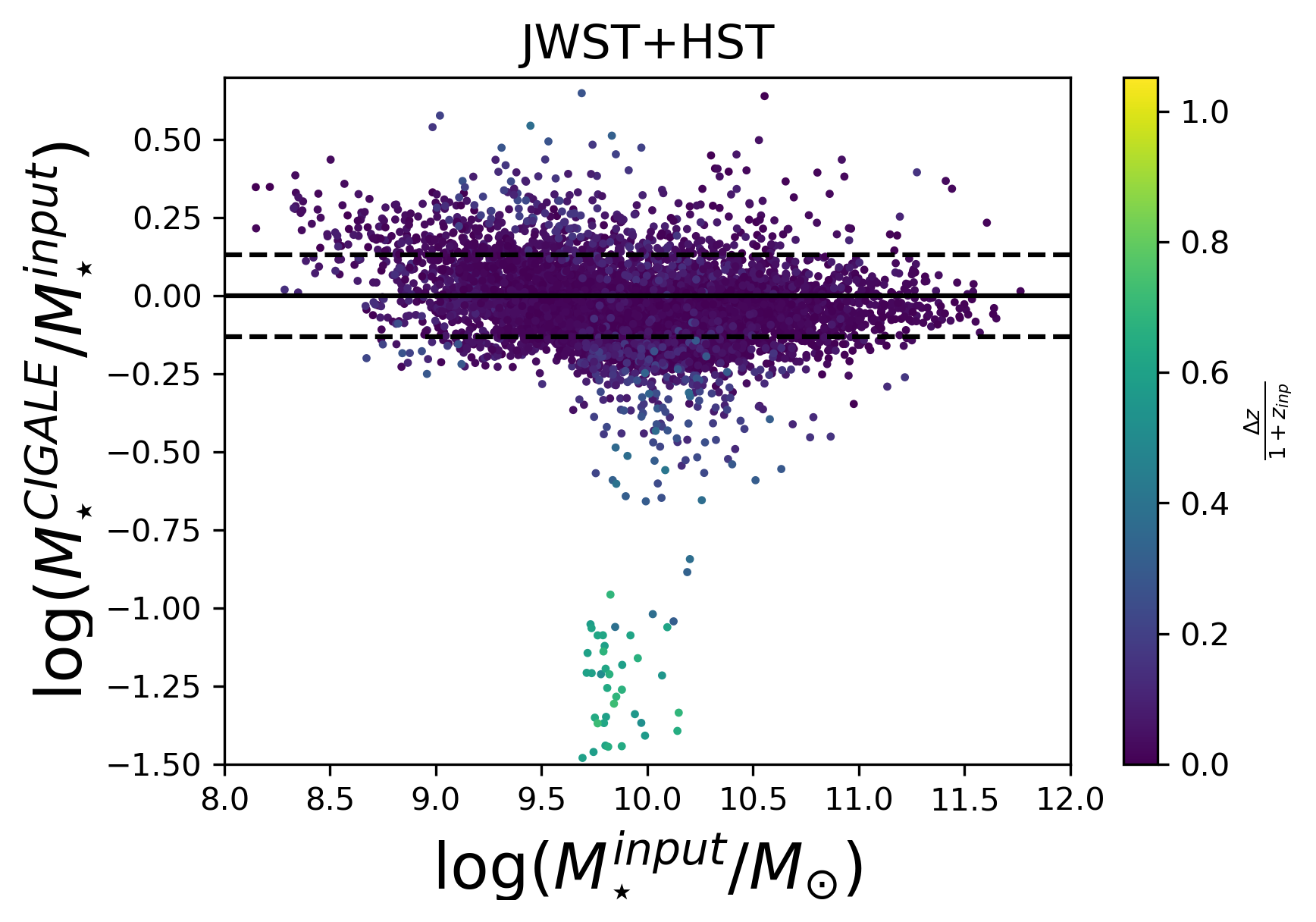}
\includegraphics[width=.4\textwidth]{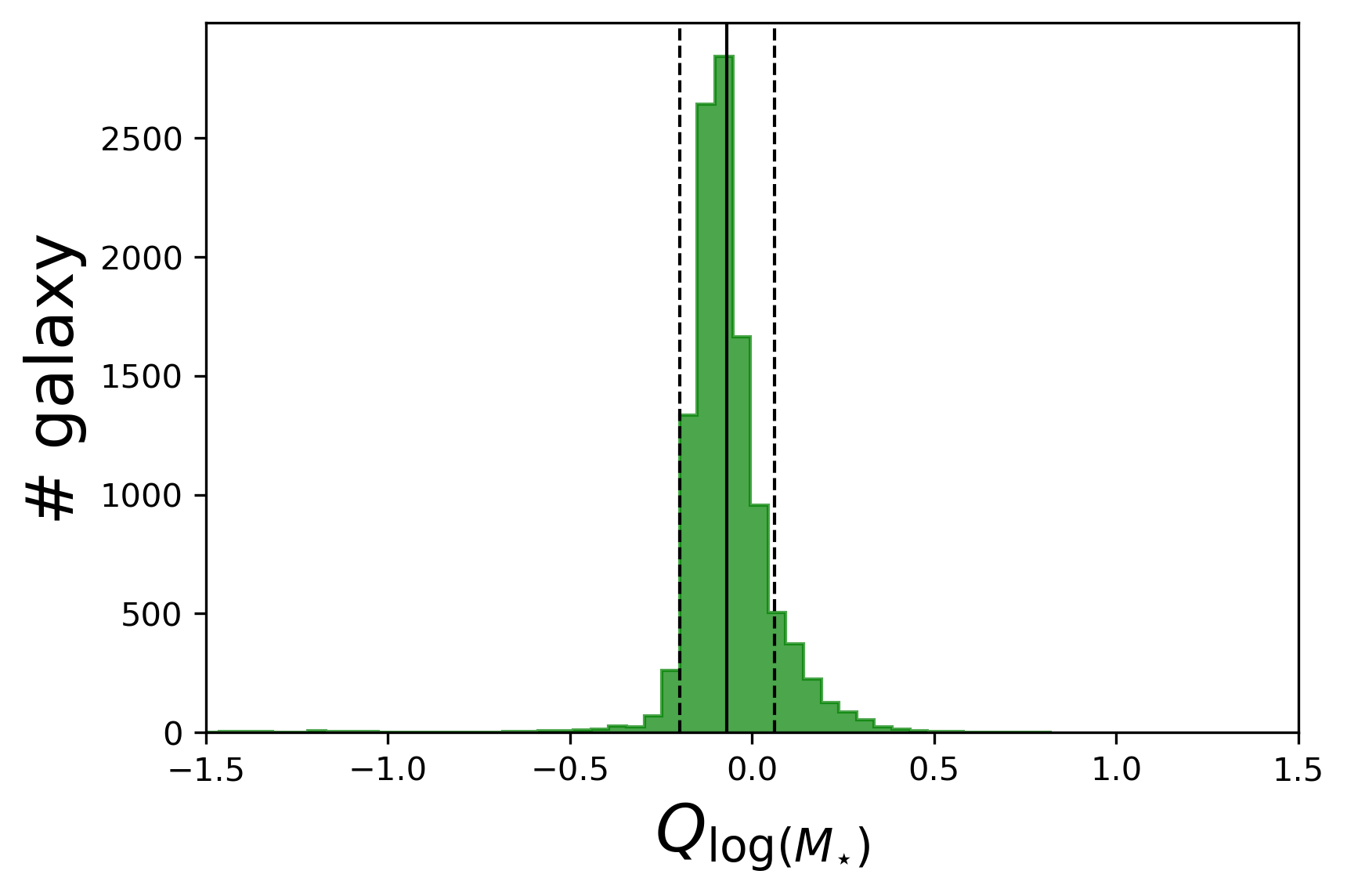}
      \caption{Scatter plot of the logarithm of the ratio between the estimated stellar mass ($M_{\star}^{\rm CIGALE}$) and the input stellar mass ($M_{\star}^{\rm input}$) as a function of $M_{\star}^{\rm input}$ for the parent sample of galaxies (i.e., $\log(M_{\rm vir}/M_{\odot})\geq11.3$) from the \textit{JWST} photometry (top row) and the \textit{JWST}+\textit{HST} photometry (bottom row). The points are colour-coded with their values of $|\Delta z|/(1+z_{\rm input})$. The solid black line marks the locus of equality between the model values and the CIGALE estimated values. The boundaries of the $1\,\sigma$ dispersion around the mean are shown by the black dashed lines. The histograms of $Q_{\log (M_{\star})}$ are shown in the right-hand panels. The black solid line marks the mean value while the dashed lines denote the $1\,\sigma$ dispersion around the mean. Overall, CIGALE leans towards a slight but significant systematic underestimate of the stellar mass. In the top left panel, the small dark blue patch of sources having $\log\left(M_{\star}^{\rm CIGALE}/M_{\star}^{\rm input}\right)<-0.5$ and $9\leq \log(M_{\star}^{\rm input}/M_{\odot})\leq11$ is due to the sources which are AGN dominated and have catastrophic photo-$z$ estimation errors ($|\Delta z|/(1+z_{\rm input})\gtrsim0.5$) by EAZY. Similarly, such outliers are present in the bottom left panel seen as yellowish green patch ($|\Delta z|/(1+z_{\rm input})\gtrsim0.6$) with $\log\left(M_{\star}^{\rm CIGALE}/M_{\star}^{\rm input}\right)<-1$ and $9.5\leq \log(M_{\star}^{\rm input}/M_{\odot})\leq10.5$.}
      \label{figcigsmall}
\end{figure*}

\begin{figure*}
    \centering
\includegraphics[width=.42\textwidth]{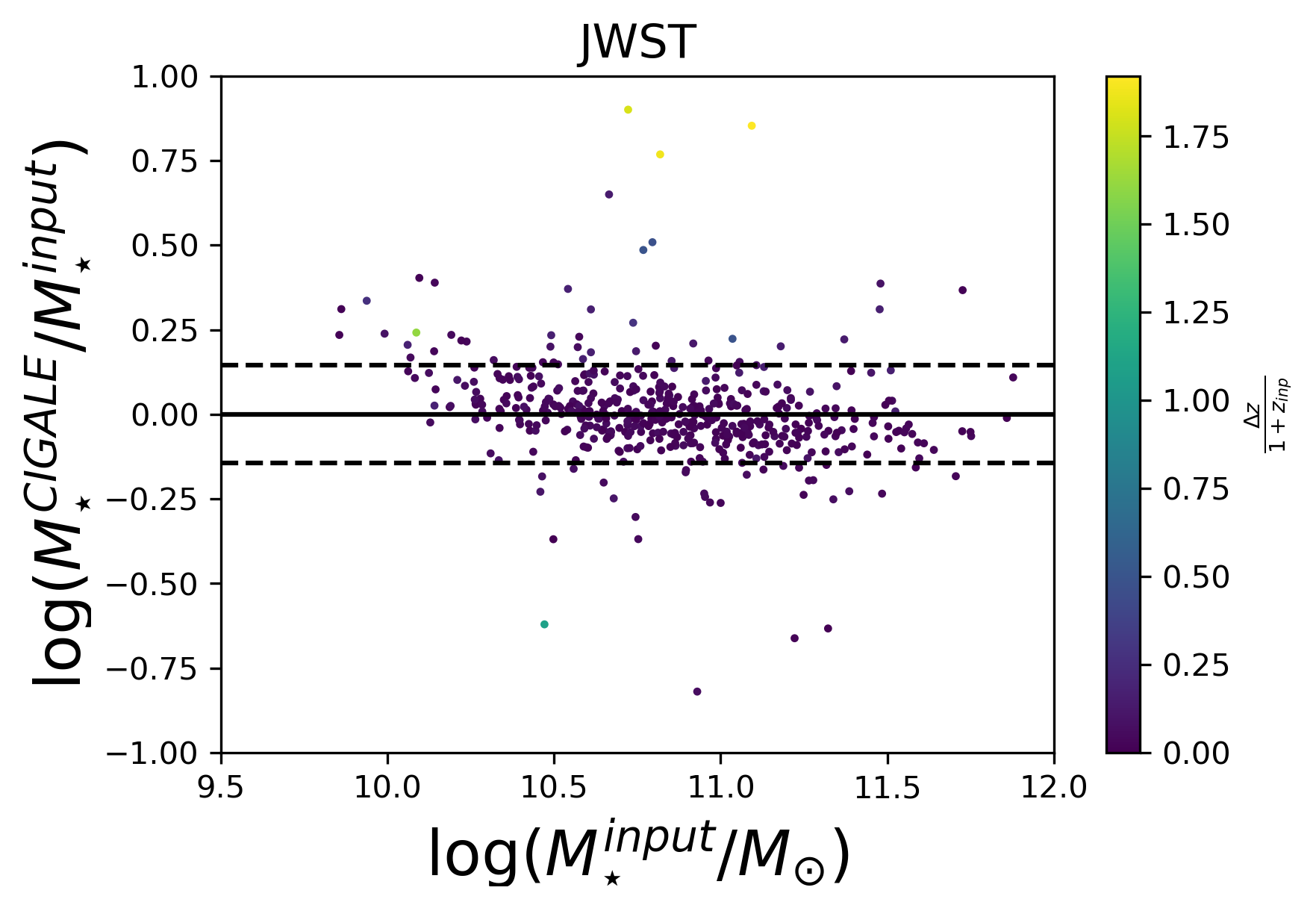}
\includegraphics[width=.4\textwidth]{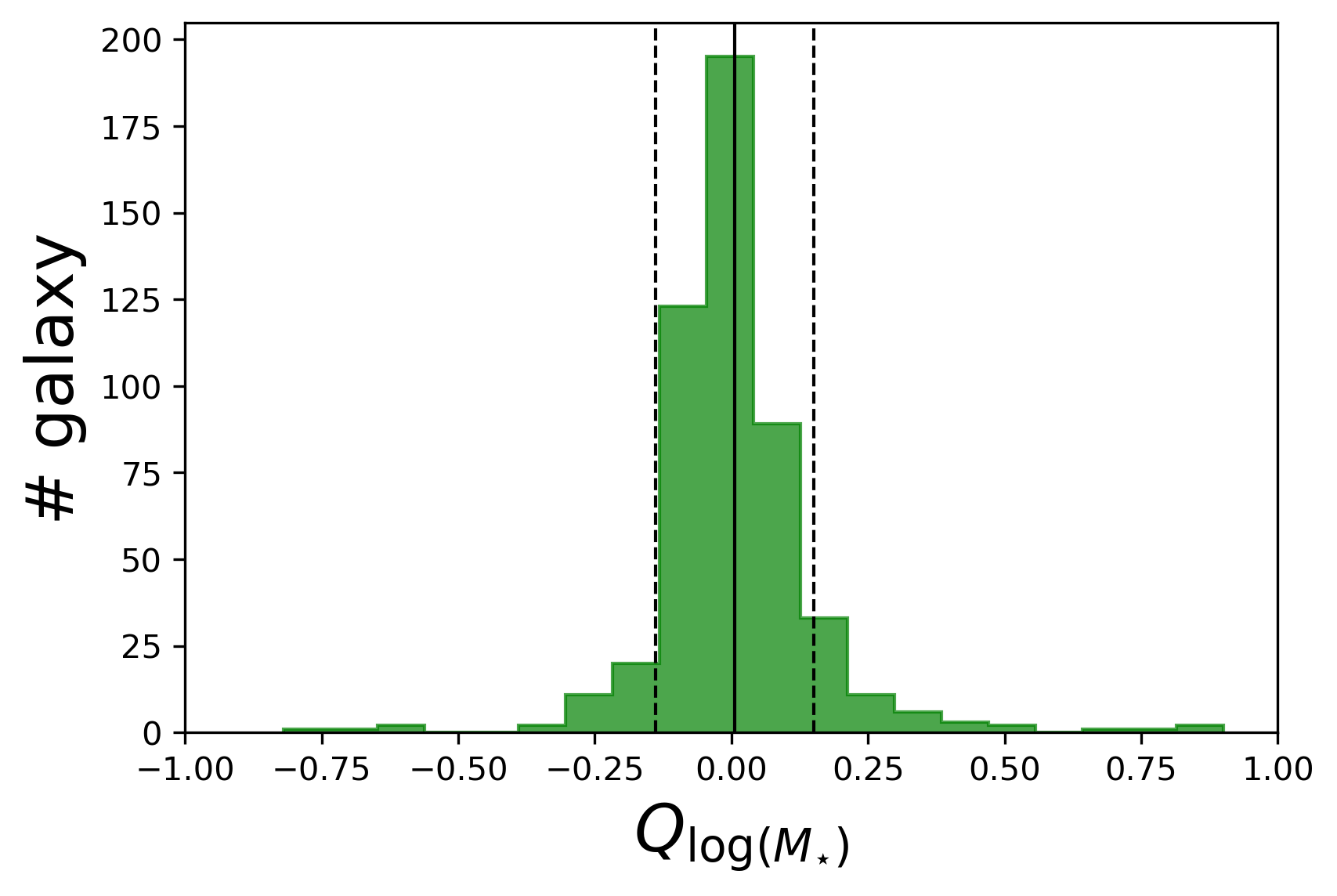}
\includegraphics[width=.42\textwidth]{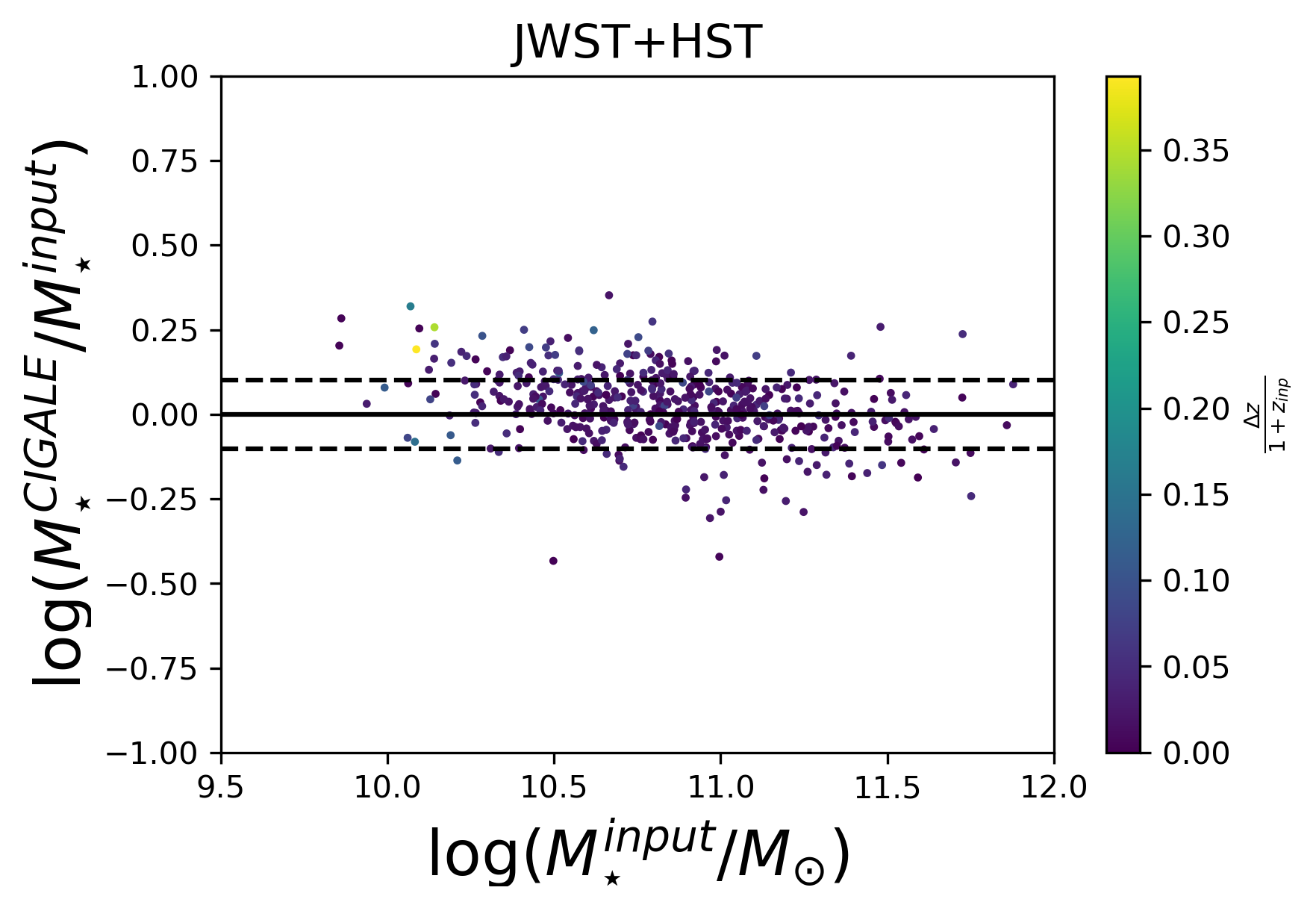}
\includegraphics[width=.4\textwidth]{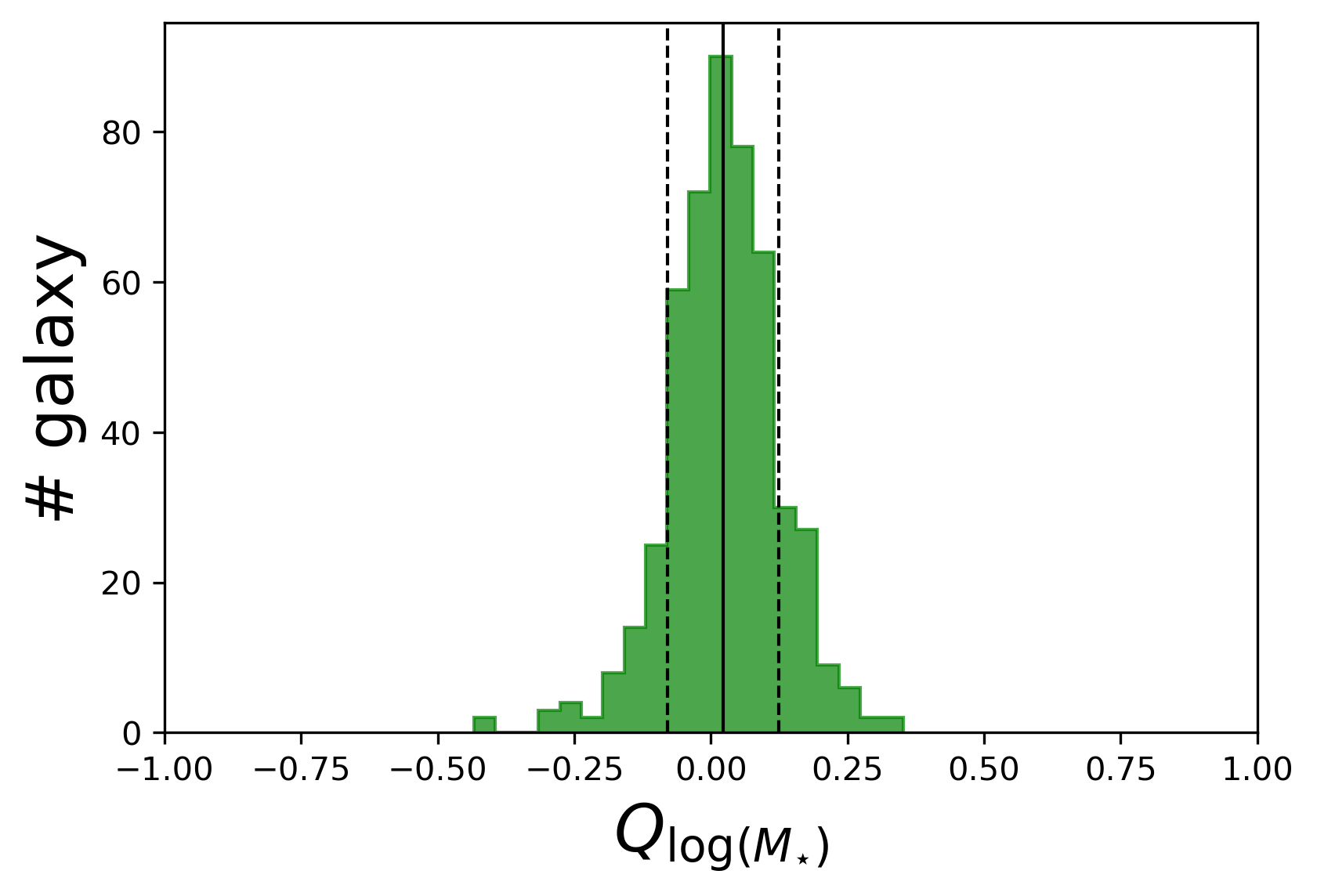}
      \caption{Scatter plot of the logarithm of the ratio of estimated stellar mass ($M_{\star}^{\rm CIGALE}$) and the input stellar mass ($M_{\star}^{\rm input}$) as a function of $M_{\star}^{\rm input}$ for the DSFG sample from the \textit{JWST} photometry (top row) and the \textit{JWST}+\textit{HST} photometry (bottom row). The points are colour-coded with their values of $|\Delta z|/(1+z_{\rm input})$. The histograms of $Q_{\log (M_{\star})}$, are shown in the right-hand panels. The meanings of the different line styles are the same as in Figure \ref{figcigsmall}.}
      \label{figcigsmsub}
\end{figure*}

The scatter plots of the logarithm of the ratio between the estimated stellar mass ($M_{\star}^{\rm CIGALE}$) and the input stellar mass ($M_{\star}^{\rm input}$) as a function of $M_{\star}^{\rm input}$ for the parent and the DSFG sample are shown in Figures \ref{figcigsmall} and \ref{figcigsmsub}, respectively. The points are colour coded with their values of $|\Delta z|/(1+z_{\rm input})$. The histogram of $Q_{\log (M_{\star})}$, is also shown in the figures. When the stellar mass is derived from the \textit{JWST} photometry alone, the $1\,\sigma$ dispersions in $Q_{\log (M_{\star})}$ are 0.2 and 0.14 for the parent catalogue and DSFG sample, respectively. CIGALE leans towards a slight but significant systematic underestimate of the stellar mass. Moreover, when using an exponentially declining SFH, stellar masses are often underestimated as was found by \cite{pforr_recovering_2012} while performing SED fitting on the semi-analytic model Galaxies In Cosmological Simulations \citep[GalICS;][]{Hatton_2003}. 

\begin{figure}
    \centering
\includegraphics[width=.47\textwidth]{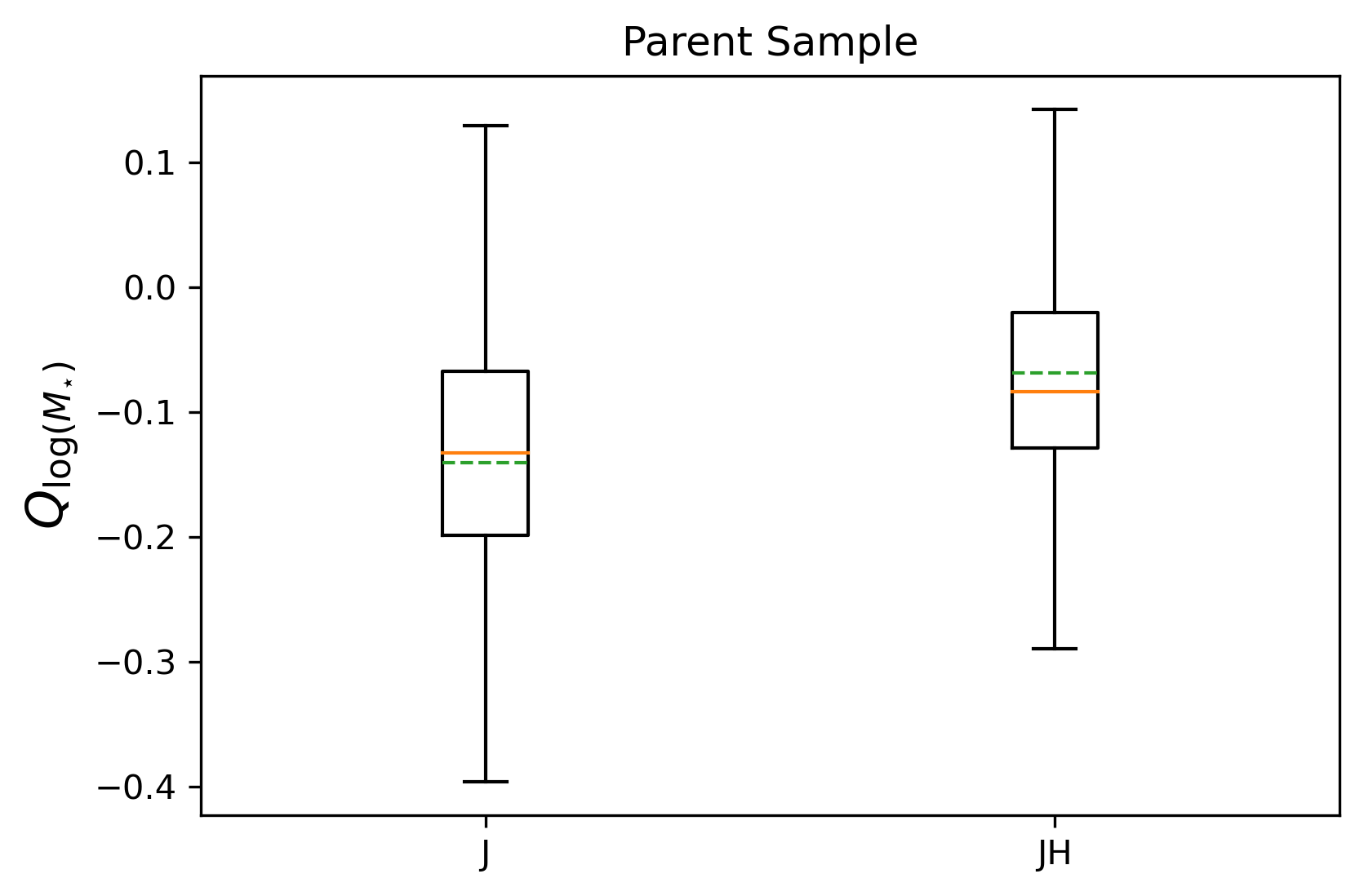}
\includegraphics[width=.47\textwidth]{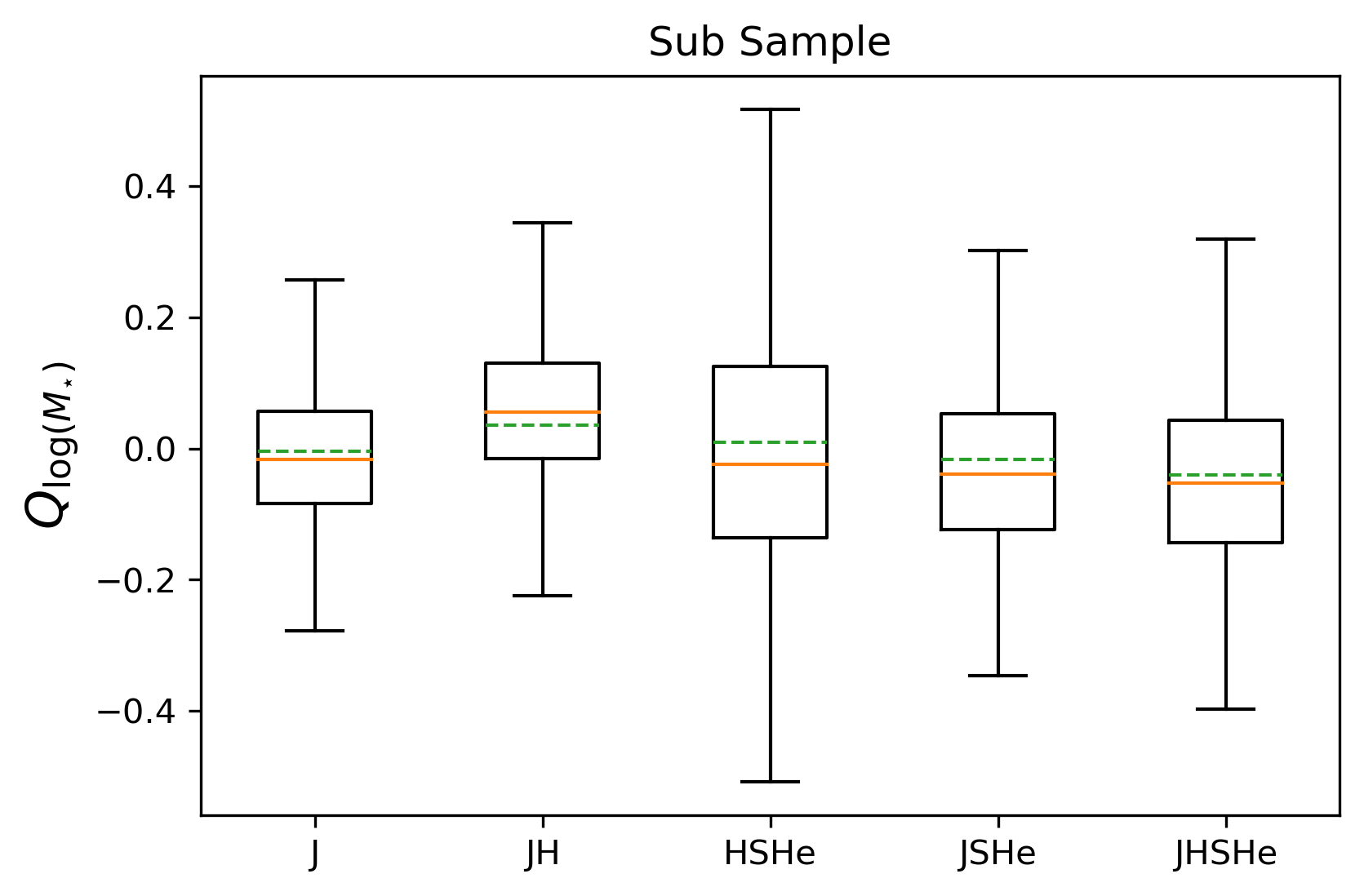}
      \caption{Boxplot showing the distribution of the $Q_{\log (M_{\star})}$, for the parent sample (top panel) using {\it JWST} and {\it JWST}+{\it HST} photometry and the DSFG sample (bottom panel) using {\it JWST}, {\it JWST}+{\it HST}, {\it JWST}+{\it Spitzer}+{\it Herschel} and {\it JWST}+{\it HST}+{\it Spitzer}+{\it Herschel} photometry. The solid orange line denotes the median, while the dashed green line represents the mean. The interquartile range and whiskers indicate the spread and outliers of the distribution. Here the abbreviations in the labels are: J - {\it JWST}, JH - {\it JWST}+{\it HST}, JSHe - {\it JWST}+{\it Spitzer}+{\it Herschel} and JHSHe - {\it JWST}+{\it HST}+{\it Spitzer}+{\it Herschel}. }
      \label{figboxsmall}
\end{figure}

We obtain the mean of $Q_{\log (M_{\star})}$ as $-0.14$ and $-0.004$ with \textit{JWST} photometry for the parent and the DSFG sample, respectively. We observe that the $1\,\sigma$ dispersions 
reduce to $0.15$ for the parent sample and $0.1$ for the DSFG sample on removing the catastrophic photo-$z$ outliers (AGN-dominated sources or power law like SEDs). Besides, the mean differences also reduce to $-0.12$ for the parent sample. As shown in Figure \ref{figboxsmall} (top panel) for the parent sample, including {\it HST} photometry alongside {\it JWST} leads to a noticeable reduction in both the bias and scatter of stellar mass estimates, underscoring the value of extended NIR coverage in improving SED fitting accuracy. Combining the \textit{JWST} with the \textit{HST} photometry, the $1\,\sigma$ dispersion in $Q_{\log (M_{\star})}$ reduces to $0.14$ for the parent sample, which after removal of outliers further decreases to $0.1$. The mean difference becomes $-0.068$ and $-0.063$ respectively. This clearly shows that the dispersion in the stellar mass estimates is mainly due to the uncertainty in the photo-$z$. However, the overall estimation of stellar mass can also affected by the difference in the slope of the dust attenuation law of the BCs used in CIGALE and the one used in our SED formalism
\citep{mitra_euclid_2024}. Besides, the assumed SFH and its parameterisation also affects the estimation of stellar mass as pointed out by \cite{michalowski_stellar_2012} and \cite{Lower_2020}.  

Moreover for the DSFG sample the $1\,\sigma$ dispersion reduces to $0.1$ upon adding {\it HST} photometry to that of {\it JWST}. The mean offset is $0.035$. Adding HST photometry in the F435W and F160W bands to {\it JWST}/NIRCam data significantly improves the accuracy of stellar mass estimates by reducing key degeneracies in the SED fitting process. The F435W filter probes the rest-frame ultraviolet at redshifts $z\sim1.5-3$, providing sensitivity to recent star formation and dust attenuation. This helps constrain the contribution of young stellar populations and mitigates the age--dust degeneracy that can otherwise bias mass estimates. The F160W filter, on the other hand, samples the rest-frame optical light and is crucial for capturing features like the Balmer and 4000{\AA} breaks, which are strong indicators of older stellar populations that dominate the total stellar mass. Together, these two HST bands fill important gaps in wavelength coverage and complement the longer-wavelength NIRCam filters by anchoring both the blue and red ends of the SED. As a result, the stellar population properties -- particularly the mass-to-light ratio -- are better constrained, leading to a measurable reduction in the $1\,\sigma$ dispersion of stellar mass estimates.


For the subsample, we also considered different combinations of photometry to estimate the stellar mass. The estimated stellar masses derived from the photometry from \textit{HST}, \textit{Spitzer} and \textit{Herschel} have mean and median deviations from the true values of $0.01$ and $0.23$ in $Q_{\log (M_{\star})}$. An rms value and a mean value of $0.16$ and $-0.13$ are obtained when using \textit{JWST}+\textit{Spitzer}+\textit{Herschel} photometry. Estimating stellar mass from \textit{HST}+\textit{JWST}+\textit{Spitzer}+\textit{Herschel} photometry also gave a $1\,\sigma$ dispersion of $0.15$ and a mean offset of $-0.04$. 
Figure~\ref{figboxsmall} (bottom panel) shows the recovery of stellar mass for the DSFG sub-sample using various photometric combinations. \textit{JWST}/NIRCam photometry alone yields minimal bias and low scatter, highlighting its strength in constraining stellar mass. Adding \textit{HST} data does not significantly improve the estimates and introduces a slight positive bias. Legacy IR data without {\it JWST} result in larger scatter, while including \textit{JWST} improves accuracy. As the evolved stellar populations dominate the stellar mass of galaxies and have most of their emission in the rest-frame optical/near-IR, the addition of far-IR photometry does not significantly improve the constraints on the stellar mass estimate. These findings underscore the remarkable capability of \textit{JWST} in recovering the stellar mass of DSFGs. 

\subsubsection{SFR}

\begin{figure*}
      \centering
\includegraphics[width=.42\textwidth]{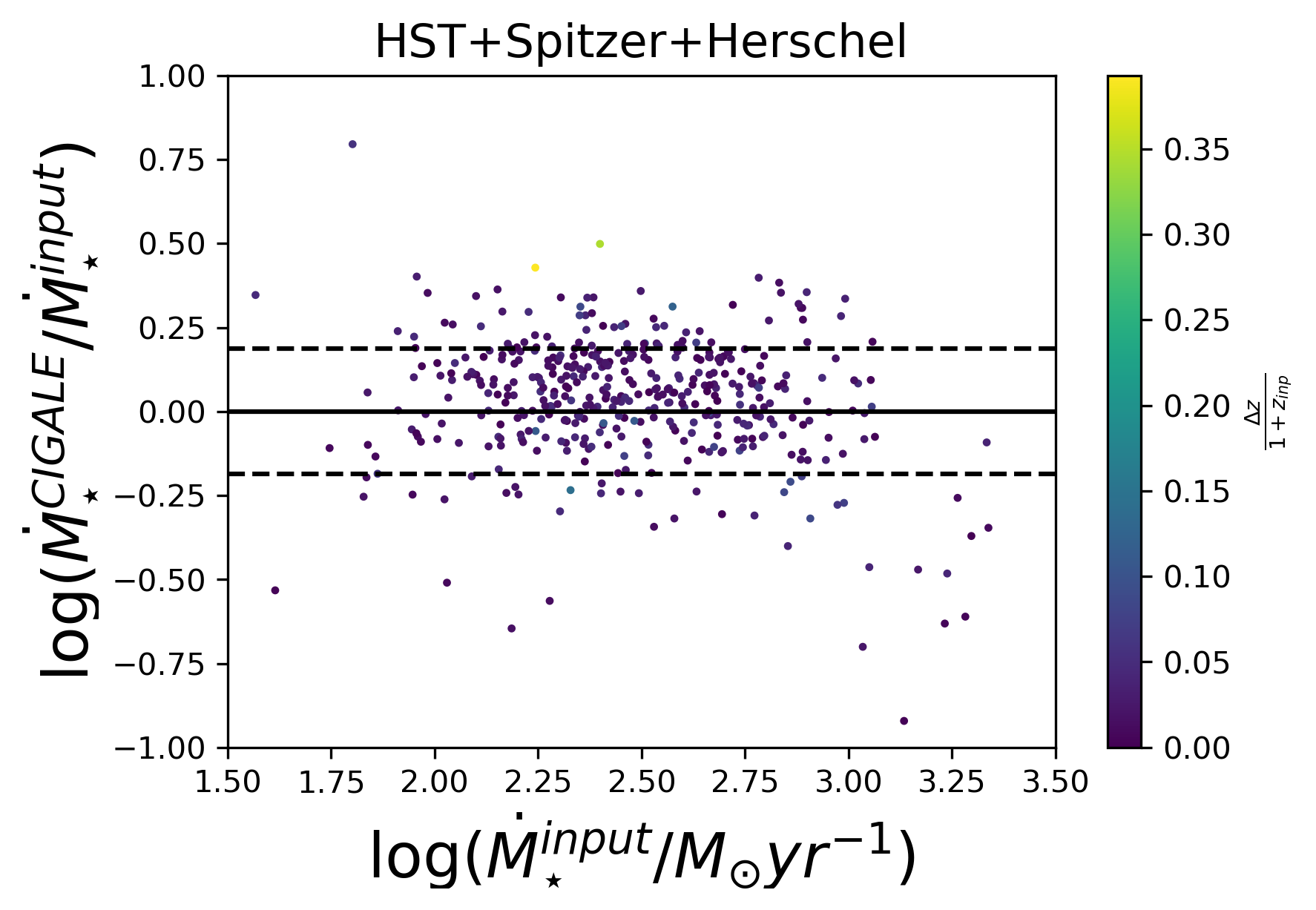}
\includegraphics[width=.4\textwidth]{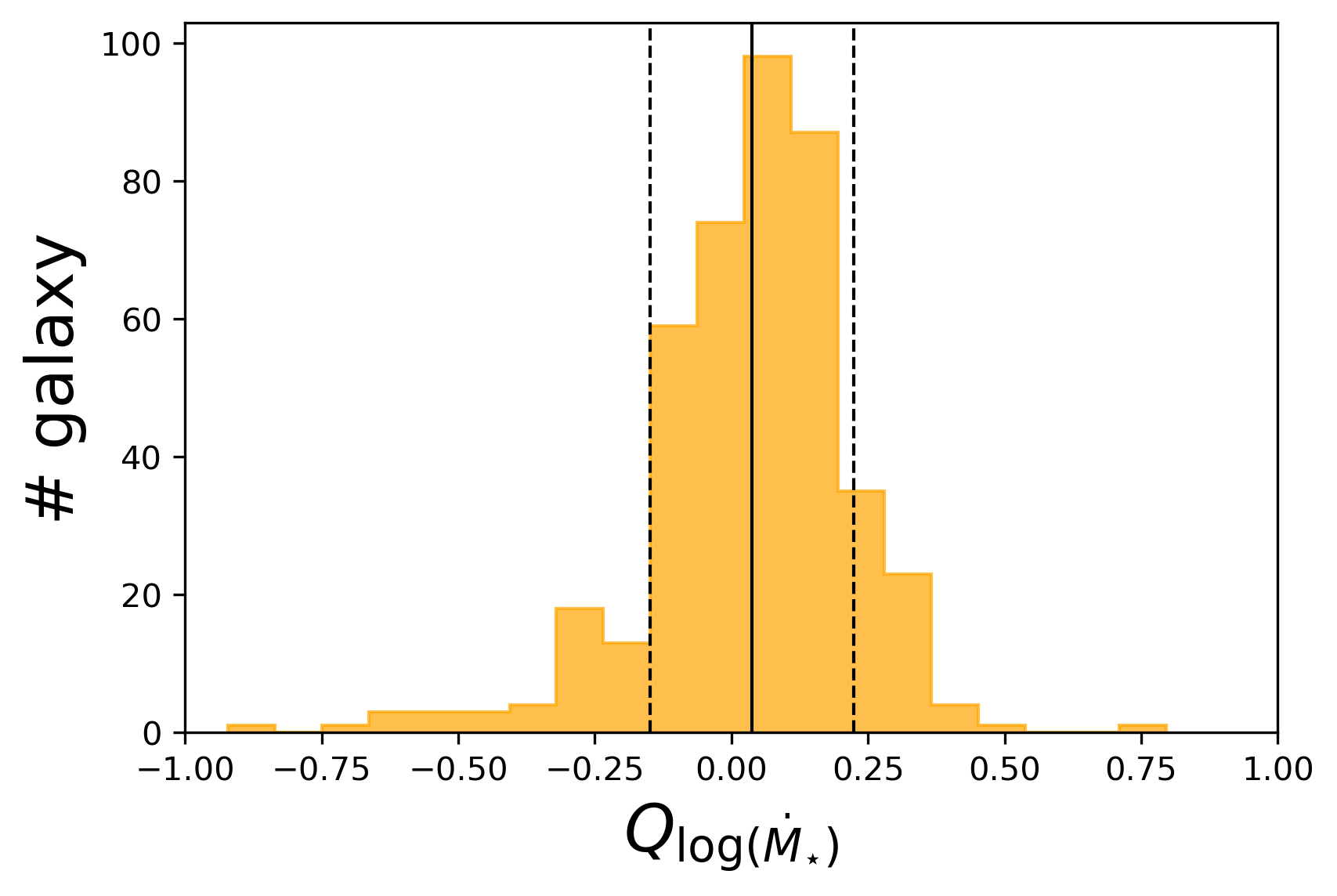}
\includegraphics[width=.42\textwidth]{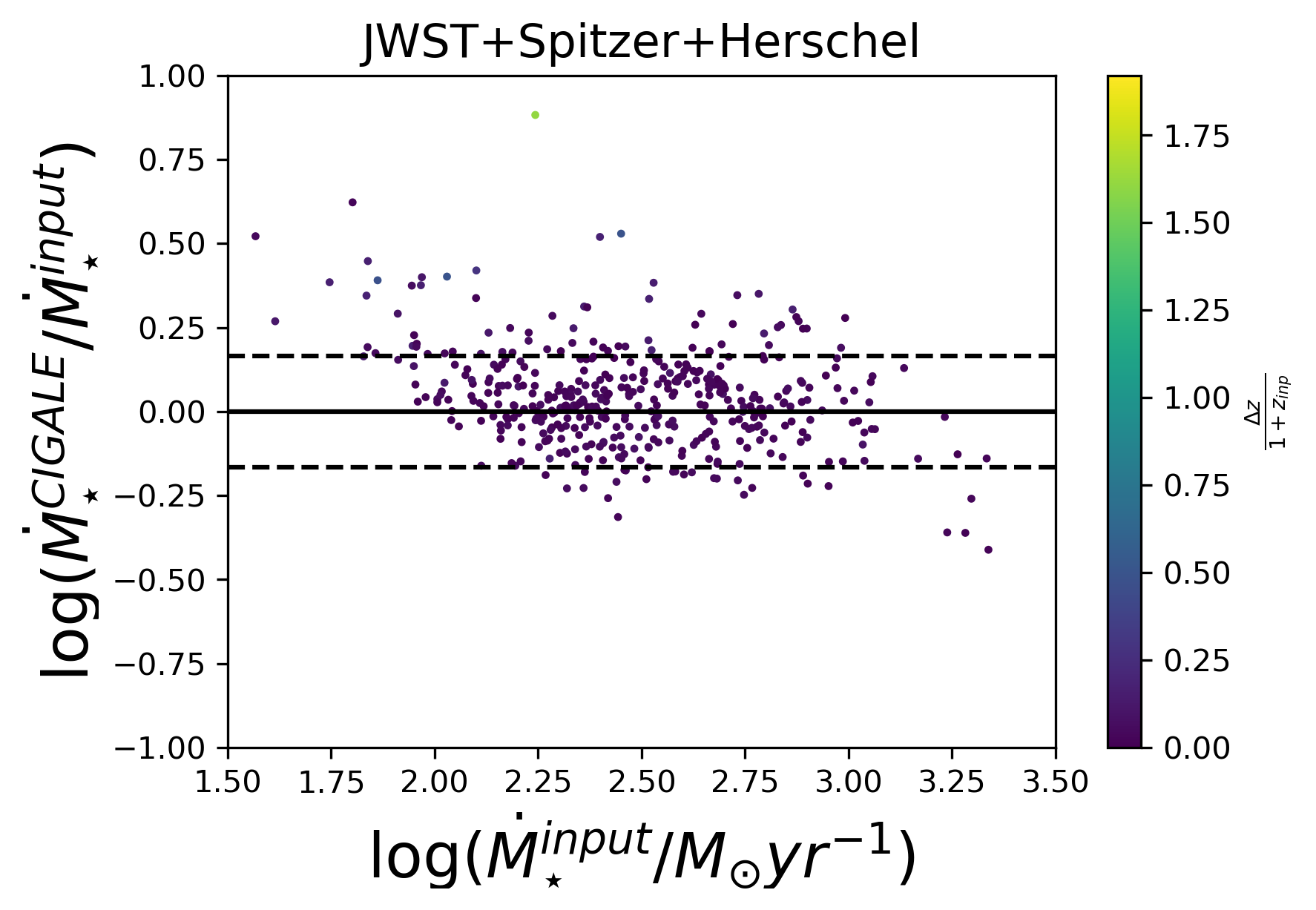}
\includegraphics[width=.4\textwidth]{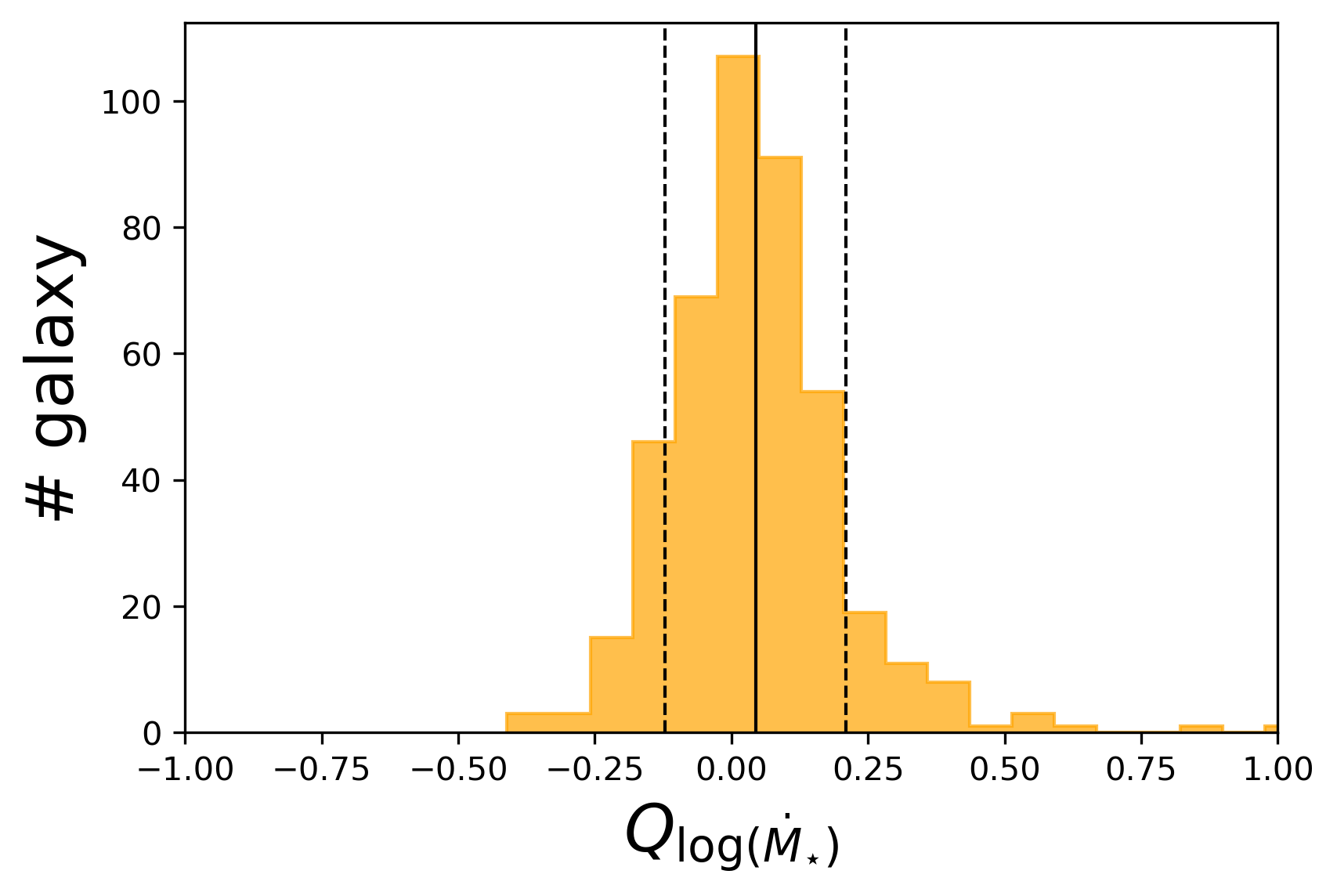}
\includegraphics[width=.42\textwidth]{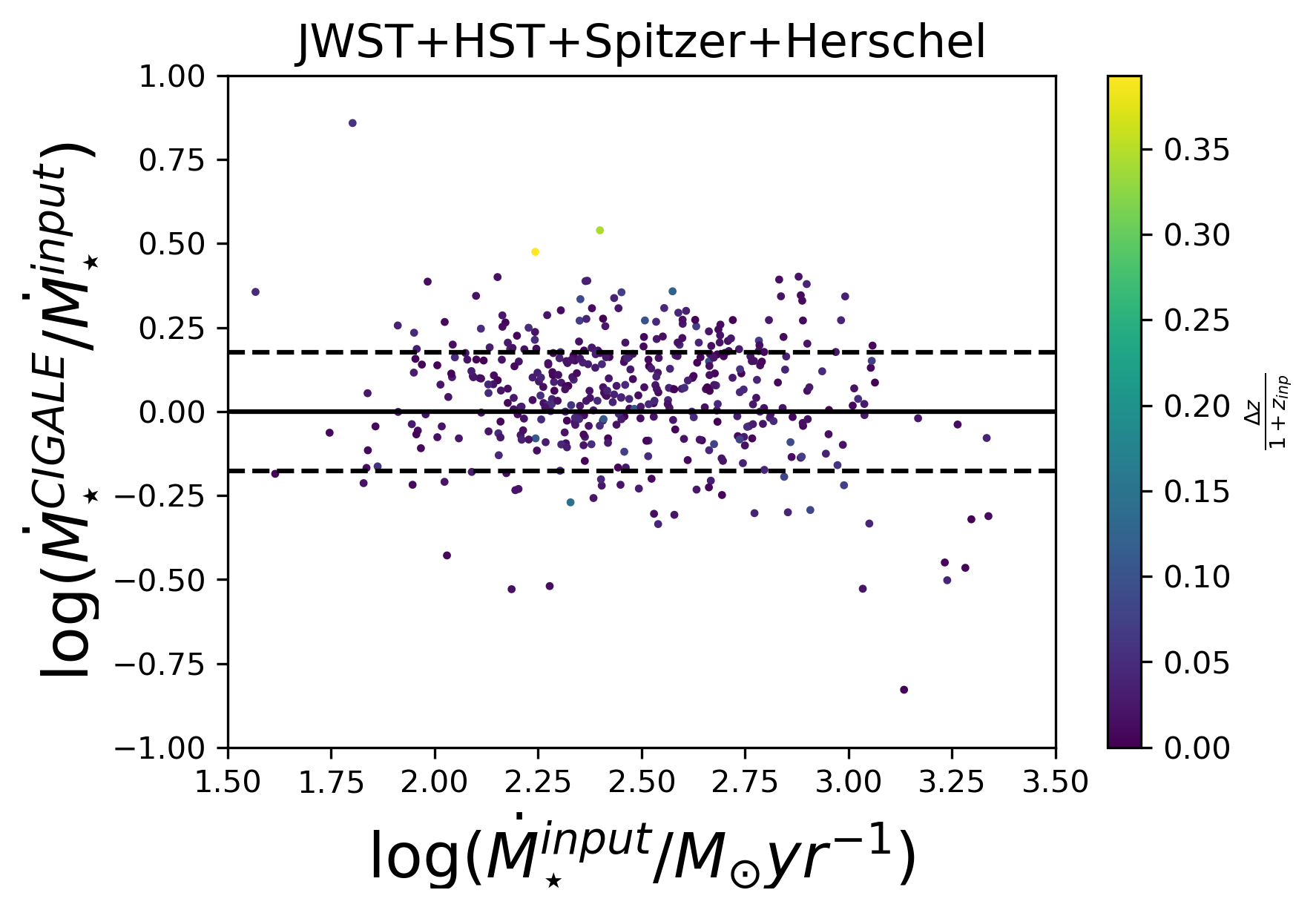}
\includegraphics[width=.4\textwidth]{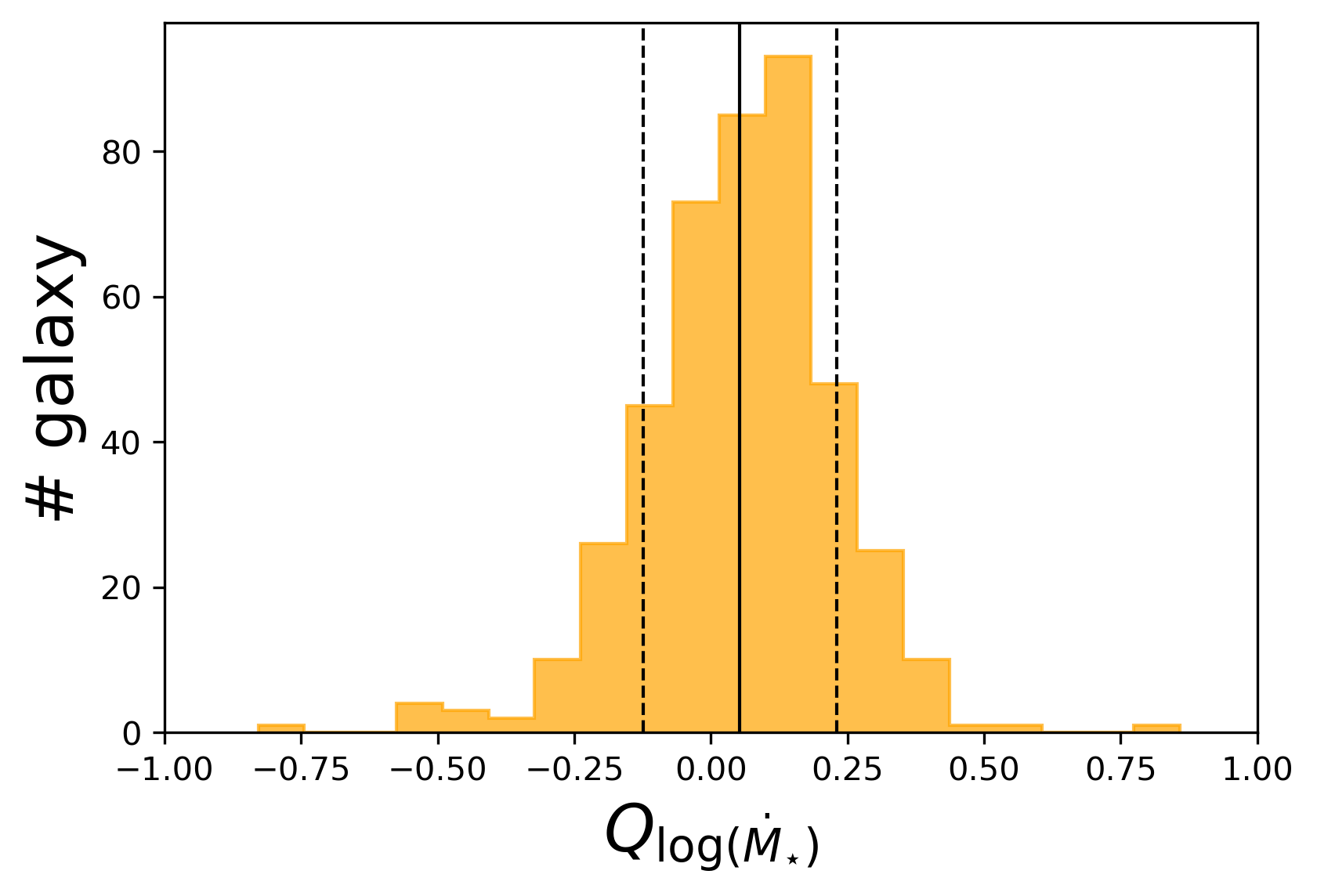}
      \caption{Scatter plot of the logarithm of the ratio of estimated ($\dot{M}_{\star}^{\rm CIGALE}$) to input ($\dot{M}_{\star}^{\rm input}$) SFR obtained for the DSFG sample from  \textit{HST}+\textit{Spitzer}+\textit{Herschel}, \textit{JWST}+\textit{Spitzer}+\textit{Herschel} and \textit{JWST}+\textit{HST}+\textit{Spitzer}+\textit{Herschel} photometry. The points are colour-coded with their values of $|\Delta z|/(1+z_{\rm input})$. The histogram of $Q_{\log (\dot{M}_{\star})}$, is also shown in the figures  The meaning of the different line styles is the same as in Figure \ref{figcigsmall}.}
      \label{figcigsfr}
\end{figure*}

\begin{figure}
    \centering
\includegraphics[width=.47\textwidth]{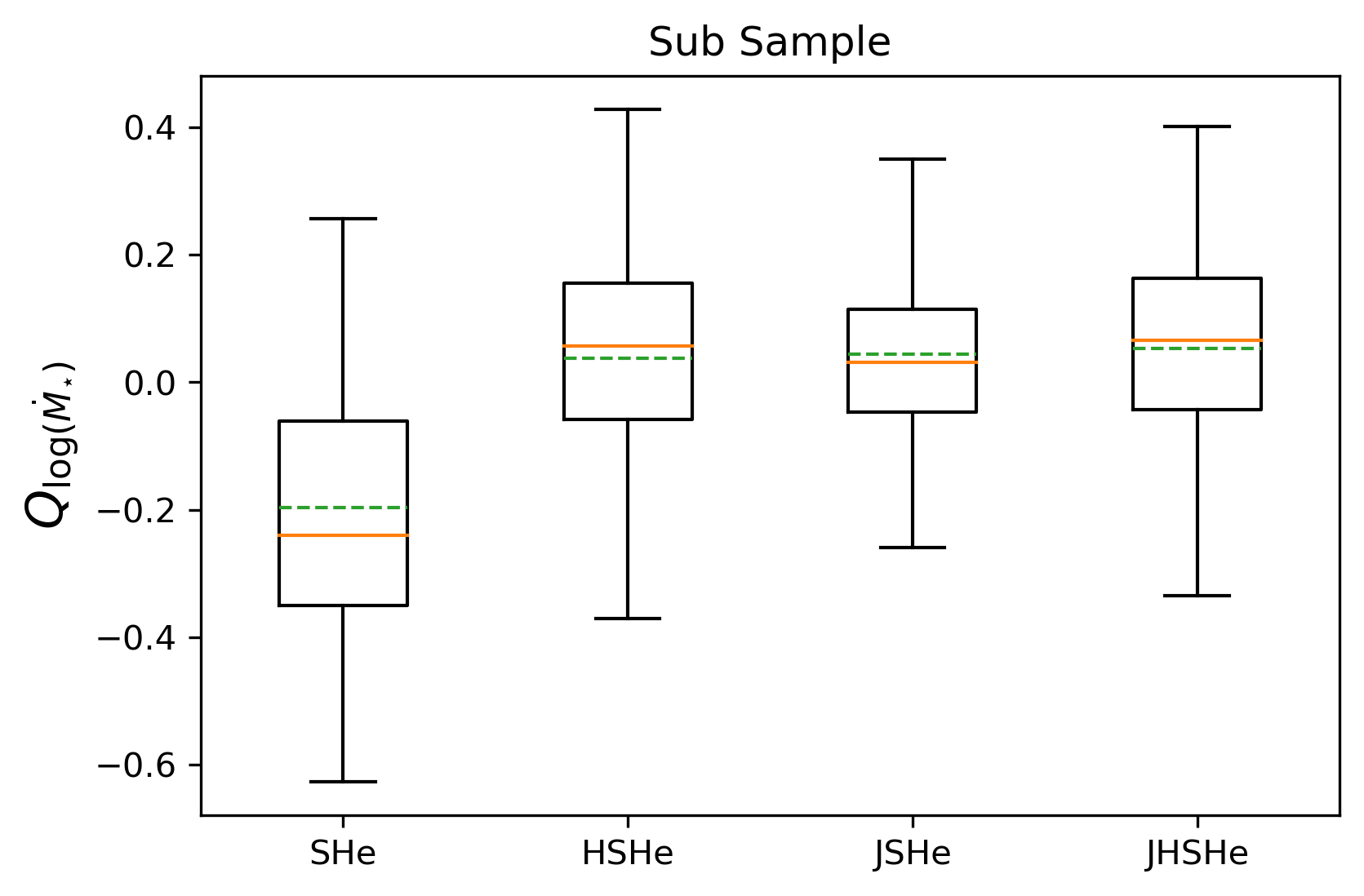}
      \caption{Boxplot showing the distribution of the $Q_{\log (\dot{M}_{\star})}$ for the DSFG sample using {\it HST}+{\it Spitzer}+{\it Herschel} (HSHe), {\it JWST}+{\it Spitzer}+{\it Herschel} (JSHe) and {\it JWST}+{\it HST}+{\it Spitzer}+{\it Herschel} (JHSHe) photometry. The solid orange line denotes the median and the dashed green line indicates the mean. The interquartile range and whiskers indicate the spread and outliers of the distribution.}
      \label{figboxsfrsub}
\end{figure}

The SFR of dust-enshrouded objects cannot be traced accurately using only the near-IR photometry \citep{pforr_recovering_2012, pforr_recovering_2013, euclid_collaboration_euclid_2023-1, mitra_euclid_2024}. In estimating the SFR, $\dot{M_\star}$, from \textit{JWST} photometry, we obtain  $1\,\sigma$ dispersions of $\approx0.8$ and $0.55$ in $Q_{\log (\dot{M_{\star}})}$ for the parent sample and sub-sample, respectively; the mean offsets are $0.14$ and $-0.05$, respectively. Adding the \textit{HST} to \textit{JWST} photometry, the $1\,\sigma$ dispersions reduce, but are still high with values of $0.36$ and $0.47$ for the two samples respectively. The mean differences are $-0.044$ and $-0.019$. Therefore, complementing the \textit{JWST} and/or the \textit{HST} photometry with photometric data from FIR bands is crucial for estimating the SFR. 

We now give estimates of SFR for different combinations of photometry for the DSFG sample only. For comparison, we also calculate the SFR from the FIR photometry alone. The mean and the $1\,\sigma$ dispersion of $Q_{\log (\dot{M_{\star}})}$ are $-0.196$ and $0.22$ for the \textit{Spitzer} plus \textit{Herschel} photometry. Adding the \textit{HST} photometry to the above improves the estimation with the $1\,\sigma$ dispersion being $0.169$, and the mean reduces to $0.037$. Figure \ref{figcigsfr} shows the scatter plot of the logarithm of the ratio of estimated  ($\dot{M}_{\star}^{\rm CIGALE}$) to input SFR ($\dot{M}_{\star}^{\rm input}$), colour coded with the values of $|\Delta z|/(1+z_{\rm input})$. Also shown are the histogram of the ratio $Q_{\log (\dot{M_{\star}})}$. Upon replacing \textit{HST} with \textit{JWST} in the above combination, we get a $1\,\sigma$ dispersion of $0.16$ and a mean of $0.044$. However, when combining all the photometry from \textit{JWST}+\textit{HST}+\textit{Spitzer}+\textit{Herschel}, the dispersion increases to $0.18$ and the mean becomes $0.053$. 

The HST F435W and F160W filters sample the rest-frame UV and optical light at high redshift, which trace only the unobscured star formation. In DSFGs, this component is heavily suppressed and variable, leading to increased uncertainties and degeneracies in CIGALE's energy balance when fitting both obscured and unobscured SFR components -- thereby increasing the dispersion in total SFR estimates. Moreover, this increase in the dispersion can also be due to the presence of some photo-$z$ outliers in the sample. 

A slight offset between the SFRs by CIGALE and the true values is expected, due to the different dust attenuation law slopes adopted by CIGALE and by  \cite{da_cunha_simple_2008}. However, our results show that the effect is minor. Figure \ref{figboxsfrsub} presents the accuracy of SFR recovery for the DSFG sub-sample using different combinations of photometric bands. The results show that using only {\it Spitzer} and {\it Herschel} data leads to higher dispersion and a slight underestimation of SFRs (we recall that in most cases, we have \textit{Herschel} photometry only at $250\,\mu$m). The inclusion of {\it JWST} photometry significantly improves the estimates, reducing both bias and scatter. Adding {\it HST} data provides minimal additional benefit and may slightly worsen the results, likely due to the limited utility of rest-frame UV data in dust-obscured galaxies. Overall, these findings emphasize the crucial role of {\it JWST} in accurately constraining the SFRs of DSFGs.


Given that these galaxies are IR bright and dusty, the availability of robust far-IR/sub-mm data plays a vital role in the estimation of the SFR. 
Restricting the analysis of the SFR to the $33\%$ sources from the DSFG sample detected by \textit{Spitzer}+\textit{Herschel} at $>5\,\sigma$ in all \textit{Spitzer}/MIPS and \textit{Herschel} (PACS and SPIRE) bands gives a $1\,\sigma$ dispersion of $0.14$, $0.13$ and $0.15$ respectively in $Q_{\log (\dot{M_{\star}})}$ for \textit{Spitzer}+\textit{Herschel}, \textit{JWST}+\textit{Spitzer}+\textit{Herschel} and \textit{JWST}+\textit{HST}+\textit{Spitzer}+\textit{Herschel} photometry. Similarly, the $1\,\sigma$ dispersions in $Q_{\log (\dot{M}_{\star,10})}$ for these $>5\,\sigma$ detected sources are $0.12$, $0.12$, and $0.15$ respectively, for the same photometry combinations. Overall, we find that the SFR estimation was very challenging since it is very sensitive to the SFH \citep{Lower_2020}. To verify this, we performed a test by keeping all the parameters of CIGALE unchanged and changing the assumed SFH from \texttt{sfhdelayed} to \texttt{sfhdelayedbq}\footnote{The \texttt{sfhdelayed} module models a smooth, rising-and-declining star formation history, while \texttt{sfhdelayedbq} extends this by allowing for a recent burst or quenching event, making it more suitable for galaxies with abrupt SFR changes.}. Using \texttt{sfhdelayedbq} to estimate SFR we found that SFR is underestimated almost by an order of magnitude. When switching from \texttt{sfhdelayed} to \texttt{sfhdelayedbq} in CIGALE, the model allows for a recent burst or quenching event. For DSFGs, which typically have high, sustained SFRs, the burst/quench parameters can cause the fit to interpret their IR-luminous phase as a post-burst decline, leading to underestimated current SFRs. This misinterpretation happens because \texttt{sfhdelayedbq} may assign much of the star formation to an earlier burst, reducing the inferred ongoing SFR compared to the smoother \texttt{sfhdelayed} model. \cite{Hunt_2019} argued that due to the sensitive nature of SFR to SFH, certain SED fitting algorithms may find it difficult to determine the most appropriate SFH due to degeneracies. Consequently, a range of different SFHs may yield comparably good SED fits.


\subsubsection{Dust Luminosity and Dust Mass}

\begin{figure*}
      \centering
\includegraphics[width=.42\textwidth]{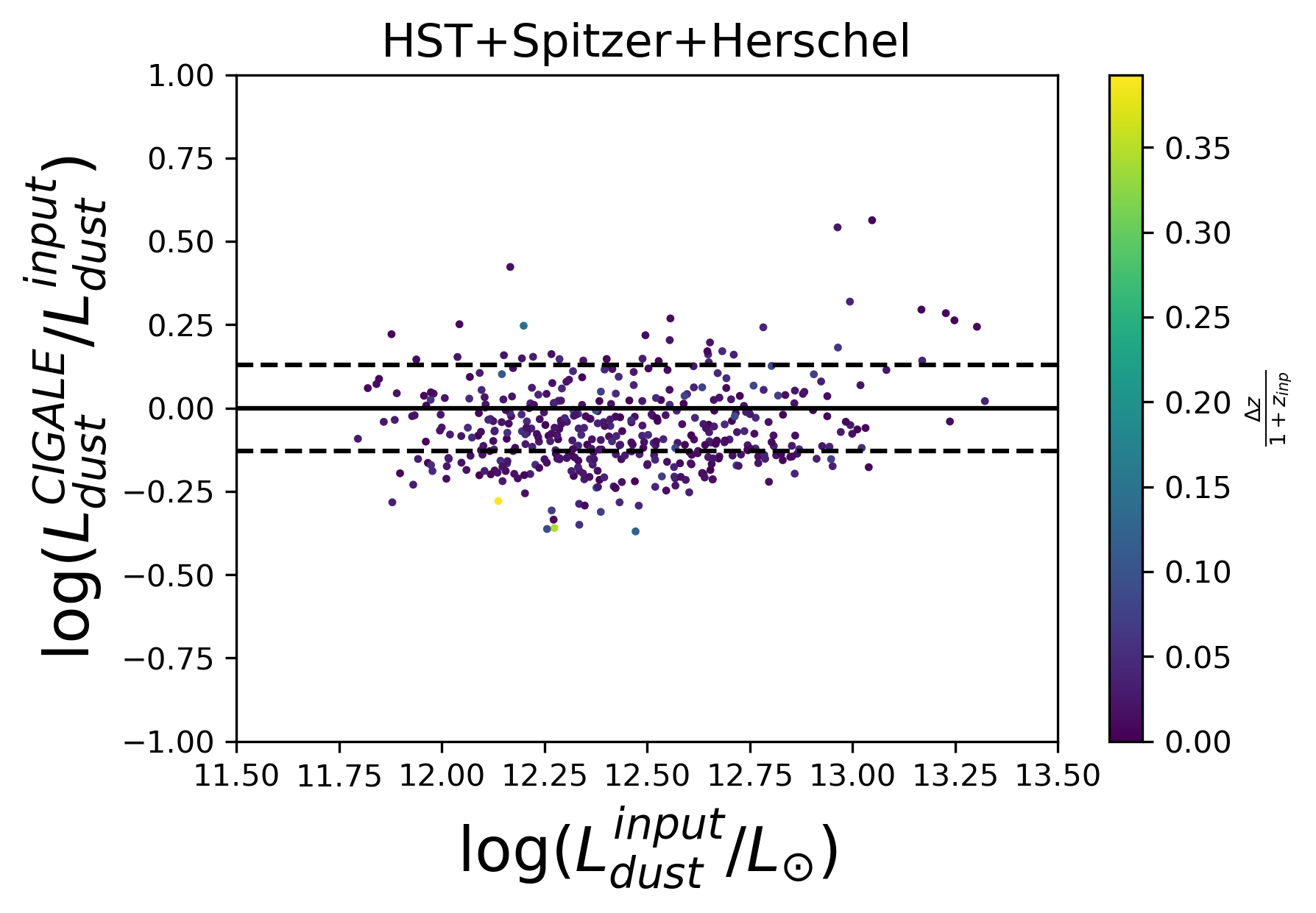}
\includegraphics[width=.4\textwidth]{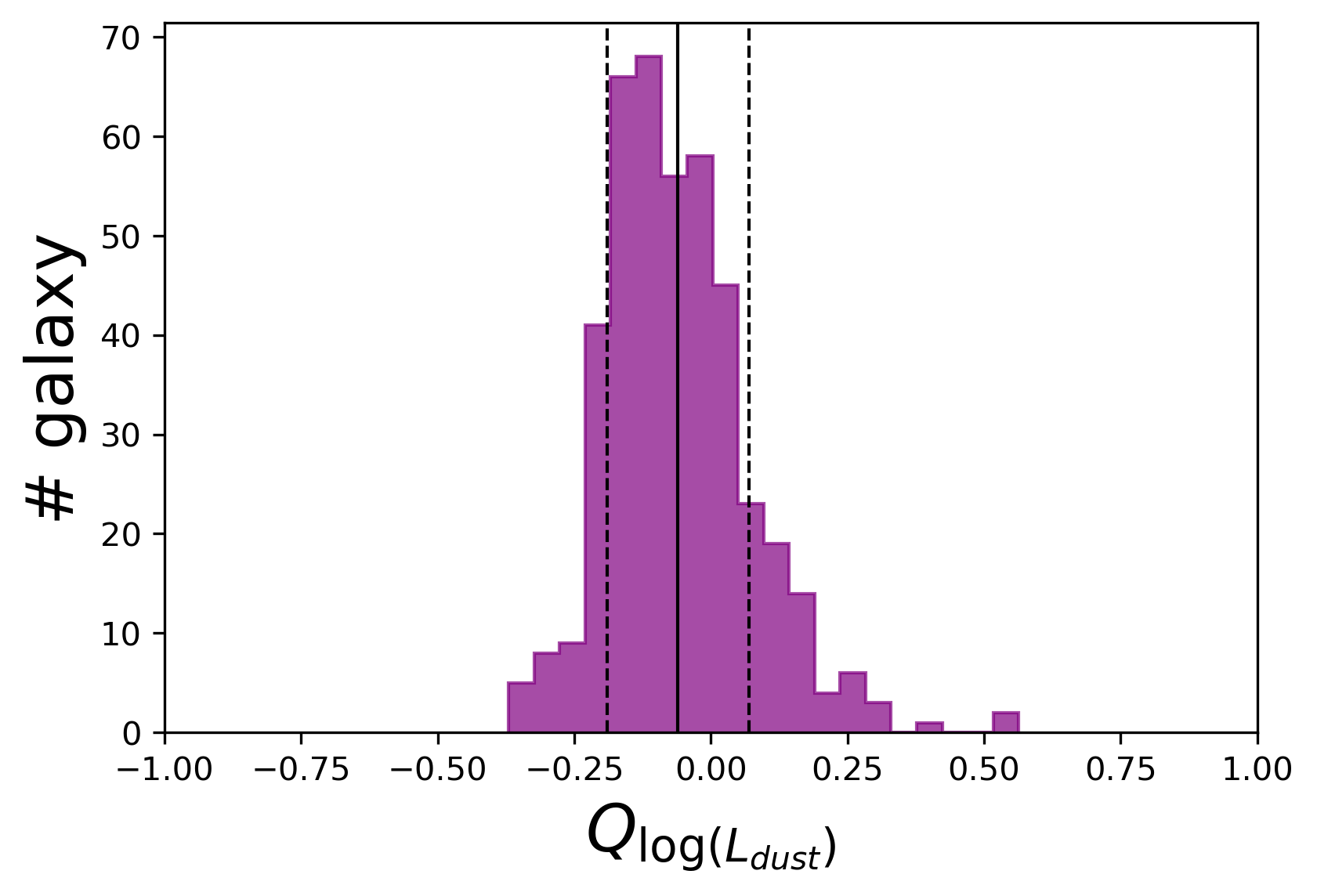}
\includegraphics[width=.42\textwidth]{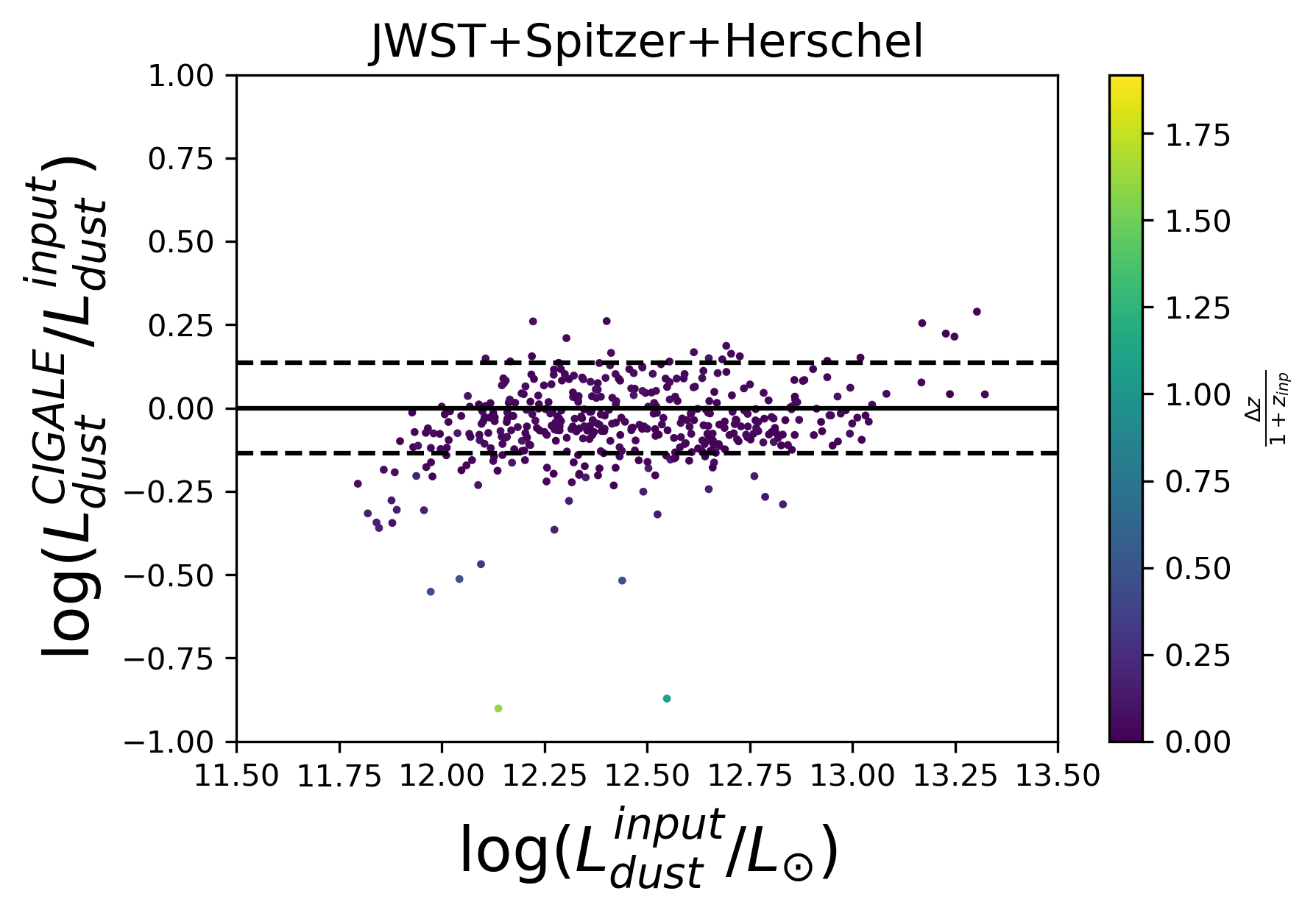}
\includegraphics[width=.4\textwidth]{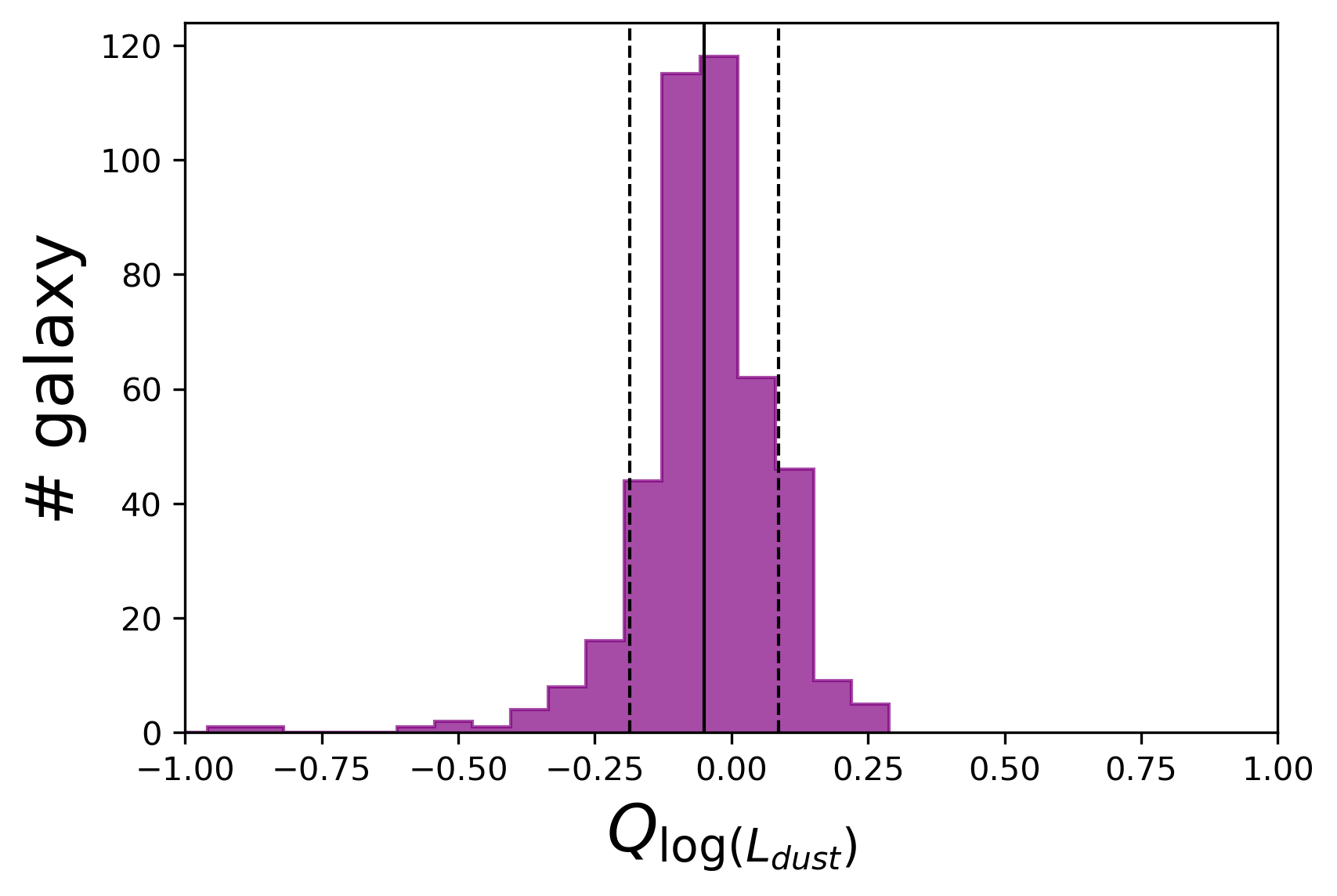}
\includegraphics[width=.42\textwidth]{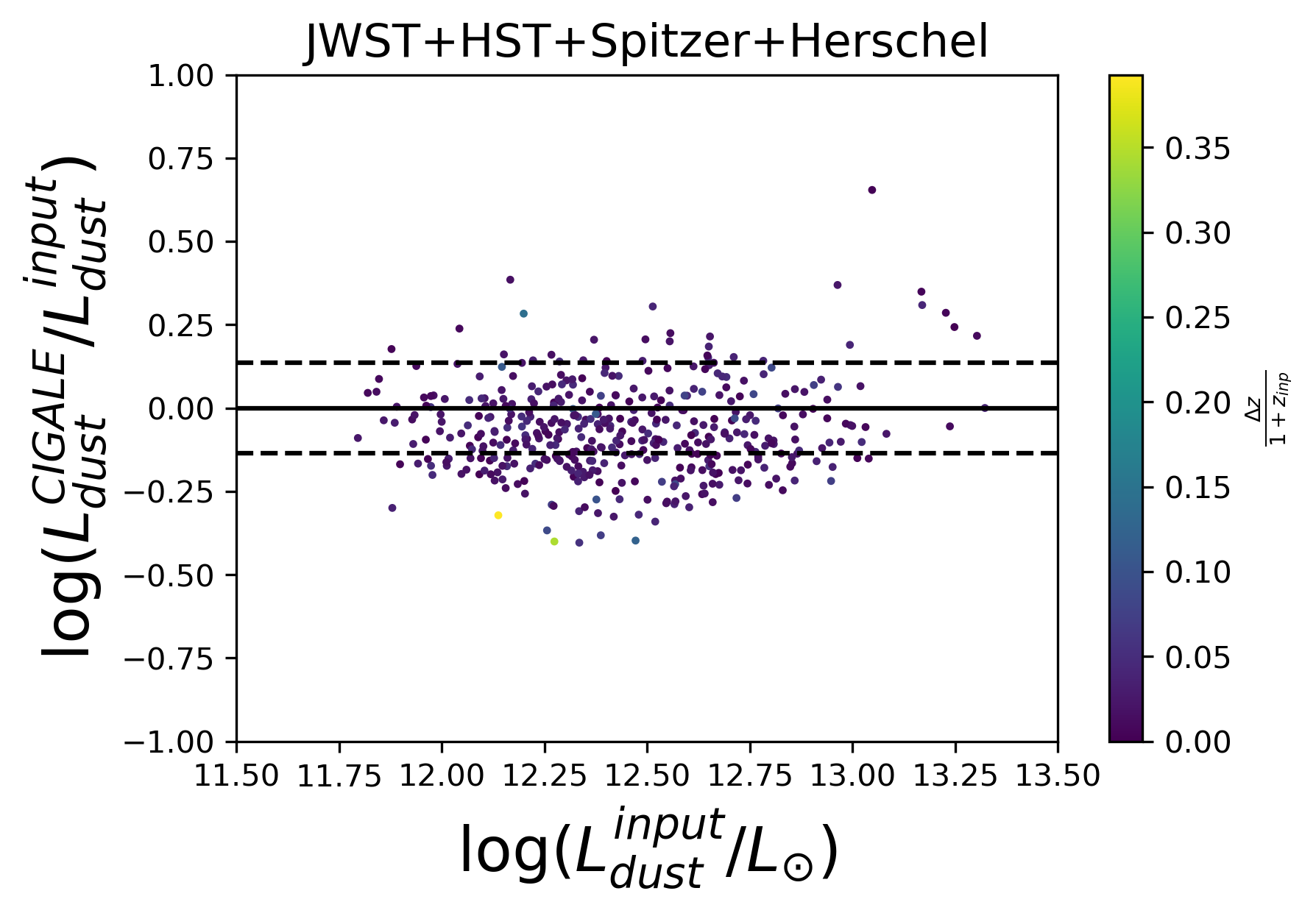}
\includegraphics[width=.4\textwidth]{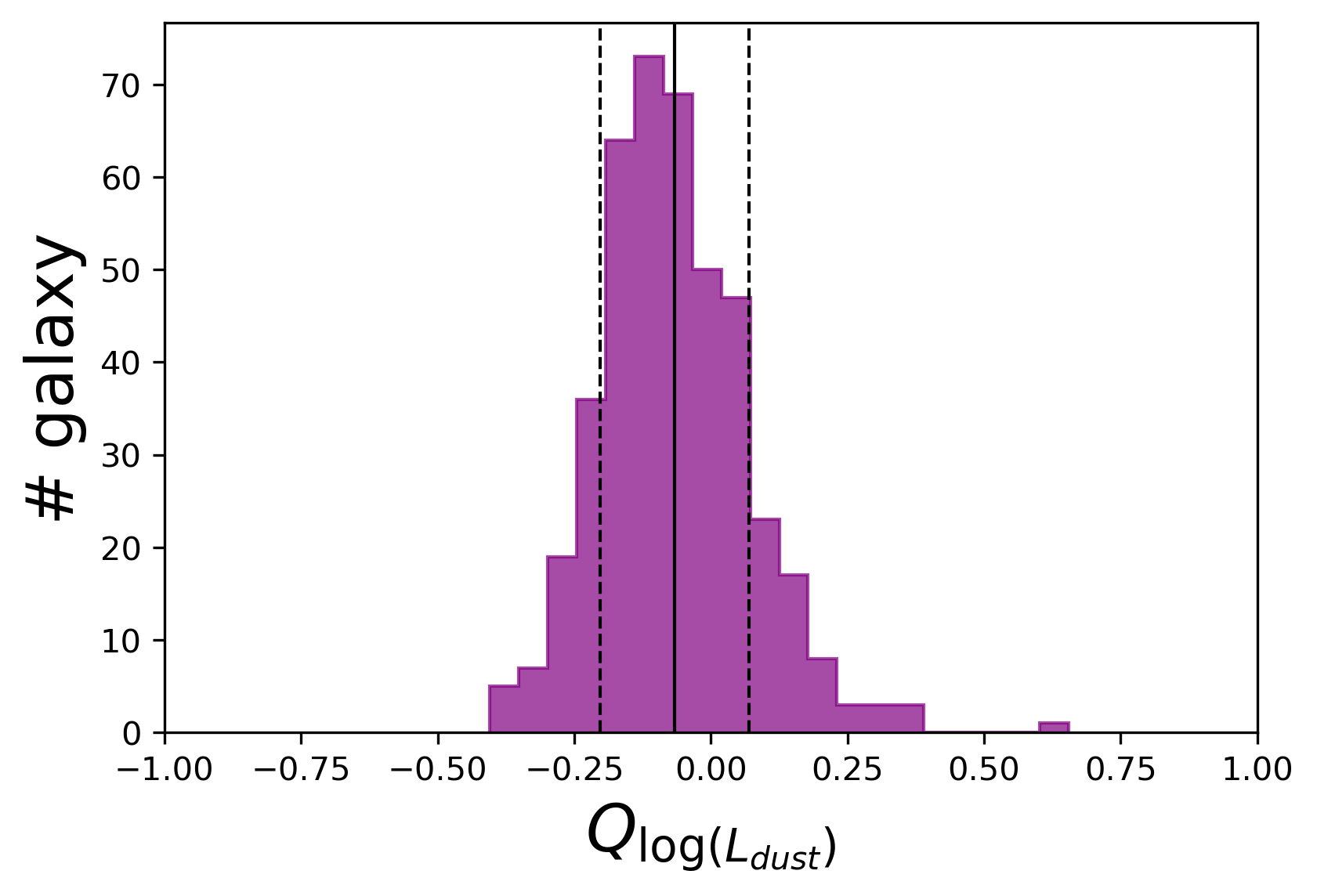}
      \caption{Same as Figure \ref{figcigsfr}, but for the dust luminosity ($L_{\rm dust}$).}
      \label{figcigdl}
\end{figure*}

\begin{figure*}
      \centering
\includegraphics[width=.42\textwidth]{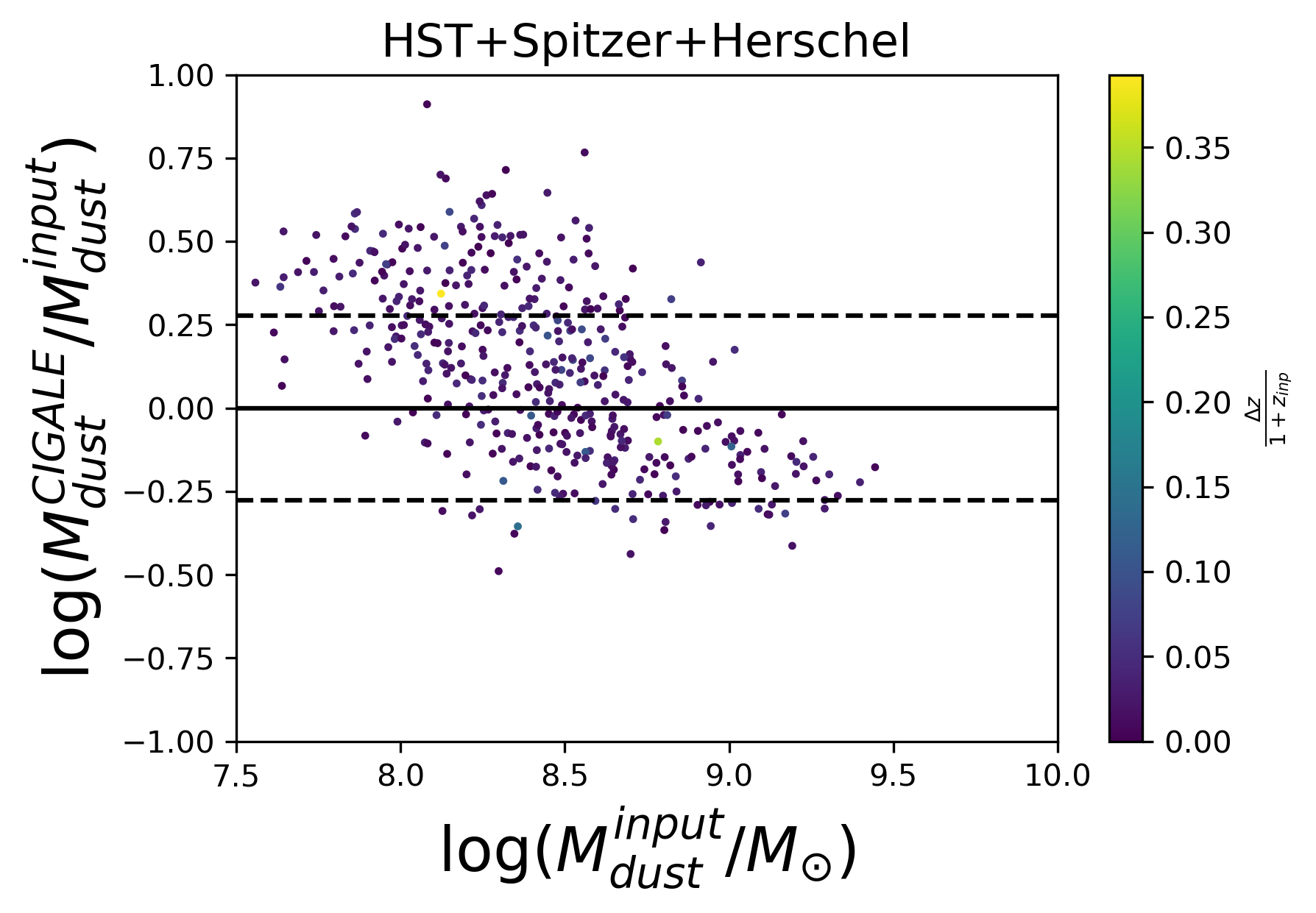}
\includegraphics[width=.4\textwidth]{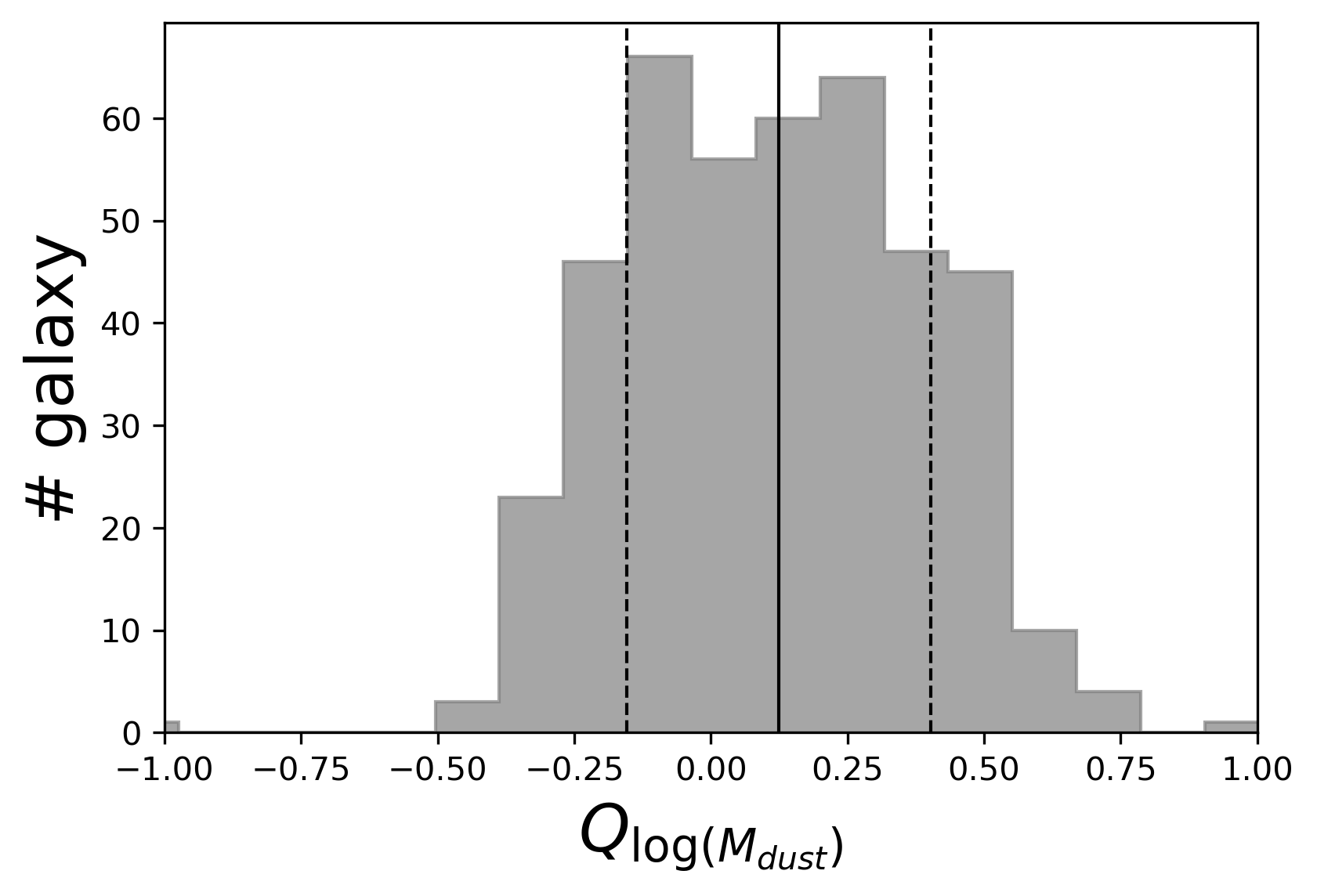}
\includegraphics[width=.42\textwidth]{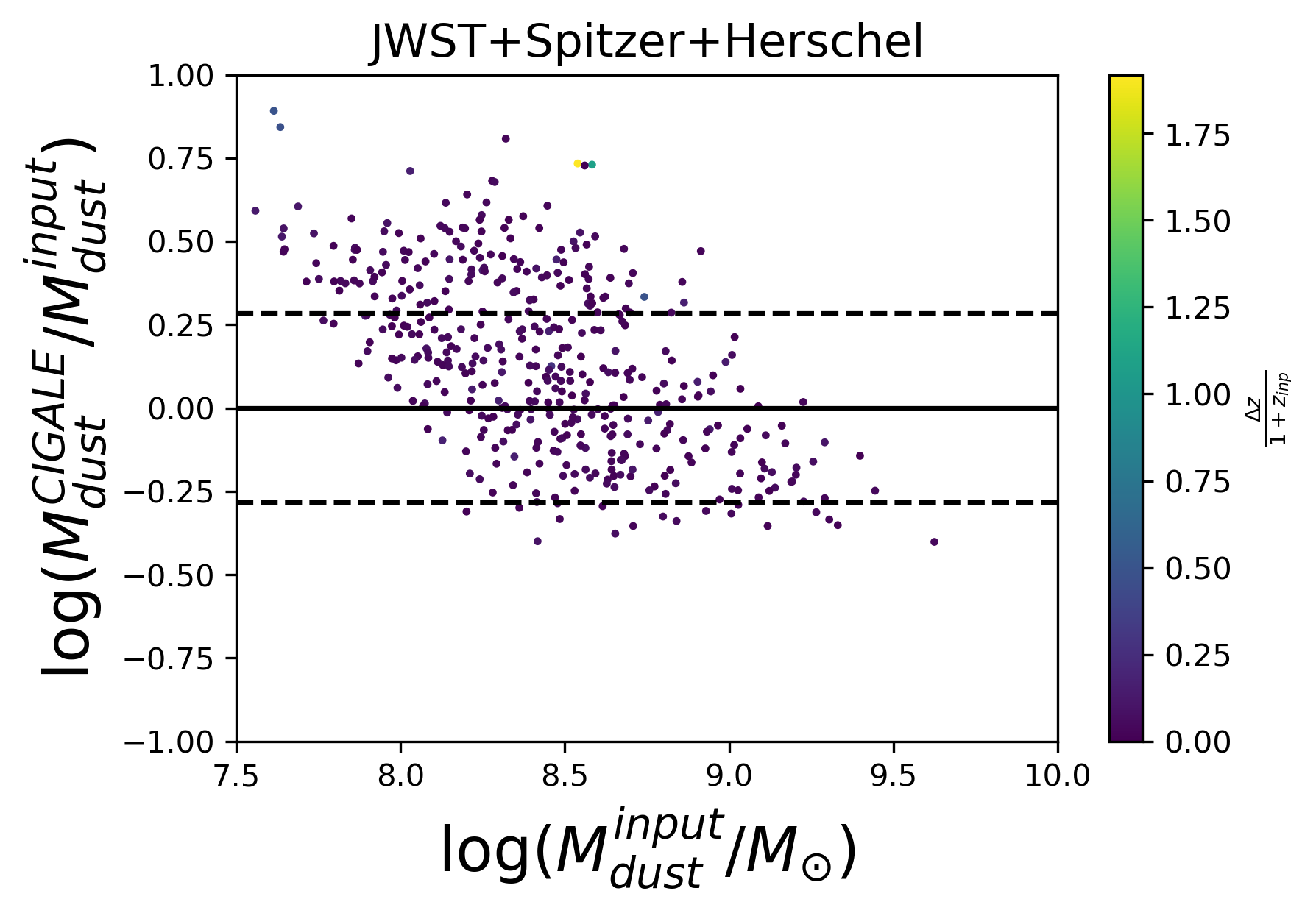}
\includegraphics[width=.4\textwidth]{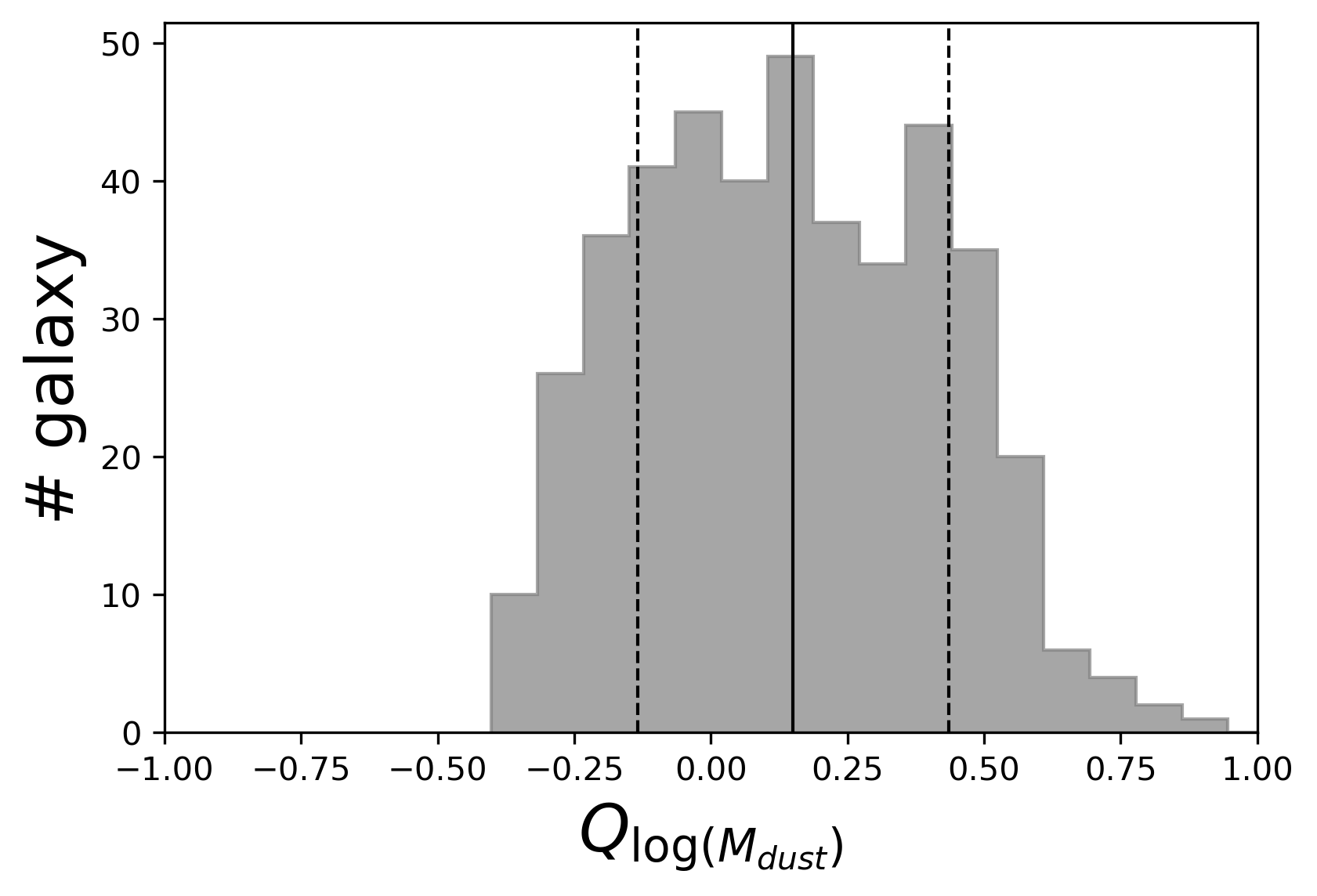}
\includegraphics[width=.42\textwidth]{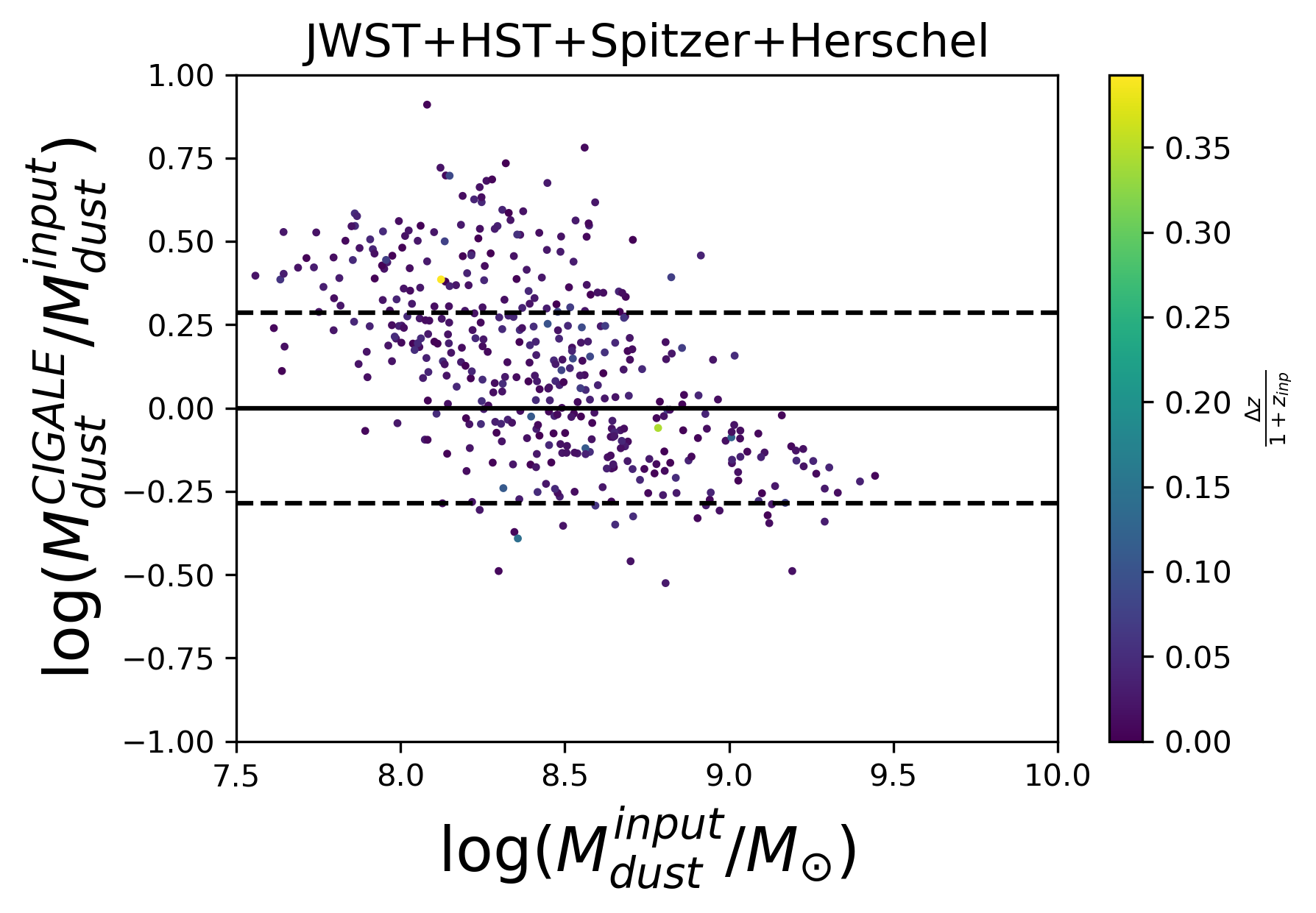}
\includegraphics[width=.4\textwidth]{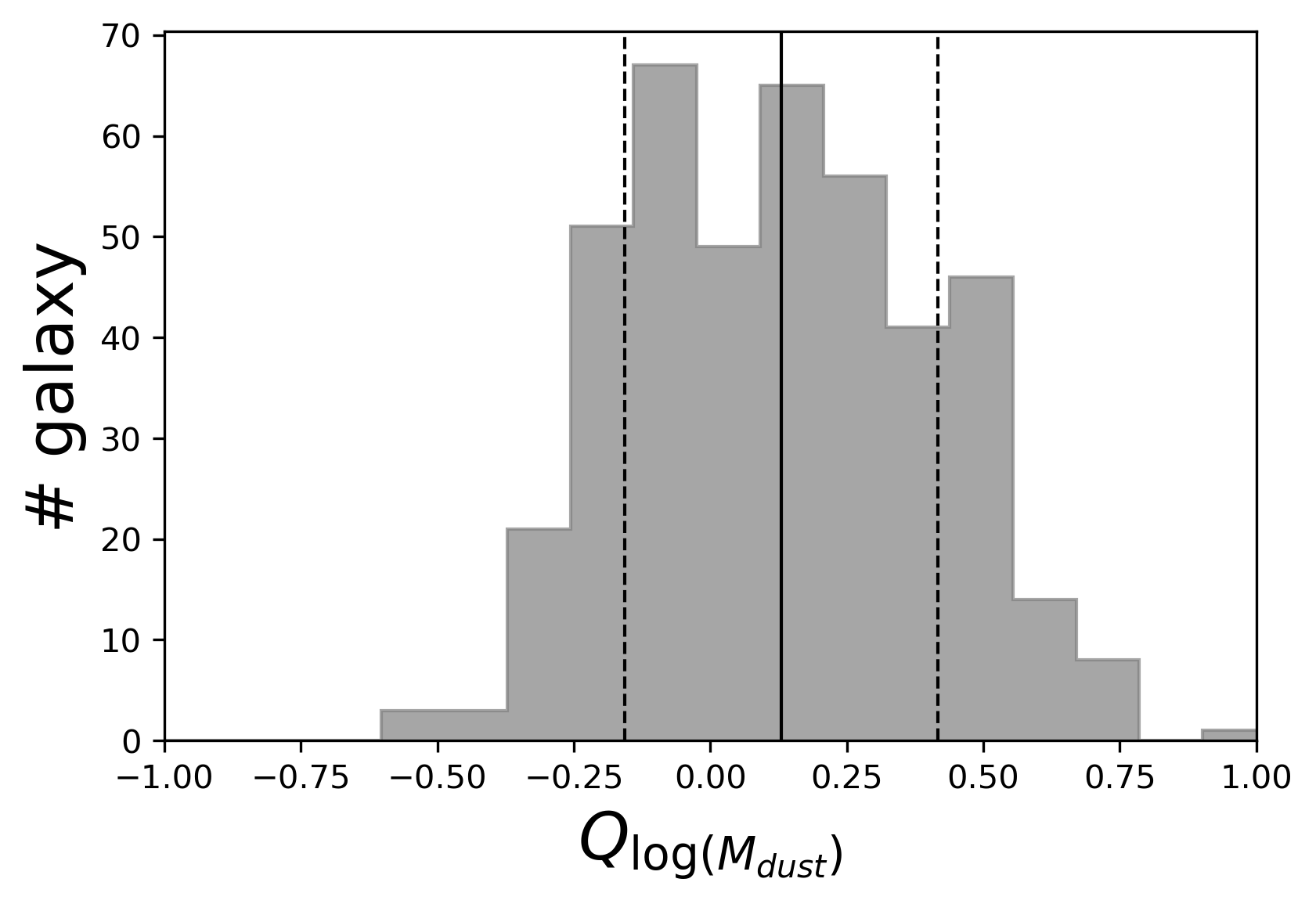}
      \caption{Same as Figure \ref{figcigsfr}, but for the dust mass ($M_{\rm dust}$).}
      \label{figcigdm}
\end{figure*}

\begin{figure}
    \centering
\includegraphics[width=.47\textwidth]{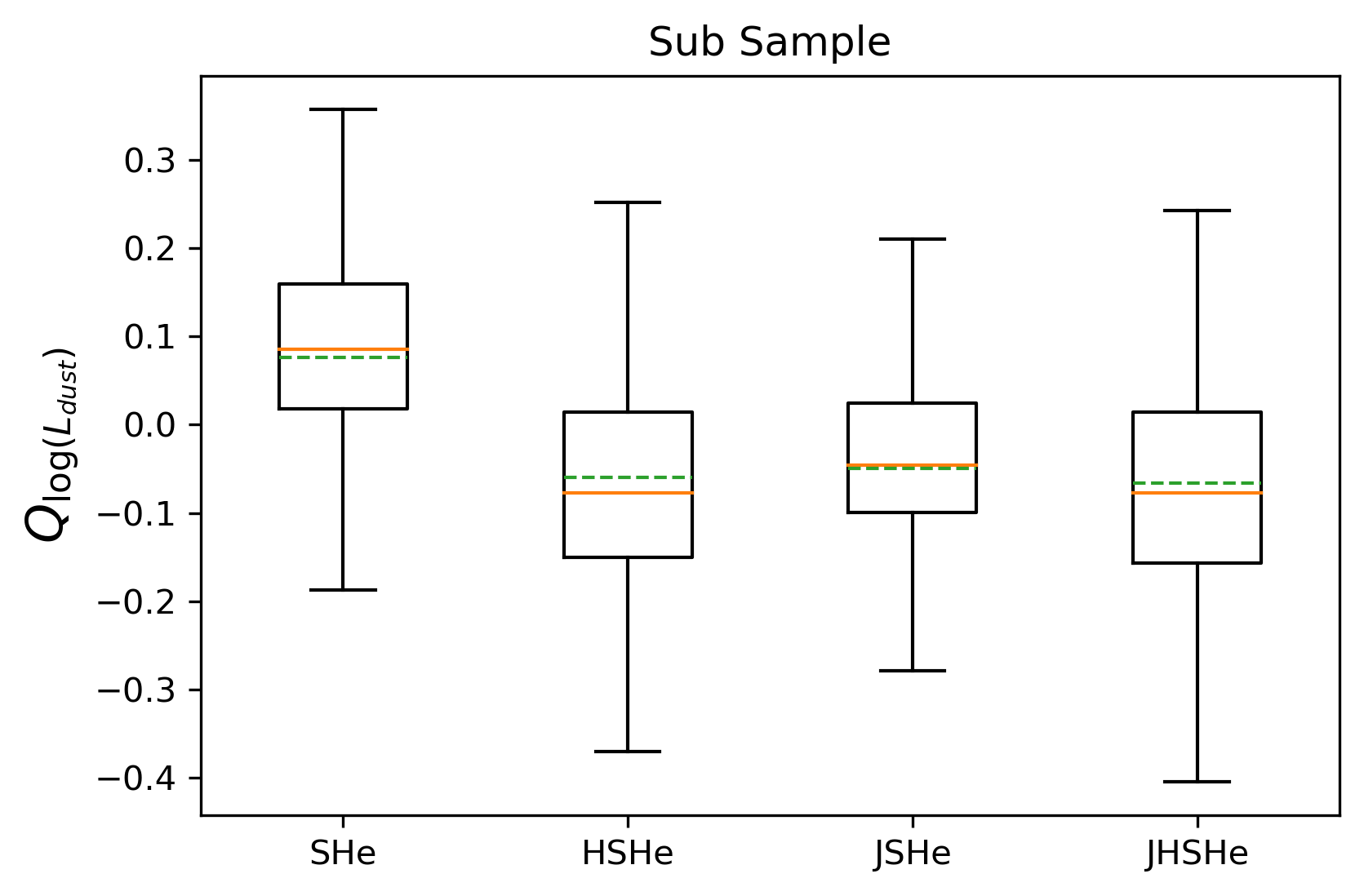}
\includegraphics[width=.47\textwidth]{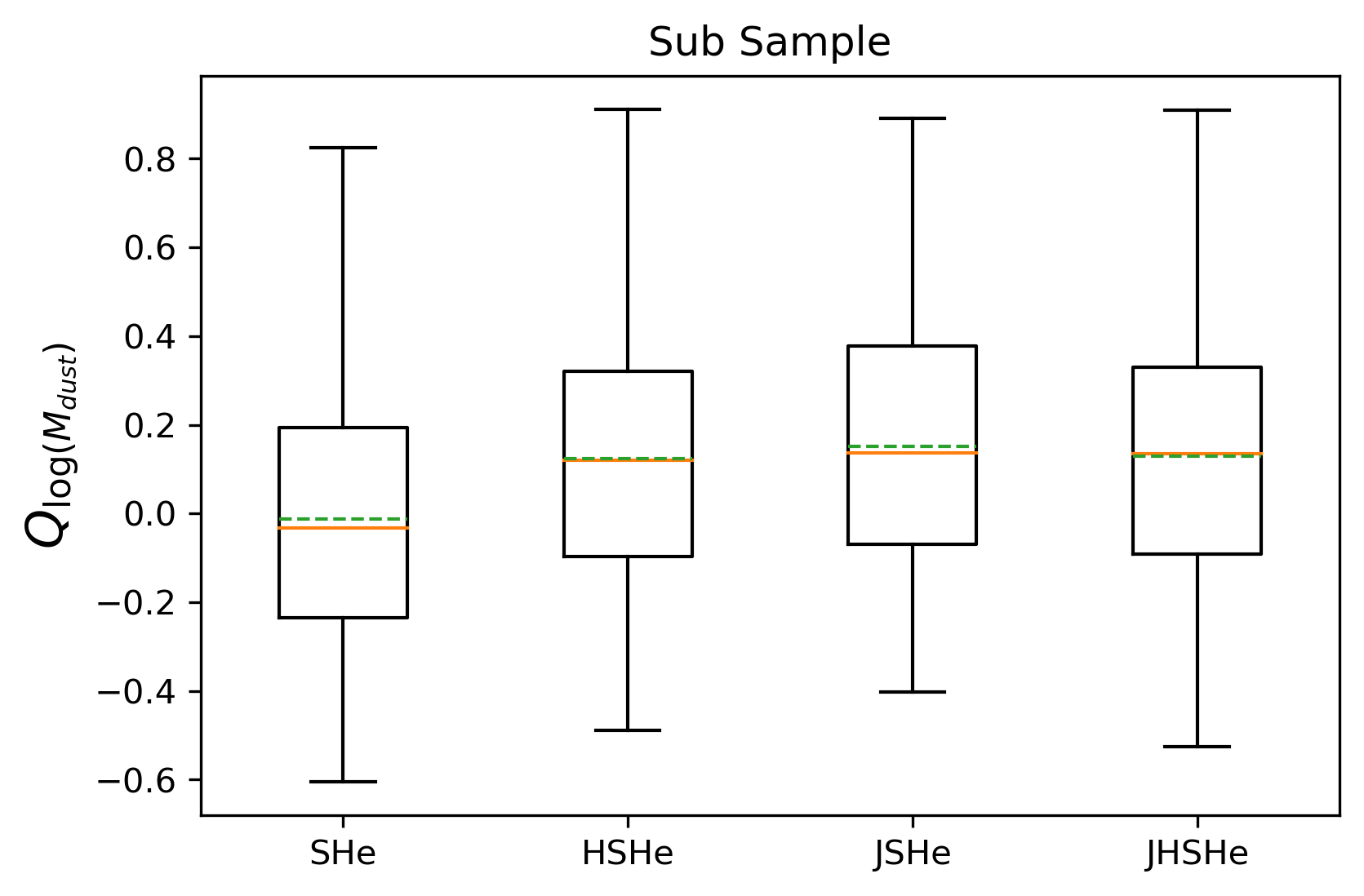}
      \caption{Same as Figure \ref{figboxsfrsub}, but for the dust luminosity ($L_{\rm dust}$, top panel) and dust mass ($M_{\rm dust}$, bottom panel).}
      \label{figboxdustsub}
\end{figure}

The dust luminosity and the dust mass of our simulated DSFGs are computed using different combinations of photometric data: \textit{Spitzer}+\textit{Herschel}, \textit{HST}+\textit{Spitzer}+\textit{Herschel}, \textit{JWST}+\textit{Spitzer}+\textit{Herschel} and \textit{JWST}+\textit{HST}+\textit{Spitzer}+\textit{Herschel}. Figure \ref{figcigdl} shows the distribution of $\log\left(L_{\rm dust}^{\rm CIGALE}/L_{\rm dust}^{\rm input}\right)$ as a function of input $L_{\rm dust}$ along with the histogram of $Q_{\log(L_{\rm dust})}$. With all the different combinations of photometry, we could recover the dust luminosity quite accurately with the $1\,\sigma$ dispersion in $Q_{\log(L_{\rm dust})}$ being the least for \textit{JWST}+\textit{Spitzer}+\textit{Herschel}, with a value of $0.12$. The mean offset in this case is $-0.049$. Upon adding the HST data to the above, the $1\,\sigma$ dispersion increases to $0.13$ due to the photo-$z$ outliers, and the mean offset becomes $-0.066$. Combining only \textit{HST} data with FIR photometry, we get the dispersion value of $0.13$ and a mean value of $-0.06$. Only \textit{Spitzer}+\textit{Herschel} produces a mean value of $0.075$ and a dispersion of $0.15$. Overall, we observe that, upon adding UV/optical/NIR photometry, the dust extinction is better constrained, thus giving a more accurate estimate of $L_{\rm dust}$ as summarised in Figure \ref{figboxdustsub} (top panel). From the values of the estimated median dust luminosity (see Table \ref{tabmedianvalues}) we can categorise these galaxies as ultra-luminous infrared galaxies (ULIRGs). 

The the distribution of $\log\left(M_{\rm dust}^{\rm CIGALE}/M_{\rm dust}^{\rm input}\right)$ as a function of input $M_{\rm dust}$ along with the histogram of $Q_{\log(M_{\rm dust})}$ is shown in Figure \ref{figcigdm}. The dust mass estimation has a larger dispersion compared to the dust luminosity (see Figure \ref{figboxdustsub}, bottom panel). The $1\,\sigma$ dispersions in $Q_{\log(M_{\rm dust})}$ are 0.29, 0.27, 0.26 and 0.28 for the \textit{Spitzer}+\textit{Herschel}, \textit{HST}+\textit{Spitzer}+\textit{Herschel}, \textit{JWST}+\textit{Spitzer}+\textit{Herschel} and \textit{JWST}+\textit{HST}+\textit{Spitzer}+\textit{Herschel} photometry respectively, while the mean offsets are $-0.012$, $0.124$, $0.150$, $0.130$. Overall, CIGALE gives an overestimation of dust mass which can be due to the difference in the value of the reference emissivity of dust grains per unit mass, $\kappa_0$, adopted by the two SED fitting codes \citep{liao_alma_2024}. Moreover, \cite{Hunt_2019} while comparing various SED fitting models by fitting far-UV to sub-mm photometry data of 61 KINGFISH galaxies found that CIGALE overestimates $M_{\rm dust}$ as compared to MAGPHYS (based on \cite{da_cunha_simple_2008} SED formalism) due to the absence of agreement regarding dust opacity. 

\begin{figure}
    \centering
\includegraphics[width=.47\textwidth]{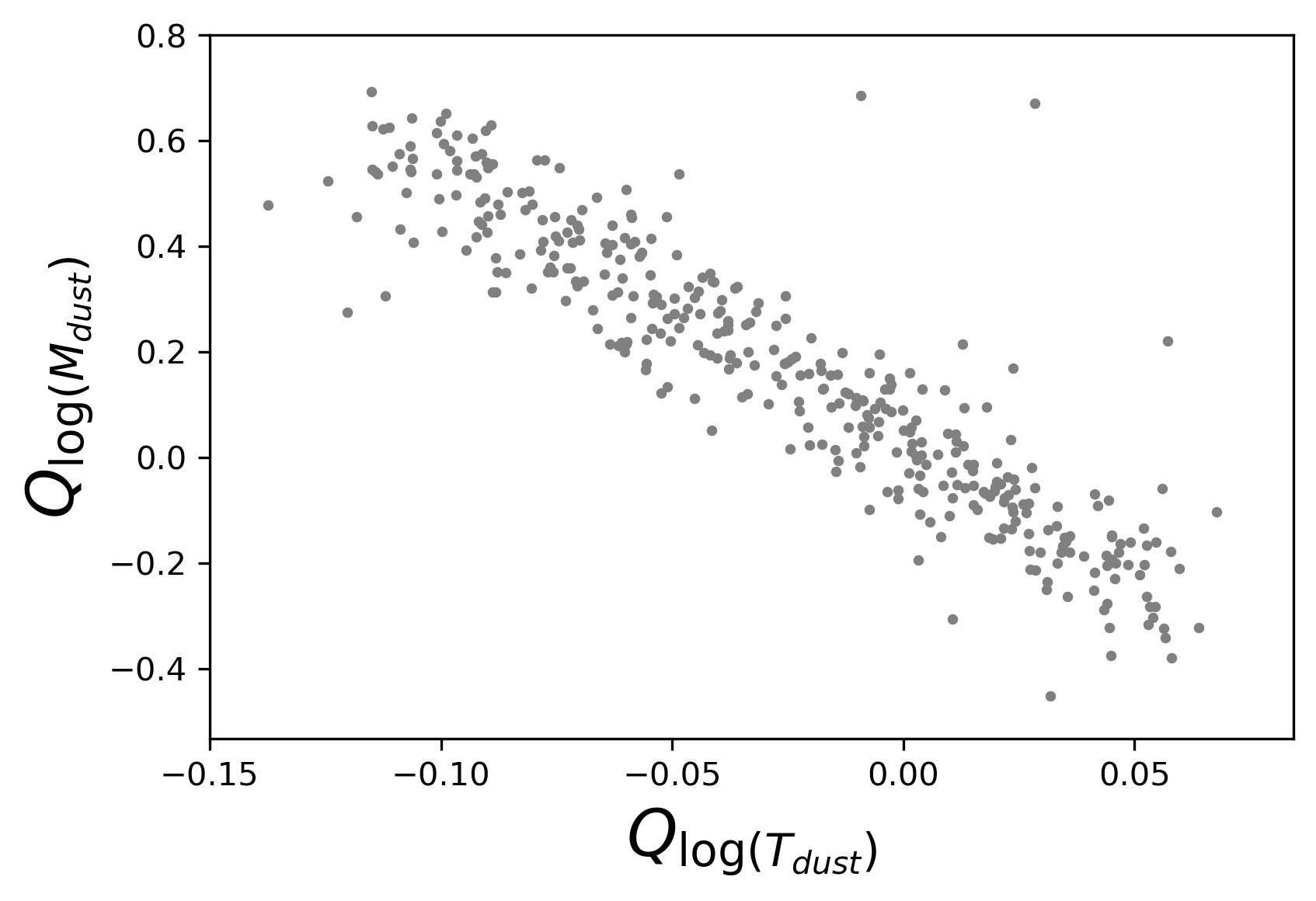}
      \caption{Scatter plot showing the correlation between $Q_{\log(M_{\rm dust})}$ and $Q_{\log(T_{\rm dust})}$. A clear negative trend is observed, highlighting the degeneracy between dust mass and dust temperature in SED fitting. This indicates that when CIGALE estimates a higher dust temperature relative to the input model, it compensates by assigning a lower dust mass, and vice versa.}
      \label{figdustmasstemp}
\end{figure}

Also, from Figure \ref{figcigdm} it is observed that there is a negative correlation between $\log\left(M_{\rm dust}^{\rm CIGALE}/M_{\rm dust}^{\rm input}\right)$ and $\log \left(M_{\rm dust}^{\rm input}\right)$. 
This negative correlation is likely due to the assumption of fixed dust absorption coefficient ($\kappa$) and emissivity index ($\beta$) by CIGALE. \cite{Bianchi_2022} pointed out the need for more realistic dust emission models that consider the temperature distribution and varying dust opacity across different environments. We note that the discrepancy between $M_{\rm dust}^{\rm CIGALE}$ and $M_{\rm dust}^{\rm input}$ is more prominent for $\log\left(M_{\rm dust}^{\rm input}\right)\lesssim8.5$, where CIGALE substantially overestimates $M_{\rm dust}$. For $\log\left(M_{\rm dust}^{\rm input}\right)\gtrsim9$ the negative correlation is not significant. To further explore this effect, we performed a test comparing the logarithmic ratios of dust mass and dust temperature estimated by CIGALE to those from our model. Our model uses a three-component dust temperature to calculate the dust mass whereas CIGALE models the dust emission using a range of interstellar radiation field intensities ($U$). To define a single dust temperature from our model, we calculate the mass-weighted dust temperature as
\begin{equation}
    T_{\rm dust}^{\rm inp} =\frac{\left(M_{\rm dust,\rm W}^{\rm BC}T_{w,\rm BC}+M_{\rm dust,\rm W}^{\rm ISM}T_{w,\rm ISM}+M_{\rm dust,\rm c}^{\rm ISM}T_{c,\rm ISM}\right)}{\left(M_{\rm dust,\rm W}^{\rm BC}+M_{\rm dust,\rm W}^{\rm ISM}+M_{\rm dust,\rm c}^{\rm ISM}\right)}
\end{equation}
For more details about the model calculation of dust mass and the meanings of the above symbols, we refer the readers to \cite{mitra_euclid_2024}. To calculate the dust temperature estimated by CIGALE, we use the best fit mean intensity $U_{\rm mean}$ in the relation from \cite{draine_andromedas_2014}
\begin{equation}
    T_{\rm dust}^{\rm CIGALE}=18U_{\rm mean}^{\frac{1}{6}}
\end{equation}
We then compute $Q_{\log(T_{\rm dust})}=\log\left(T_{\rm dust}^{\rm CIGALE}/T_{\rm dust}^{\rm inp}\right)$ and compute its correlation with $Q_{\log(M_{\rm dust})}$. We observe a clear negative correlation between $Q_{\log(M_{\rm dust})}$ and $Q_{\log(T_{\rm dust})}$ as shown in Figure \ref{figdustmasstemp}, consistent with the known degeneracy in SED fitting. This anti-correlation indicates that when CIGALE infers a higher dust temperature relative to our model, it tends to infer a lower dust mass, and vice versa. This is because, for a given FIR luminosity, higher temperatures require less dust to produce the same emission and models must compensate accordingly. This trend highlights that differences in dust temperature assumptions can significantly impact the inferred dust masses, even when the overall FIR luminosities are matched.

The estimated dust mass from the different photometry combinations of our simulated sources has a median value of $\log(M_{\rm dust}/M_{\odot})\gtrsim8.5$. Similar dust mass estimates were obtained by \cite{da_cunha_alma_2015} while studying with \textit{ALMA} a sample of 122 $870\,\mu$m-detected sources in the Extended Chandra Deep Field South. \cite{swinbank_alma_2014} also reported dust mass estimates consistent with ours while studying the far-IR properties of a sample of 99 SMGs in the Extended Chandra Deep Field South at $870$ $\mu\rm m$ with \textit{ALMA}.  In general, these galaxies have a higher dust content than the low-$z$ ($z<1$) galaxies, as was shown by \cite{rowlands_herschel-atlas_2014}.

\subsection{Comparison with physical properties of DSFG sources in the ASTRODEEP-\textit{JWST} catalog}
\label{astrodeep}

To enable a realistic comparison between observed and simulated DSFGs, we constructed a catalog of DSFG counterparts based on the ASTRODEEP-{\it JWST} photometric catalog by applying near-infrared colour selection criteria optimized for identifying dusty galaxies, following the method described by \cite{Barger_2023}. These criteria are tailored to select DSFGs using only NIRCam bands, providing a practical strategy for identifying heavily dust-obscured galaxies in {\it JWST} surveys where far-IR data may be limited. In parallel, we applied the same selection criteria to our simulated DSFG subsample to generate a corresponding mock catalog that mimics the observational selection. For both the observed and simulated samples, we performed SED fitting using CIGALE, incorporating the same model assumptions and wavelength coverage, in order to derive physical parameters such as stellar mass and SFR. This procedure ensures a fair comparison between the simulated and observed populations under identical selection and fitting conditions, allowing us to assess the reliability of our modeling framework in reproducing the physical properties of {\it JWST}-selected DSFGs.

The ASTRODEEP-\textit{JWST} catalog provides photometric redshift estimates using EAZY and ZPHOT \citep{Fontana_2000} based on \textit{JWST} NIRCam and \textit{HST} photometry for sources in 6 extragalactic deep fields including GOODS-S, aimed at studying the high redshift Universe. The catalog comprises all sources detected at 5$\sigma$ in the NIRCam F444W band. The GOODS-S catalog consists of 73\,638 sources in the redshift range $0\leq z\leq 20$. As our simulations are limited to the range $1\leq z\leq 8$ (for more details see \cite{mitra_euclid_2024}), we considered only the 55\,028 ASTRODEEP sources within this range. 

\begin{figure}
    \centering
\includegraphics[width=.47\textwidth]{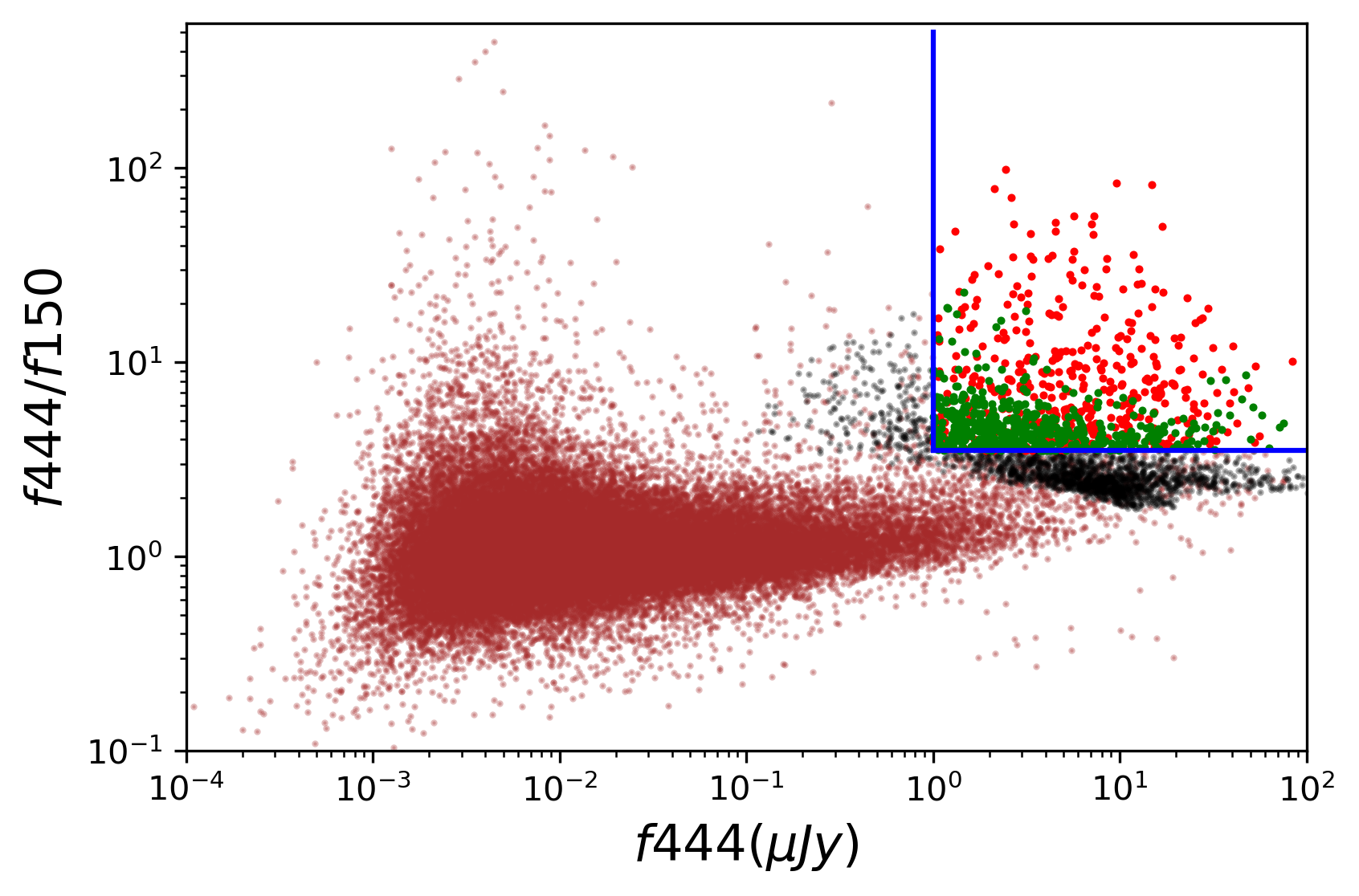}
      \caption{Color–flux diagram for NIRCam DSFG selection showing the ratio $f444/f150$ vs $f444$ (in $\mu$Jy) for galaxies in the simulated (black points) and ASTRODEEP (brown points) sample. Red points correspond to NIRCam DSFG candidates from ASTRODEEP sample, while green points represent NIRCam DSFG candidates from the simulated sample. The \cite{Barger_2023} selection criteria is shown in blue lines.}
      \label{figredNIRCamselection}
\end{figure}

The NIRCam \cite{Barger_2023} photometric criteria for selecting DSFG is defined as follows:
\begin{equation}
    S_{\rm F444W}/S_{\rm F150W}>3.5,\hspace{1ex}\\
    S_{F444W}> 1\,\mu{\rm Jy}
\end{equation}
By applying the above selection criteria to both the ASTRODEEP-{\it JWST} catalog and the simulated sample we create a sample of NIRCam DSFGs having 866 and 516 galaxies respectively. This colour selection along with the distribution of sources both in the ASTRODEEP-{\it JWST} catalog and the simulated sample is shown in Figure \ref{figredNIRCamselection}. 

For the estimation of physical properties of the ASTRODEEP NIRCam DSFGs sources we used the same modules of CIGALE specified in subsect.~\ref{secsoftawres}. For these ASTRODEEP sources, the estimated stellar mass and SFR have a median value of $\log(M_{\star}/M_{\odot})=10.12\pm0.3$ and $\log(\dot{M}_{\star}/M_{\odot}\hbox{yr}^{-1})=2.4\pm0.4$ respectively. For the simulated NIRCam DSFGs, the estimated stellar mass and SFR have a median value of $\log(M_{\star}/M_{\odot})=10.3\pm0.3$ and $\log(\dot{M}_{\star}/M_{\odot}\hbox{yr}^{-1})=2.0\pm0.3$ respectively.

\begin{figure*}

\centering
\includegraphics[width=.47\textwidth]{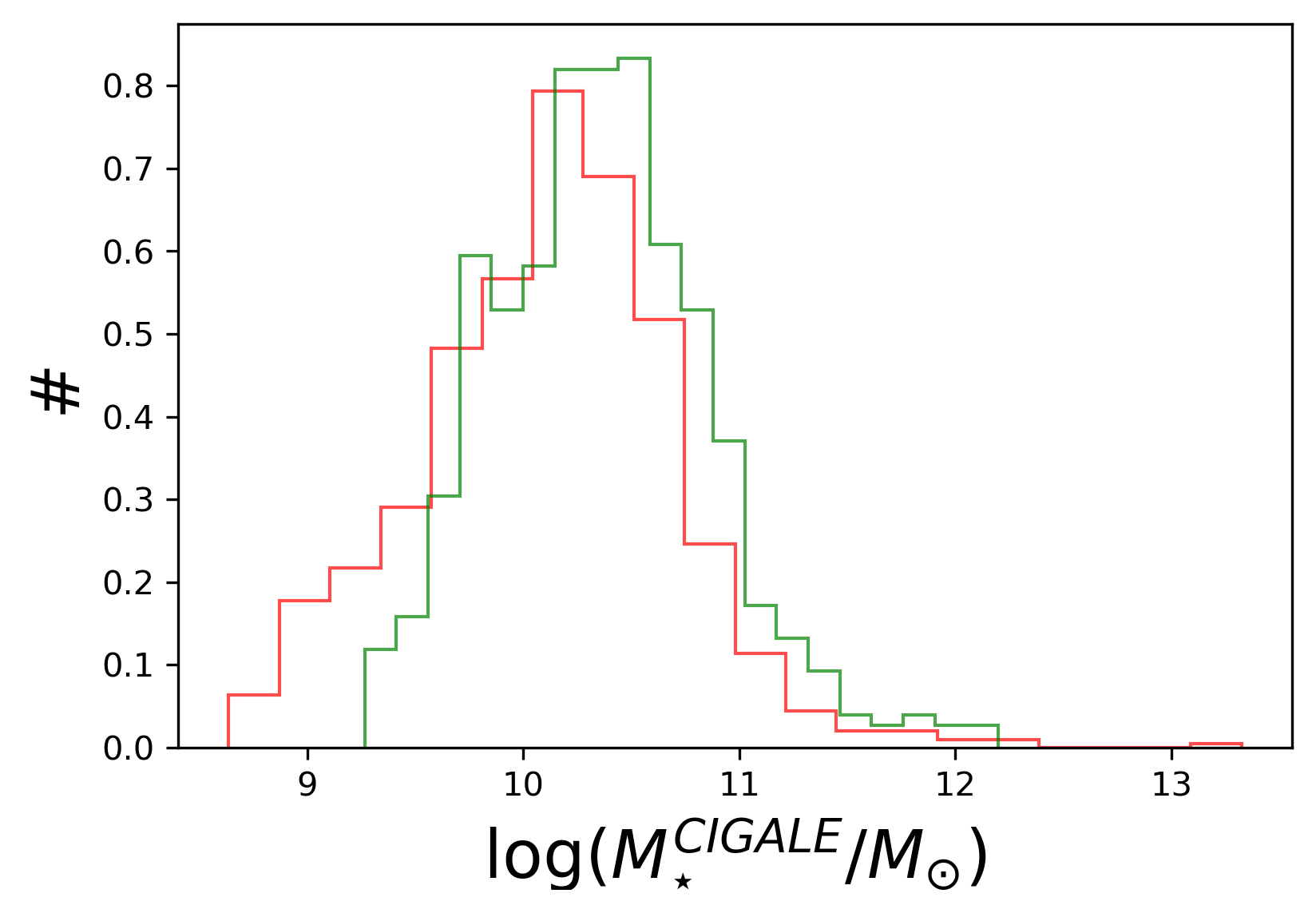}
\includegraphics[width=.47\textwidth]{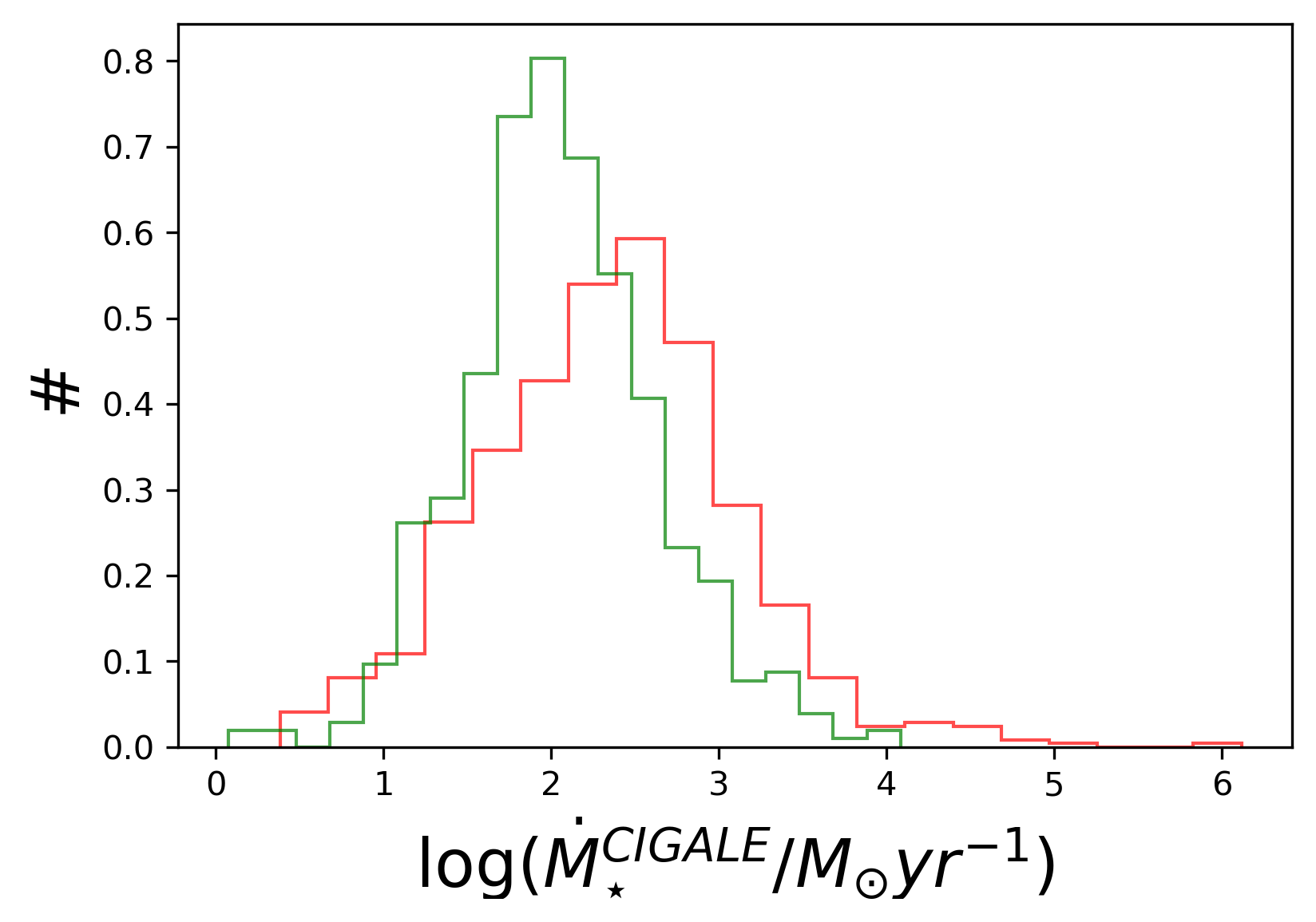}

\caption{Normalized distribution of stellar masses (left) and SFR (right) derived from CIGALE for the simulated (green) and the ASTRODEEP (red) NIRCam-selected DSFGs.}
\label{figastrosimcomp}
\end{figure*}

Figure \ref{figastrosimcomp} shows the histograms of stellar masses and SFRs of both the ASTRODEEP and the simulated DSFGs as a function of redshift. The distributions of stellar mass and SFR for the ASTRODEEP and the simulated sources are in good agreement.
This overall agreement between the simulated DSFG sample and the NIRCam-selected DSFGs from the ASTRODEEP catalog highlights the consistency of the simulation framework with observational data. Furthermore, this comparison underscores the efficacy of {\it JWST} in detecting DSFGs with stellar masses as low as $\sim10^{10}M_{\odot}$. 

\section{Summary and Conclusions}
\label{sec5}

We have investigated the ability of the \textit{JWST} to study the properties of $z>1.5$ progenitors of present-day massive spheroidal galaxies, using the JADES survey strategy in the GOODS-S field.
We also analysed the effect of complementing \textit{JWST} data with the data from previous surveys with \textit{HST}, \textit{Spitzer} and \textit{Herschel}. Our analysis is based on the physical model by \cite{cai_hybrid_2013} for the evolution of proto-spheroidal galaxies, as upgraded by \cite{mitra_euclid_2024}. We have found that our model is consistent with the recent observations of MIR LFs of these high$-z$ galaxies. 

Using the model, we simulated a sample of proto-spheroidal galaxies, which we called the parent sample, with $\log(M_{\rm vir}/M_{\odot})\gtrsim11.3$ and $z\ge 1.5$, over a survey area of $87.5$ sq. arcmin., similar to the area covered by JADES-Medium in the GOODS-S field. According to the model, at the $5\,\sigma$ detection limit in the 9 NIRCam filters chosen for JADES, \textit{JWST} detects 27748 galaxies of which about $41\%$ are also detected by \textit{HST} above $5\,\sigma$ in the F435W and F160W bands. In this sample, only about $\sim2\%$ of the \textit{JWST} sources are detected at $>5\,\sigma$ by \textit{Spitzer} and \textit{Herschel}. So, to include the photometry from the above IR instruments, we focused on the sub-sample of galaxies with $\log(M_{\rm vir}/M_{\odot})\gtrsim12$, as \textit{Herschel} did not detect DSFGs below that threshold anyway. From this sub-sample, we constructed a DSFG sample by selecting galaxies that are detected at $\geq 5\sigma$ at $250\,\mu$m with \textit{Herschel}. This yielded 507 simulated dusty galaxies. Among these, 503 galaxies ($\sim 99\%$) are also detected at $5\sigma$ in all nine NIRCam bands, and approximately 98.8\% are detected by \textit{HST}. Furthermore, 434 galaxies ($\sim 86\%$) in this DSFG sample are simultaneously detected at $>5\sigma$ in both \textit{Spitzer}/MIPS bands.

The first thing we estimated for our simulated DSFGs was the photometric redshift. For that, we used EAZY on the \textit{JWST} photometry alone as well as on the \textit{JWST}+\textit{HST} photometry. Then we estimated the physical properties using the code CIGALE and different combinations of photometric data: \textit{JWST}, \textit{JWST}+\textit{HST}, \textit{Spitzer}+\textit{Herschel}, \textit{HST}+\textit{Spitzer}+\textit{Herschel}, \textit{JWST}+\textit{Spitzer}+\textit{Herschel} and \textit{JWST}+\textit{HST}+\textit{Spitzer}+\textit{Herschel}. Our findings are as follows
\begin{enumerate}

\item We found that for $90\%$ of the sources detected by \textit{JWST} at $5\,\sigma$, EAZY gave a good estimate of the photometric redshift with the accuracy in $(1+z)$ being better than $15\%$. Catastrophic outliers in the $4\leq z \leq 6$ range were mostly due to the degeneracy between the Lyman break and the 4000 Å break. Upon addition of the HST photometry to the \textit{JWST} data, most catastrophic outliers were removed; in fact, by adding filters blueward of $0.7$ $\mu\rm m$, the Lyman break could be properly sampled. We got a more accurate estimate of redshift, with $<15\%$ discrepancy in (1+z) for $\gtrsim98\%$ sources detected by both \textit{JWST} and \textit{HST}. Catastrophic error estimates are then limited to the 1--2\% AGN-dominated sources. For the DSFG sample, {\it JWST} provided estimates of the photometric redshift with the accuracy in $(1+z)$ being better than $15\%$ for $\gtrsim95\%$ sources, which further increased to $\sim99\%$ upon adding {\it HST} photometry. 

\item Using the \textit{JWST} photometry alone, $\log\left(M_{\star}^{\rm CIGALE}/M_{\star}^{\rm input}\right)$ has a $1\,\sigma$ dispersion of $\approx0.2$ for the parent sample and $0.14$ for the sub-sample respectively. On removing the photo-$z$ outliers, 
the $1\,\sigma$ dispersion reduces to $\approx0.15$ and $0.1$ respectively. This clearly shows the direct effect of wrong photo-$z$ estimation on the estimation of stellar mass. As expected, adding the sub-mm photometry from \textit{Spitzer} and \textit{Herschel} produced little improvement on the stellar mass estimate. Adding the \textit{HST} photometry to that of \textit{JWST}, the stellar mass could be recovered relatively well with a $1\,\sigma$ dispersion as low as $0.14$ for the parent sample and $0.1$ for the sub-sample; also the mean offset between true and estimated stellar masses decreases. This shows that, as expected, the stellar mass is more sensitive to rest-frame UV/optical/near-IR wavelengths than to the far-IR/sub-mm part of the SED.


\item The SFR cannot be accurately derived from the \textit{JWST} photometry alone: the rms dispersion of  $Q_{\log(\dot{M}_{\star})}=\log(\dot{M}_\star^{\rm CIGALE}/\dot{M}_{\star}^{\rm input})$, is $0.55$. The addition of the \textit{HST} photometry leads to a modest improvement, but the dispersion remains high. Only FIR/sub-mm data from \textit{Spitzer}/MIPS and \textit{Herschel} (PACS and SPIRE) allowed us to decrease the dispersion in $Q_{\log (\dot{M}_{\star})}$ to 0.16--0.18. As expected, the availability of robust FIR/sub-mm photometry is crucial for constraining the SFR accurately. Unfortunately, these data are available only for a tiny fraction of the sample. On the other hand, we found that the FIR/submm photometry alone yields a dispersion of $0.22$, demonstrating that the addition of \textit{JWST} data significantly improves the accuracy. 
On the whole, the sampled sources have a median SFR $\log(\dot{M_{\star}}/M_{\odot}\hbox{yr}^{-1})\sim2.5$, clearly showing that these galaxies are going through intense star formation activity.

\item Dust luminosity and dust mass could only be estimated for sources having \textit{Spitzer} and \textit{Herschel} photometry. The estimated dust luminosity has a median value of $\log(L_{\rm dust}/L_{\odot})\sim12.5$, while the median value of the estimated dust mass is $\log(M_{\rm dust}/M_{\odot})\sim8.5$, showing that these DSFGs at $z\gtrsim1.5$ are highly luminous and are more heavily dust-enshrouded than $z\sim1$ galaxies. Again, the addition of \textit{JWST} data improves the accuracy. Estimates of dust masses are affected by the substantial uncertainty on the value of the reference emissivity of dust grains per unit mass. It is also affected by dust temperature assumptions.

\item Our estimates of stellar mass and SFR for the ASTRODEEP-\textit{JWST} sources demonstrate the capability of \textit{JWST} to detect high-$z$ galaxies with stellar mass almost an order of magnitude lower than possible in the pre-\textit{JWST} era. The comparison of the estimated physical properties of the ASTRODEEP sources and our simulated sources shows a good level of consistency between the two.
\end{enumerate}

Therefore, in this work, we have shown that, in the GOODS-S field, \textit{JWST} alone can well constrain the stellar mass of massive DSFGs. The estimates of the physical properties of DSFGs can be improved by exploiting the rich \textit{HST}, \textit{Spitzer} and \textit{Herschel} data available from previous surveys like CANDELS. To recover the IR properties of DSFGs, the \textit{JWST} and \textit{HST} data should be complemented with those from \textit{Spitzer} and \textit{Herschel}; however, deeper far-IR/sub-mm data are badly needed. Overall, it can be concluded that \textit{JWST} can enrich the study of DSFGs by producing tighter constraints on the physical properties and revealing fainter, lower-mass dusty galaxies previously inaccessible to {\it Herschel} or {\it Spitzer}.

This work clearly demonstrates the transformative role of \textit{JWST} in advancing our understanding of dusty galaxy populations at high redshift. Nonetheless, it also highlights the limitations imposed by the lack of sufficiently deep far-infrared and submillimetre observations. Future facilities such as the proposed \textit{PRIMA} or next-generation ground-based sub-mm observatories will be crucial for probing the peak of dust emission and disentangling the star formation and AGN components more robustly. Continued development of sensitive, high-resolution IR probes will further complement the capabilities of \textit{JWST}, enabling a more complete census of the dust-obscured universe and enhancing our ability to trace galaxy evolution across cosmic time. A critical problem for space-borne far-IR/sub-mm probes is angular resolution. A promising perspective is offered by far-IR interferometry, that can achieve sub-arcsec resolution \citep{Leisawitz2008, Bonato2024}.




\begin{acknowledgments}
DM acknowledges the postgraduate studentship provided by the UK Science and Technology Facilities Council (STFC). In this work, we have used the following Python packages: \textit{Astropy}\footnote{\url{https://www.astropy.org/}}\citep{2013astropy, 2018astropy, The_Astropy_Collaboration_2022}, \textit{Scipy}\footnote{\url{https://scipy.org/index.html}}\citep{jones2001scipy}, \textit{Numpy}\footnote{\url{https://numpy.org/}}\citep{van_der_Walt_2011, Harris_2020}, \textit{Joblib}\footnote{\url{https://joblib.readthedocs.io/en/stable/}}, \textit{COLOSSUS}\footnote{\url{https://pypi.org/project/colossus/}} \citep{diemer_colossus_2018}, \textit{Seaborn}\footnote{\url{https://seaborn.pydata.org/index.html}}\citep{2021JOSS....6.3021W} and \textit{Matplotlib}\footnote{\url{https://matplotlib.org/}}\citep{2007CSE.....9...90H}. The authors would like to thank the anonymous referee for the thorough review and constructive feedback, which helped to refine the manuscript.

\end{acknowledgments}

\begin{contribution}
DM was responsible for the analysis and modelling as well as the preparation and presentation of this manuscript. MN and GDZ contributed to the interpretation and the overall development of the manuscript. 
\end{contribution}

\bibliography{manuscript}{}
\bibliographystyle{aasjournal}

\end{document}